\documentclass[%
 reprint,
 amsmath,amssymb,
 aps,
]{revtex4-2}

\usepackage{graphicx}
\usepackage{dcolumn}
\usepackage{bm}
\usepackage{comment}
\usepackage{booktabs}
\usepackage{multirow}
\usepackage{color}

\begin{document}

\preprint{APS/123-QED}

\title{Solitary-wave loads on a three-dimensional submerged horizontal plate:\\Numerical computations and comparison with experiments}

\author{Tian Geng}
\email{geng.tian@ucdconnect.ie}
\affiliation{School of Mathematics and Statistics, University College Dublin, Belfield, Dublin 4, Ireland}

\author{Hua Liu}
\email{hliu@sjtu.edu.cn}
\affiliation{Department of Engineering Mechanics, Shanghai Jiao Tong University, Shanghai 200240, China}

\author{Frederic Dias}
\email{frederic.dias@ucd.ie}
\affiliation{ENS Paris-Saclay, CNRS, Centre Borelli, Universit\'e Paris-Saclay, Gif-sur-Yvette, 91190, France\\
School of Mathematics and Statistics, University College Dublin, Belfield, Dublin 4, Ireland}

\date{\today}

\begin{abstract}
A parallelized three-dimensional (3D) boundary element method is used to simulate the interaction between an incoming solitary wave and a 3D submerged horizontal plate under the assumption of potential flow. The numerical setup follows closely the setup of laboratory experiments recently performed at Shanghai Jiao Tong University. The numerical results are compared with the experimental results. An overall good agreement is found for the two-dimensional wave elevation, the horizontal force and the vertical force exerted on the plate, and the pitching moment. Even though there are some discrepancies, the comparison shows that a model solving the fully nonlinear potential flow equations with a free surface using a 3D boundary element method can satisfactorily capture the main features of the interaction between nonlinear waves and a submerged horizontal plate. 
\end{abstract}

\maketitle


\section{\label{sec:intro}Introduction}

\noindent Submerged horizontal plates are common coastal engineering structures that are used for several purposes. When located close to or at the mean water surface, they can act as effective breakwaters for offshore wave control and harbour protection as discussed by Yu \cite{Yu2002}. Meanwhile, a pulsating reverse flow may occur below these breakwaters under certain circumstances. Turbines can be put under the plate (Graw \cite{Graw1993}; Carter \cite{Carter2005}) to convert wave energy. A lot of other coastal structures, such as bridges, docks and very large floating structures, can be modeled as submerged plates in order to study the effects of storm surges, tsunamis and other extreme wave events.

Periodic wave scattering by submerged plates has been widely studied. Siew and Hurley \cite{Siew1977} provided first-order reflection and transmission coefficients for the scattering of long waves by a submerged plate. Patarapanich \cite{Patarapanich1984} studied forces and moments exerted on plates both experimentally and numerically using a finite element method. The effects of various parameters such as the ratio of plate length to wavelength and the submergence depth were investigated as well. Cheong \textit{et al}. \cite{Cheong1996} extended the eigenfunction expansion method to the complete range of water depths and compared reflection and transmission results with finite-element simulations. Dong \textit{et al.} \cite{Dong2018} used a modified matched eigenfunction expansion method to analyse wave scattering on a submerged horizontal plate over variable topography.

Due to their simplicity for experimental studies, solitary waves have been used extensively by researchers. Lo and Liu \cite{Lo2013} conducted shallow water experiments of solitary waves incident on a submerged plate. Experimental results were compared with numerical simulations and analytical solutions based on linear long wave theory. Strong vortices were observed near the trailing edges using particle image velocimetry (PIV). Seiffert \textit{et al}. \cite{Seiffert2014} conducted a series of laboratory experiments to investigate the forces exerted on a submerged plate by solitary waves.  Hayatdavoodi and Ertekin \cite{Ertekin2015} studied wave-induced loads due to solitary and cnoidal waves using Green-Naghdi theory and the influence of several parameters was discussed. Dong \textit{et al}. \cite{Dong2016} conducted experiments and simulated solitary-wave interactions with a submerged horizontal plate both on a flat bottom and on a sloping beach. Christou \textit{et al}. \cite{Christou2020} studied the influence of the angle of attack when a solitary wave propagates over a thin finite square plate. They used Hydro3D, an open source Large Eddy Simulation code. Xie \textit{et al}. \cite{Xie2020} used a multiphase flow model combined with the large-eddy simulation approach to investigate the interaction of a solitary wave with a thin submerged plate. Wang \textit{et al}. \cite{Wang2020} performed three-dimensional (3D) experiments and measured the spatial and temporal variation of the two-dimensional (2D) free-surface deformation using a multi-lens stereo reconstruction system. The hydrodynamic loads were measured by underwater load cells. Wave focusing induced by the plate led to an increased maximum elevation along the streamwise centerline of the plate. A 6-stage loading process based on the maxima of the vertical wave force and the pitching moment was proposed. One of the conclusions is that the vertical wave force on the plate is reduced compared to that obtained in previous 2D experiments. Although strong vortices were observed at the trailing edges of the plate, it is legitimate to ask the following question: can this problem only be solved by Computational Fluid Dynamics (CFD) or can fully nonlinear potential flow theory still be applied to this problem? Of course, it depends on the Reynolds number. In Xie \textit{et al}. \cite{Xie2020}, the Reynolds number based on the wave speed and plate thickness is approximately equal to $10^4$, which justifies the use of CFD methods. In the experiments of Wang \textit{et al}. \cite{Wang2020}, the Reynolds number is one order of magnitude larger ($\approx 10^{5}$) because of a thicker plate and a larger water depth. 

In the present paper, we first describe the numerical method in Section 2. The laboratory experiment is reviewed in Section 3.  Numerical results for the wave elevation, the vertical force and the moment are provided in Section 4.  They are compared with experimental results. Velocity fields are also shown. The effect of vortices that is neglected in the numerical simulations is discussed. In Section 5, we investigate the influence of the thickness of the plate. Additional results are provided as Supplementary Material. 



\section{Numerical Method}

The  fully nonlinear potential flow model with a free surface is used to solve the problem of a solitary wave impacting on a submerged horizontal plate. The fluid domain is denoted by $\Omega$, with boundary $\Gamma$. The boundary includes the free surface, the wavemaker, the bottom, the submerged plate and a vertical wall far downstream of the plate. 

\subsection{Mathematical formulation}

The velocity potential $\phi(\boldsymbol{x},t)$, where $\boldsymbol{x}=(x,y,z)$ is the vector of spatial coordinates with $z$ the vertical coordinate and $t$ is the time, is used to represent inviscid irrotational flows. The continuity equation in the fluid domain is Laplace's equation for $\phi$:
\begin{equation}
\nabla^2\phi=0 \,.
\end{equation}
We follow the approach described in Grilli \textit{et al}.'s \cite{Grilli2010}. 
The three-dimensional free space Green's function is defined as
\begin{equation}
G(\boldsymbol{x},\boldsymbol{x}_l)=\frac{1}{4\pi r} \,, \quad \frac{\partial G}{\partial n}(\boldsymbol{x},\boldsymbol{x}_l)=-\frac{1}{4\pi}\frac{\boldsymbol{r} \cdot \boldsymbol{n}}{r^3} \,,
\end{equation}
where $\boldsymbol{r}=\boldsymbol{x}-\boldsymbol{x}_l$ with $r=|\boldsymbol{r}|$ being the distance from the source point $\boldsymbol{x}$ to the collocation point $\boldsymbol{x}_l$, and $\boldsymbol{n}$ representing the normal unit vector pointing out of the domain at point $\boldsymbol{x}$.

\indent Green's second identity transforms Laplace's equation (1) into a integral equation on the boundary:
\begin{equation}
\alpha(\boldsymbol{x}_l)\phi(\boldsymbol{x}_l)=\int_\Gamma \left \{\frac{\partial\phi}{\partial n}(\boldsymbol{x})G(\boldsymbol{x},\boldsymbol{x}_l)-\phi(\boldsymbol{x})\frac{\partial G}{\partial n}(\boldsymbol{x},\boldsymbol{x}_l) \right \} d\Gamma \,,
\end{equation}

where $\alpha(\boldsymbol{x}_l)$ is proportional to the exterior solid angle made by the boundary at the collocation point $\boldsymbol{x}_l$.

\indent On the free surface, $\phi$ satisfies the nonlinear kinematic and dynamic boundary conditions, written in a mixed Eulerian-Lagrangian form, with the material derivative $D/Dt \equiv \partial/\partial t + \nabla \phi \cdot \nabla$:
\begin{align}
	&\frac{D\boldsymbol{x}}{Dt}=\nabla \phi \,, \\
	&\frac{D\phi}{Dt}=-gz+\frac{1}{2}\nabla \phi \cdot \nabla \phi \,,
\end{align}
where $\boldsymbol{x}$ is the position vector of a free-surface fluid particle and \textit{g} the acceleration due to gravity. The atmospheric pressure has been set equal to 0. In the case of wave generation by a wavemaker moving with velocity $\boldsymbol{U}$, the normal velocity is continuous over the surface of the wavemaker:

\begin{equation}
\frac{\partial \phi}{\partial n}=\boldsymbol{U} \cdot \boldsymbol{n} \,.
\end{equation}
At the bottom $\Gamma_{b} (t)$ and along other fixed parts of the boundary, the no-flow condition $\partial \phi / \partial n=0 $ is prescribed.

\subsection{Time integration}
Following the method implemented in Grilli \textit{et al}.'s \cite{Grilli2001} 3D model, second-order explicit Taylor series expansions are used to express both the new position and potential on the free surface. First-order coefficients are given by the boundary conditions (4) and (5). The pairs $\partial \phi / \partial t, \partial ^2 \phi /\partial t \partial n$ that are needed to obtain second-order coefficients are computed by solving another integral equation similar to equation (3). For the evaluation of the tangential derivatives, a fourth-order interpolation scheme is employed. 

The time-step is adapted by finding the minimum distance between two nodes on the free surface. Grilli \textit{et al.} \cite{Grilli1996} found an optimal value for the constant Courant number $C_0$ of roughly 0.4. In order to maintain the stability when strong nonlinear free surface deformations occur, an equally-spaced regridding method is adopted every 10 time steps, starting when the crest of the solitary wave arrives at the front edge of the plate. Lagrangian points would otherwise concentrate and eventually lead to a crash of the computations. In the literature, researchers use similar smoothing techniques to remove instabilities. For instance, Longuet-Higgins \textit{et al}. \cite{Longuet1976} used a filter every 5, 10 or 20 time-steps. Ming Xue \textit{et al}. \cite{Ming2001} applied a similar technique every $N_s$ ($N_s$ typically 3 or 6) steps. Grilli \textit{et al}. \cite{Grilli2001} and Fochesato and Dias \cite{Fochesato2006} also used a free surface node regridding method.

\subsection{Spatial discretization}
In this subsection, we follow closely Fochesato and Dias \cite{Fochesato2006}. The boundary is discretized into \textit{N} collocation nodes and \textit{M} high-order elements are used to interpolate in between \textit{q} of these nodes. Within each element, the boundary geometry and the field variables are discretized using polynomial shape functions $N_j(\xi,\eta)$, where $\xi$ denotes the intrinsic coordinates of the reference element:
\begin{equation}
	 \boldsymbol{x}(\xi,\eta)=\sum_{j=1}^q N_j(\xi,\eta)\boldsymbol{x}_j \,, \\
\end{equation}
\begin{equation}
	 \phi(\xi,\eta)=\sum_{j=1}^q N_j(\xi,\eta)\phi _j \,, \quad \frac{\partial \phi}{\partial n}(\xi,\eta)=\sum_{j=1}^q N_j(\xi,\eta)\frac{\partial \phi}{\partial n_j} \,,
\end{equation}
where the indices $j=1,\dots,q$ locally number the nodes within each element. We choose cubic reference elements $(q=16)$,  which provides $C_2$ continuity in between elements.

The integral equation (3) is transformed into a sum of integrals over the boundary elements, each one being calculated on the reference element. The change of variables is described by the Jacobian of the transformation $\boldsymbol{J}^i$ for the \textit{i}th element. Thus, the discretized form of the integrals can be written as
\begin{widetext}
\begin{equation}
	\int_\Gamma \frac{\partial \phi}{\partial n}(\boldsymbol{x})  G(\boldsymbol{x},\boldsymbol{x}_l) d\Gamma =
	\sum_{i=1}^M 
	\left \{ \int_{\Gamma_\xi,_\eta} \sum_{j=1}^q   \frac{\partial \phi}{\partial n}(\boldsymbol{x}_j) N_j(\xi,\eta) 
	G(\boldsymbol{x}(\xi,\eta),\boldsymbol{x}_l)|\boldsymbol{J}^i(\xi,\eta)| d\xi\ d\eta \right \} \,, 
\end{equation}
\begin{equation}
	\int_\Gamma \phi(\boldsymbol{x})  \frac{\partial G}{\partial n}(\boldsymbol{x},\boldsymbol{x}_l) d\Gamma =
	\sum_{i=1}^M 
	\left \{ \int_{\Gamma_\xi,_\eta} \sum_{j=1}^q   \phi(\boldsymbol{x}_j) N_j(\xi,\eta) 
	\frac{\partial G}{\partial n}(\boldsymbol{x}(\xi,\eta),\boldsymbol{x}_l)|\boldsymbol{J}^i(\xi,\eta)| d\xi d\eta\right \} \,.
\end{equation}
\end{widetext}
The associated discretized boundary integral equation leads to a sum on the \textit{N} boundary nodes,
\begin{equation}
	\alpha(\boldsymbol{x}_l)\phi(\boldsymbol{x}_l)=\sum_{j=1}^N \left (K_{lj}^D \frac{\partial \phi}{\partial n}(\boldsymbol{x}_j) - K_{lj}^N \phi(\boldsymbol{x}_j)   \right ) \,,
\end{equation}
where $l=1,\dots,N$ and $K_{lj}^D$, resp. $K_{lj}^N$, are Dirichlet, resp. Neumann, global matrices.

When the collocation node \textit{l} doesn't belong to the integrated element, a standard Gauss-Legendre quadrature is used. When it does belong to the element, \textit{r} becomes zero at one of the nodes and a self-adaptive singular quadrature \cite{Guiggiani1988} is implemented to handle the presence of the singularity on the boundary.

Instead of computing the diagonal coefficient of the Neumann matrix $K_{ll}^N$, the rigid mode technique is used:
\begin{equation}
\alpha(\boldsymbol{x}_l)+K_{ll}^N=-\sum_{j=1(\ne l)}^N K_{lj}^N \,, \quad l = 1, \dots, N, 
\end{equation}
which provides the diagonal term of a row by minus the sum of its off-diagonal coefficients. In terms of the intersecting parts of different boundaries, such as between the free surface and the lateral boundaries, the boundary conditions and the normal directions are generally different. Therefore, double-nodes are used to represent these corners and the continuity of the potential is imposed.

\subsection{Domain decomposition}
In order for the simulations to mimic as closely as possible the experiments, domain decomposition \cite{Wang1995} \cite{Haas1996} \cite{Bai2007} is used to boost the efficiency. The fundamental idea of domain decomposition is to divide the computational region into two or more sub-domains. The boundary integral equation is solved independently in each sub-domain. One major issue in domain decomposition is to satisfy continuity between adjacent sub-domains. 

Following the so-called D/D-N/N scheme introduced by De Haas \textit{et al}. \cite{Haas1996}, the computational domain is decomposed into the sub-domains $\varOmega _1$ and $\varOmega _2$ which are separated by an interface $\varGamma$. On the interface, the potential and its derivative are unknown, and an initial guess needs to be imposed. An iterative procedure is then used to get the exact potential or its derivative on the interface. The properties of Laplace's equation lead to continuous potential values and of their derivatives on the interface. The scheme can be extended straightforwardly to the case of more sub-domains.

The only issue in this iterative scheme is to deal with the case when the interface has a Dirichlet boundary. Because the double nodes share the same geometry and boundary condition between the free surface and the interface, their coefficients in the matrix will be exactly the same resulting in singular algebraic equations. To deal with this difficulty, the so-called semi-discontinuous elements were proposed in \cite{Bai2007}. The idea is to use different geometrical points to do the integration based on a discontinuity coefficient $\gamma$. It brings extra error in the simulation. As stated in \cite{Bai2007}, this method provides slightly better conditioned matrix equations, but the convergence is still slow when using an iterative solver. Here, we propose another way to resolve this difficulty.

The singularity occurs because the double nodes share the same Dirichlet boundary condition. Thus if we can somehow change one of the double nodes into a Neumann boundary condition, then the continuity of potential can be imposed again. Although the normal derivatives are unknown on the free surface, in the iterative procedure we can always get an inaccurate solution from the previous step. Therefore, when the interface is of Dirichlet boundary type, we can change one of the double nodes on the free surface into a Neumann boundary condition using the normal derivative from the previous step. This slightly modified scheme has been satisfactorily implemented in the numerical code.

In order to illustrate this new scheme, we first recapitulate the original algorithm \cite{Bai2007}:

(0) Choose an initial guess $\phi^k$ on the interface $\Gamma$, $(k=0)$;

(i) Solve Laplace's equation in each sub-domain to get ${\partial \phi_1^k}/{\partial n_1}$ and ${\partial \phi_2^k}/{\partial n_2}$ on $\Gamma$;

(ii) Take an average of the solutions (the normal vectors $n_1$ and $n_2$ are opposite):
\begin{align}
	\frac{\partial \phi^{k+1}}{\partial n}=\frac{1}{2}\left(\frac{\partial \phi_1^k}{\partial n_1}-\frac{\partial \phi_2^k}{\partial n_2}\right);
\end{align}

(iii) Solve Laplace's equation with Neumann boundary condition on the interface to get $\phi_1^{k+1}$ and $\phi_2^{k+1}$;

(iv) Take an average of the solutions:
\begin{align}
	\phi^{k+2}=\frac{1}{2} \left(\phi_1^{k+1}+\phi_2^{k+1}\right);
\end{align}

(v) Calculate the maximum error $\epsilon^{k+2}$ on $\Gamma$:
\begin{align}
	\epsilon^{k+2}=\max \left|\phi_1^{k+1}-\phi_2^{k+1}\right|;
\end{align}

(vi) If $\epsilon^{k+2}$ satisfies a prescribed criterion, then exit the iteration. Otherwise, go to step (1) and repeat with $k=k+2$. 

\begin{figure}[ht]
\includegraphics[width=0.65\columnwidth]{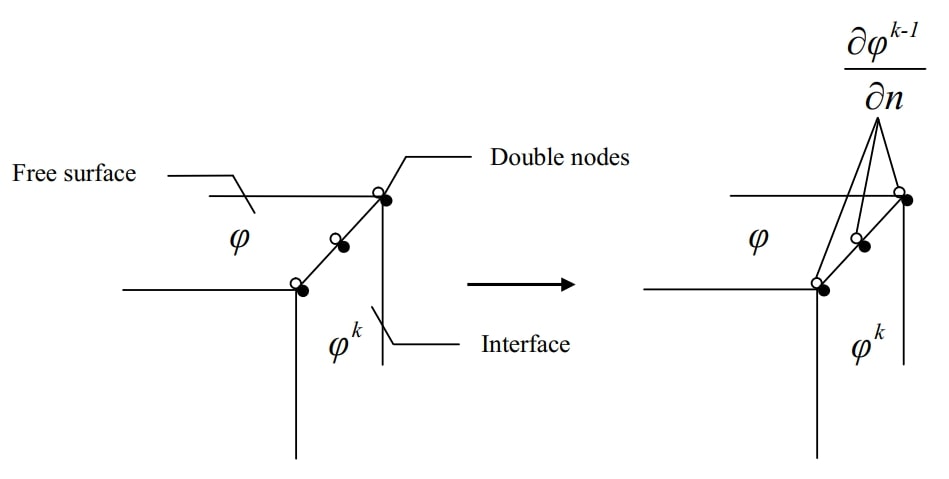}
\caption{\label{fig_dd} Illustration of the new scheme implemented in the numerical code. The interface is the common surface shared by adjacent sub-domains. Double nodes are applied on the intersection line between the interface and the free surface. Hollow circles denote the nodes on the free surface and filled circles represent the nodes on the interface. Here a Dirichlet boundary condition is applied on the interface. Therefore, the double nodes on both surfaces have the same Dirichlet boundary condition, which leads to a singularity. The new scheme resolves this issue by imposing a Neumann boundary condition on the double nodes on the free surface, using the values from the previous time step.}
\end{figure}

The problem mentioned above occurs at step (i) since the interface and the free surface both share Dirichlet boundary conditions. As shown in Fig \ref{fig_dd}, double nodes are used on the intersection of the free surface and the interface. We can change those double nodes on the free surface into a Neumann boundary condition. In the initial step (0), the normal derivative we need can be obtained from the solution of the previous time step. In step (iii), when we solve Laplace's equation with a Neumann boundary condition on the interface, we get normal derivatives on the free surface. Those values can be used during the iteration. 

\begin{figure}[ht]
\includegraphics[width=0.65\columnwidth]{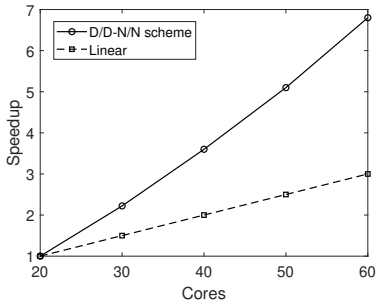}
\caption{\label{fig_speedup} Speed up of the domain decomposition method. A linear speed up is indicated for comparison.}
\end{figure}

The main advantage of this domain decomposition method is that it is superlinear -- see Table \ref{table_speedup} and Fig \ref{fig_speedup}. The reason is that the assembly of the full matrix is $O(N^2)$ and here we use a direct solver which is $O(N^3)$. We need to mention that in this scheme a few extra nodes are added on the interface.

\begin{table}[ht]
 \caption{Cores information and CPU time related to the domain decomposition method}
  \centering
  \begin{tabular}{llllll}
    \toprule
               
    Cores & 20 & 30 & 40 & 50 & 60  \\
	Time(s) & 66957 & 30075 & 18649 & 12921 & 9791 \\
	Speedup & 1 & 2.22 & 3.59 & 5.18 & 6.83 \\
    \bottomrule
  \end{tabular}
  \label{table_speedup}
\end{table}

\section{Description of the experiment}
\subsection{Experimental setup}
\begin{figure*}[ht]
\includegraphics[width=1.6\columnwidth]{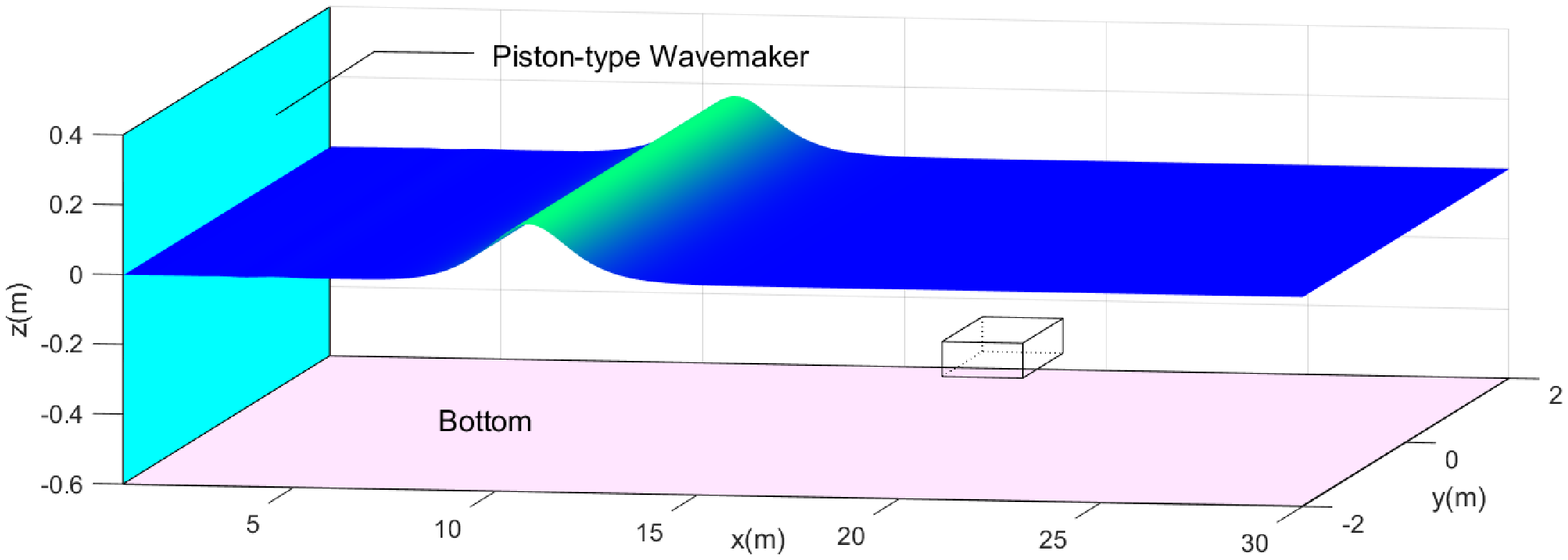}
\caption{\label{fig_overview} Overview of the wave flume and plate used for the laboratory experiments. A solitary wave is generated by the wavemaker. It then propagates along the flume and passes the plate.}
\end{figure*}

\begin{figure}[ht]
\includegraphics[width=0.65\columnwidth]{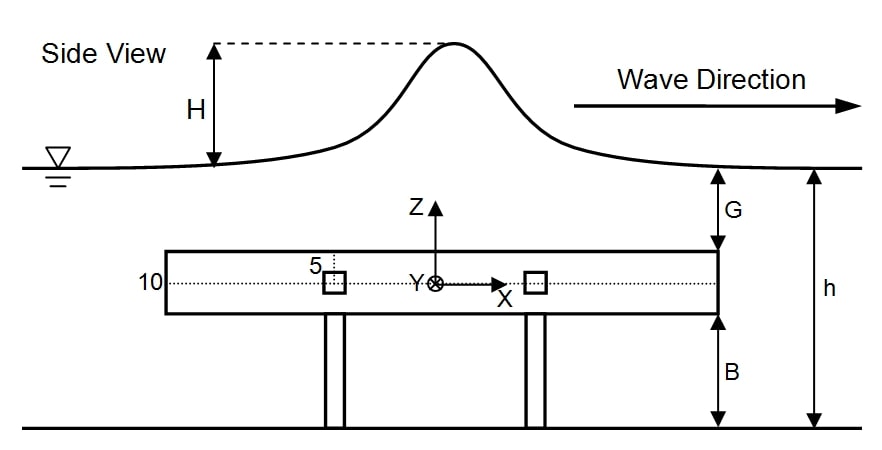}
\caption{\label{fig_sideview} Side view of the wave flume and plate used for the laboratory experiments. The plate is 200 cm long, 78 cm wide and 10 cm thick. The parameters that can be adjusted in the experiments are the wave height $H$, the water depth $h$, the distance from the bottom to the lower surface of the plate $B$ and the depth of submergence of the plate $G$.}
\end{figure}

The experiments were conducted in the Tsunami Basin for Offshore Regions in Shanghai Jiao Tong University \cite{Wang2020}. The wave flume is 42 m long and 4 m wide with a piston type wavemaker installed at one end. The plate-type structure is made of organic glass. It is 200 cm long, 78 cm wide, 10 cm thick and mounted on the bottom in the middle region of the flume (Fig \ref{fig_overview} and Fig \ref{fig_sideview}). Four piezoelectric force balance units are installed inside the plate. A dynamometric system is used to measure wave loads on the structure. The constant water depth is denoted by $h$. 

\begin{figure}[ht]
\includegraphics[width=0.65\columnwidth]{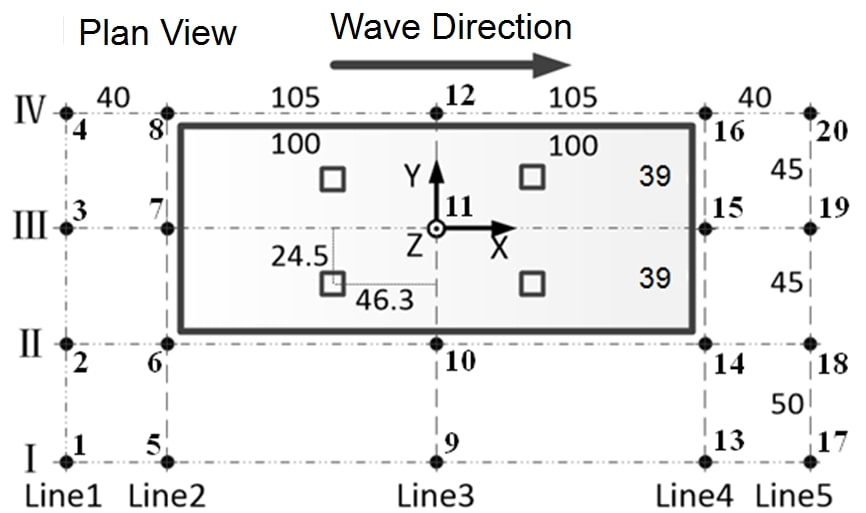}
\caption{\label{fig_topview} Top view of the wave flume and plate used for the laboratory experiments (all distances are in cm). The locations of the twenty wave gauges, labelled 1 to 20, are shown. They lie along four lines in the direction of the wave (I, II. III, IV) and five lines across the wave flume (1, 2, 3, 4, 5).}
\end{figure}

The free-surface elevation is measured by resistance-type wave gauges. Twenty wave gauges are spread around the plate. Their exact locations and labels are shown in Fig \ref{fig_topview}. Taking the center of the plate as the origin (0,0) of the $(x,y)-$plane, we list the $(x,y)-$coordinates of all the wave gauges (WG) in Table 1. s

\begin{table*}[ht]
 \caption{Coordinates of the twenty wave gauges (in cm). The coordinates $(0,0)$ correspond to the center of the plate.}
  \centering
  \begin{tabular}{llllllllll}
    \toprule
               
    WG1 & ($-$145,$-$95) & WG5 & ($-$105,-95) & WG9 & (0,$-$95) & WG13 & (105,$-$95) & WG17 & (145,$-$95) \\
	WG2 & ($-$145,$-$45) & WG6 & ($-$105,$-$45) & WG10 & (0,$-$45) & WG14 & (105,$-$45) & WG18 & (145,$-$45) \\
  	WG3 & ($-$145,0) & WG7 & ($-$105,0) & WG11 & (0,0) & WG15 & (105,0) & WG19 & (145,0) \\
	WG4 & ($-$145,45) & WG8 & ($-$105,45) & WG12 & (0,45) & WG16 & (105,45) & WG20 & (145,45) \\
    \bottomrule
  \end{tabular}
  \label{table1}
\end{table*}

\subsection{Wave generation}

Solitary waves are generated by a piston-type wavemaker on the left side of the wave tank. A third-order solitary wave profile \cite{Grimshaw1971} is defined by (note that the wave profile is uniform across the flume, so it is given as a function of $x$ only)
\begin{eqnarray}
    \eta (x,t)=&&H \, \rm{sech}^2 [  1-\frac{3}{4} \alpha \, \tanh^2 \nonumber\\
    && + \alpha ^2 \left (\frac{5}{8} \, \tanh^2-\frac{101}{80} \, \rm{sech}^2 \, \tanh^2\right )  ] \,,
\end{eqnarray}

\begin{align}
	&k=\sqrt{\frac{3 \alpha}{4h^2}} \left (1-\frac{5}{8} \alpha +\frac{71}{128} \alpha ^2    \right ) \,, \\
	&c=\sqrt{gh} \left ( 1+\frac{\alpha}{2}-\frac{3}{20} \alpha ^2 +\frac{3}{56}\alpha ^3   \right ) \,,
\end{align}
where $\eta$ is the wave elevation, $H$ the wave height, $h$ the still water depth, $\alpha=H/h$, $g$ the acceleration due to gravity and $c$ the wave celerity. The symbols sech, resp. tanh, denote sech$(k(ct-x))$, resp. tanh$(k(ct-x))$.

In reference to the improved Goring \textit{et al.} \cite{Goring1980} wave generation method introduced by Malek-Mohammadi \textit{et al.} \cite{Malek2010}, the horizontal velocity of the piston is determined by
\begin{equation}
	U(x,t)=\frac{c_w \eta(x,t)}{h+\eta(x,t)} \,,
\end{equation}
where $c_w=\sqrt{gh} \left [ 1+\frac{1}{2}(\eta/h)-\frac{3}{20} ({\eta}/{h}) ^2 +\frac{3}{56} ({\eta}/{h}) ^3   \right ] $. This wave generation method combined with the third-order solitary wave profile has the capability of generating steady solitary waves of dimensionless wave amplitude up to $H/h=0.5$ \cite{Xuan2013}. It is implemented in the numerical code.

\section{Results}

\begin{figure}[htbp]{
    \includegraphics[width=0.5\columnwidth]{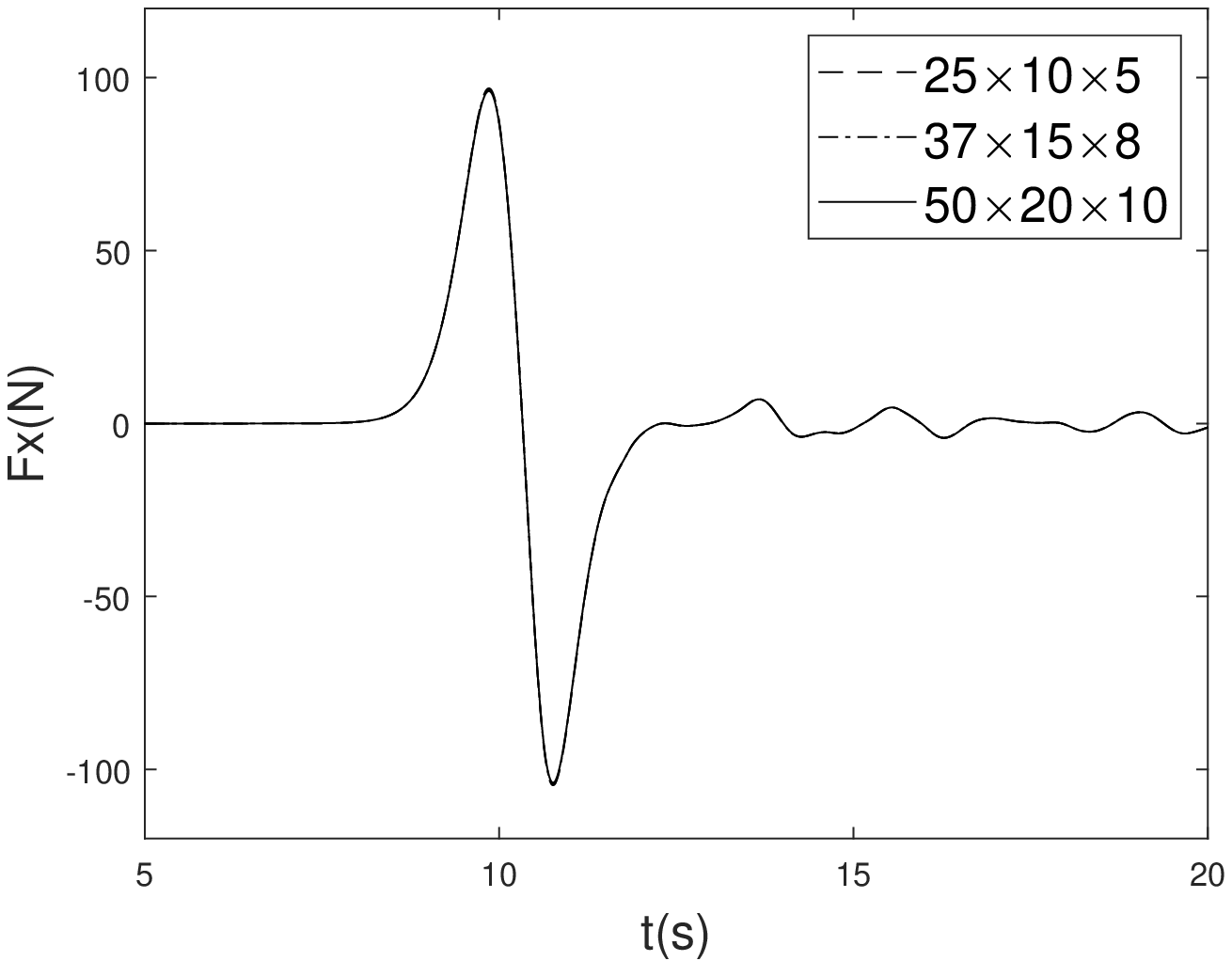}
    \includegraphics[width=0.5\columnwidth]{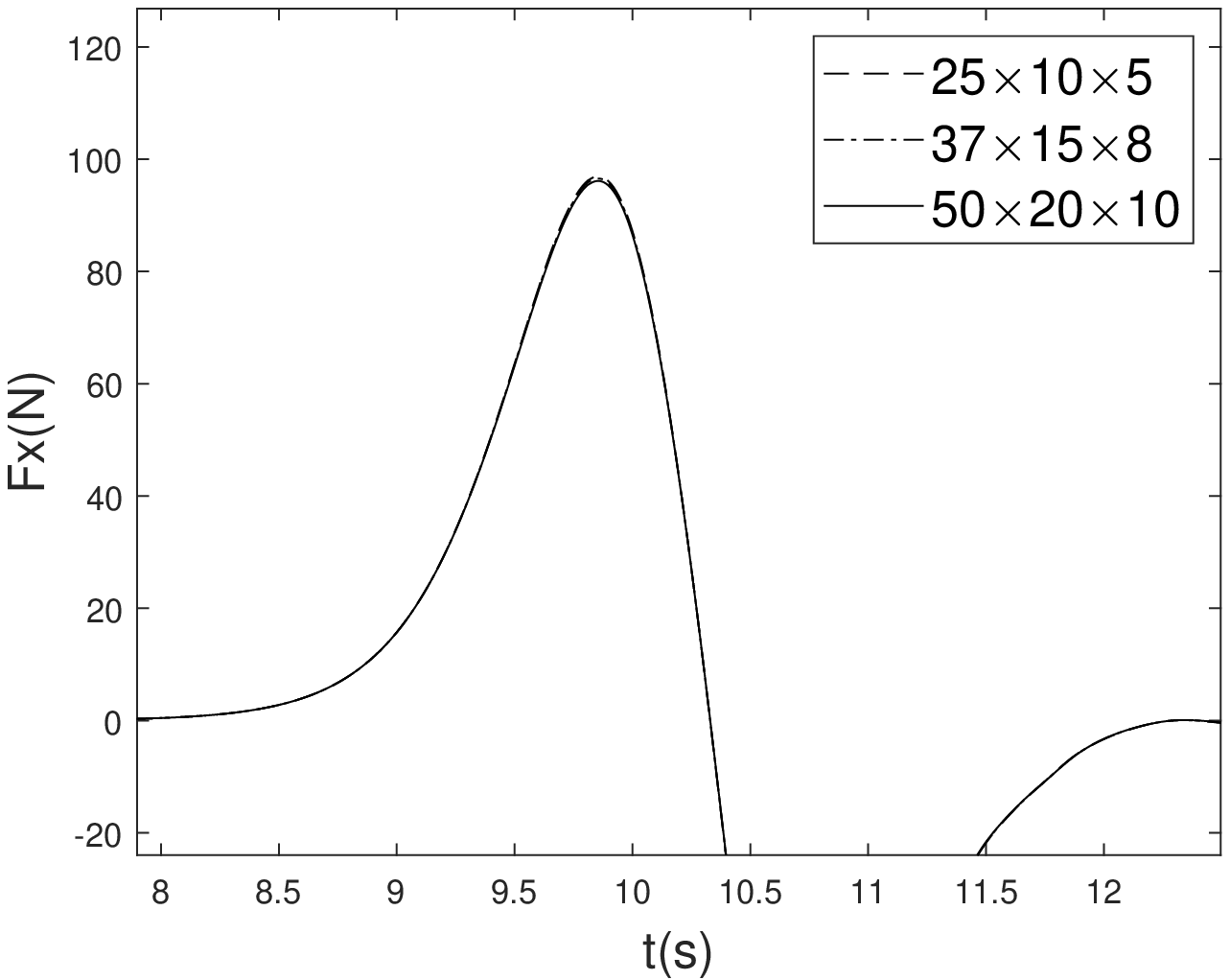}%
}
\caption{\label{fig_conv} Convergence tests for the mesh size along the plate when $B=40$ cm, $h=60$ cm, $H/h=0.3$. The horizontal force $F_x$  is compared for three sets of mesh sizes (an enlargement of the first peak of the left plot is shown in the right plot).}
\end{figure}

The size of the computational domain is exactly the same as that of the wave flume: 42 m long and 4 m wide with a moving boundary condition at the left end. At the bottom and along other fixed boundaries, the impermeable condition is applied. The right end is far way from the plate so that the reflected wave has no influence. The discretization used in the simulations is $300\times30\times10$ elements on the tank boundaries and $25\times10\times5$ on the plate boundaries along the $x$, $y$ and $z$ directions, respectively (i.e. 25450 elements and 26982 nodes).  The numbers of elements on the tank boundaries are chosen based on the finest mesh used in \cite{Grilli2001}. Additional convergence tests, shown in Fig \ref{fig_conv}, have been performed to check the level of refinement needed along the plate. As can be seen, refining the mesh further has a negligible effect on the hydrodynamic load. In Xie \textit{et al}. \cite{Xie2020}, where a CFD code was used, the computational domain is discretised by a uniform mesh $1600 \times 96 \times 160$. 

As shown in Fig \ref{fig_sideview}, $G$ is the submerged depth and $B$ the distance from the bottom to the lower surface of the plate. For the comparison between experiments and numerical simulations, 41 cases have been investigated. They correspond to various combinations of water depth, wave height and submerged depth that are listed in Table 3. In five of these combinations, breaking waves were observed in the experiments.

\begin{table}[ht]
\caption{List of the 41 cases investigated experimentally. They correspond to various combinations of water depth $h$, submerged depth $B$ and dimensionless wave height $H/h$. All dimensional quantities are expressed in cm.}
  \centering
\begin{tabular}{lll}
\hline
{\it B}(cm)               & {\it h}(cm) & {\it H/h}                       \\ \hline
\multirow{4}{*}{20} & 35    & 0.1, 0.2, 0.3, 0.4, 0.5 (break)       \\ 
                    & 40    & 0.1, 0.2, 0.3, 0.4, 0.5       \\ 
                    & 50    & 0.1, 0.2, 0.3, 0.4, 0.5       \\ 
                    & 60    & 0.1, 0.2, 0.3, 0.4           \\  \hline
\multirow{3}{*}{30} & 45    & 0.1, 0.2, 0.3  (break), 0.4  (break), 0.5  (break)   \\
                    & 50    & 0.1, 0.2, 0.3, 0.4, 0.5       \\
                    & 60    & 0.1, 0.2, 0.3, 0.4           \\ \hline
\multirow{2}{*}{40} & 55    & 0.1, 0.2, 0.3  (break), 0.4 (break) \\
                    & 60    & 0.1, 0.2, 0.3, 0.4           \\ \hline

\end{tabular}
\end{table}

We first compare the numerical results for the surface elevations, horizontal and vertical forces, and pitching moments with the experimental results. Then we present results for the velocity fields. Finally, the basic hydrodynamic loading process that occurs when the solitary wave propagates past the plate is analyzed. Numerical results for all cases can be found in the Supplementary Material. 

\subsection{Surface elevation}

From the top view of the numerical channel shown in Fig \ref{fig_topview} together with the locations of the wave gauges, it is seen that the wave gauges WG3, 7, 11, 15 and 19 lie on the middle line of the plate. The wave gauges WG2, 6, 10, 14, 18 and WG4, 8, 12, 16, 20 are symmetrically located on both sides of the plate. The wave gauges WG1, 5, 9, 13 and 17 lie further away from the plate. Although there are small discrepancies, the results of WG4, 8, 12, 16 and 20 are close to those of WG2, 6, 10, 14 and 18. Therefore they are not shown here.

\begin{figure*}[htbp]{
    \includegraphics[width=0.65\columnwidth]{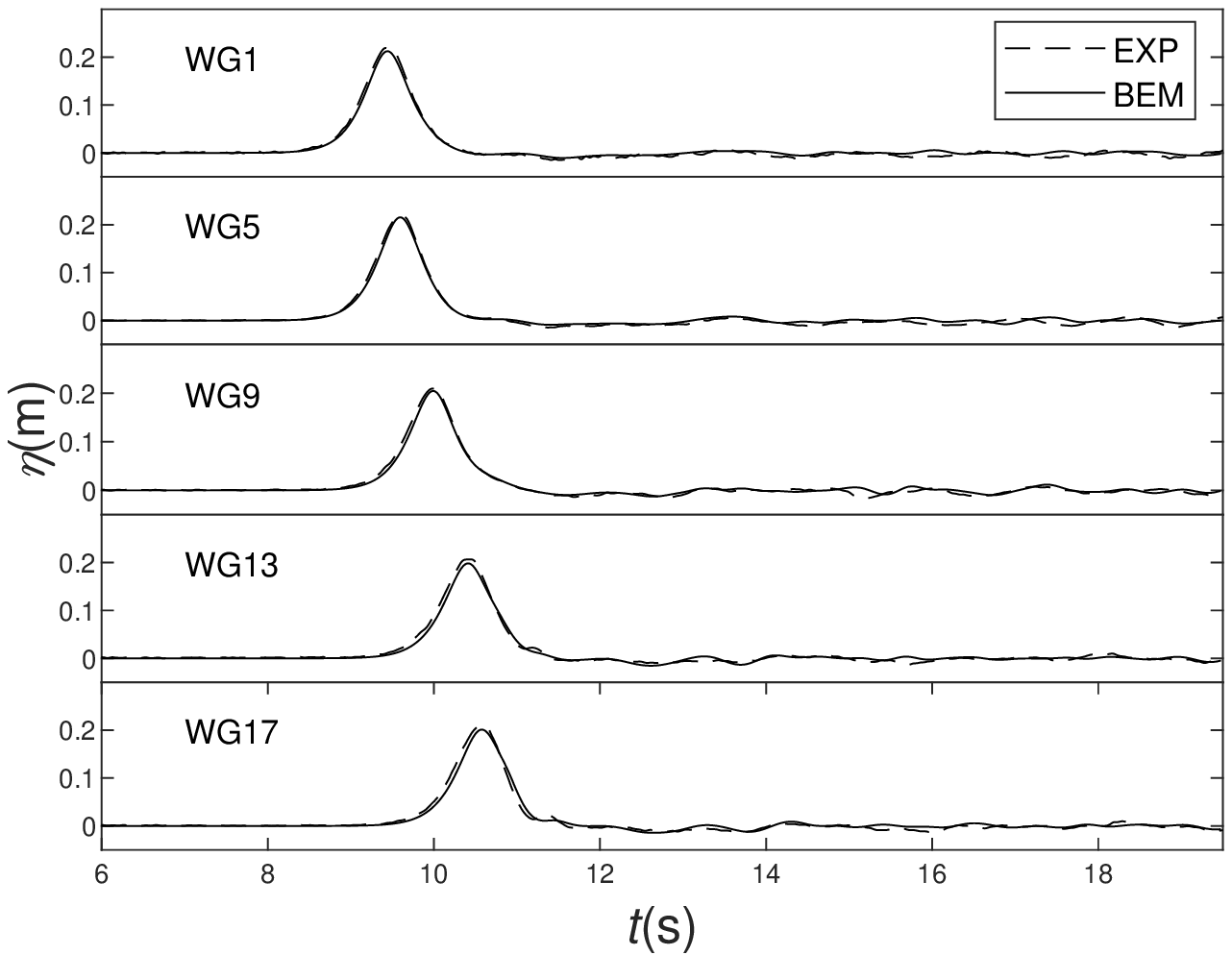}
    \includegraphics[width=0.65\columnwidth]{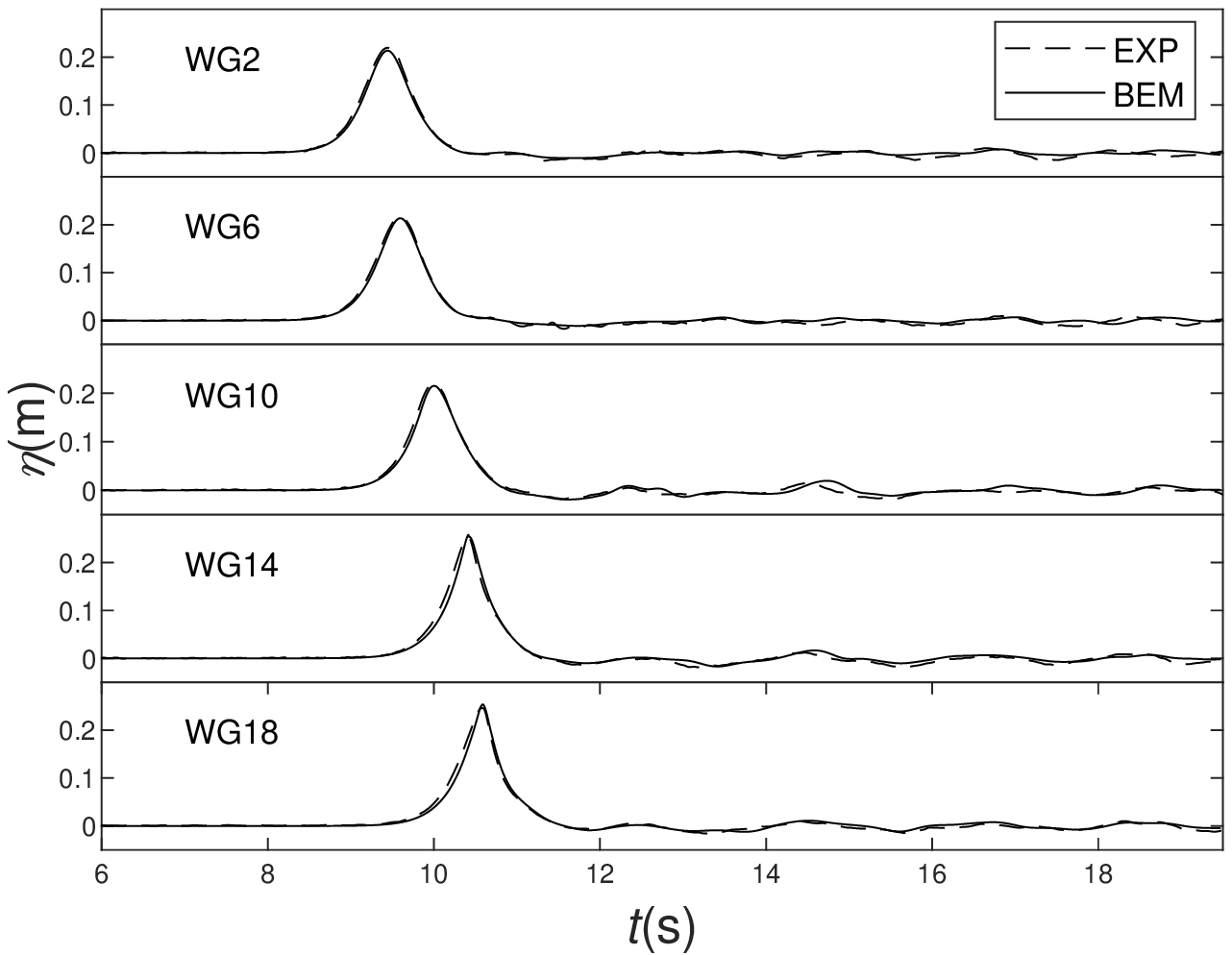}%
    \includegraphics[width=0.65\columnwidth]{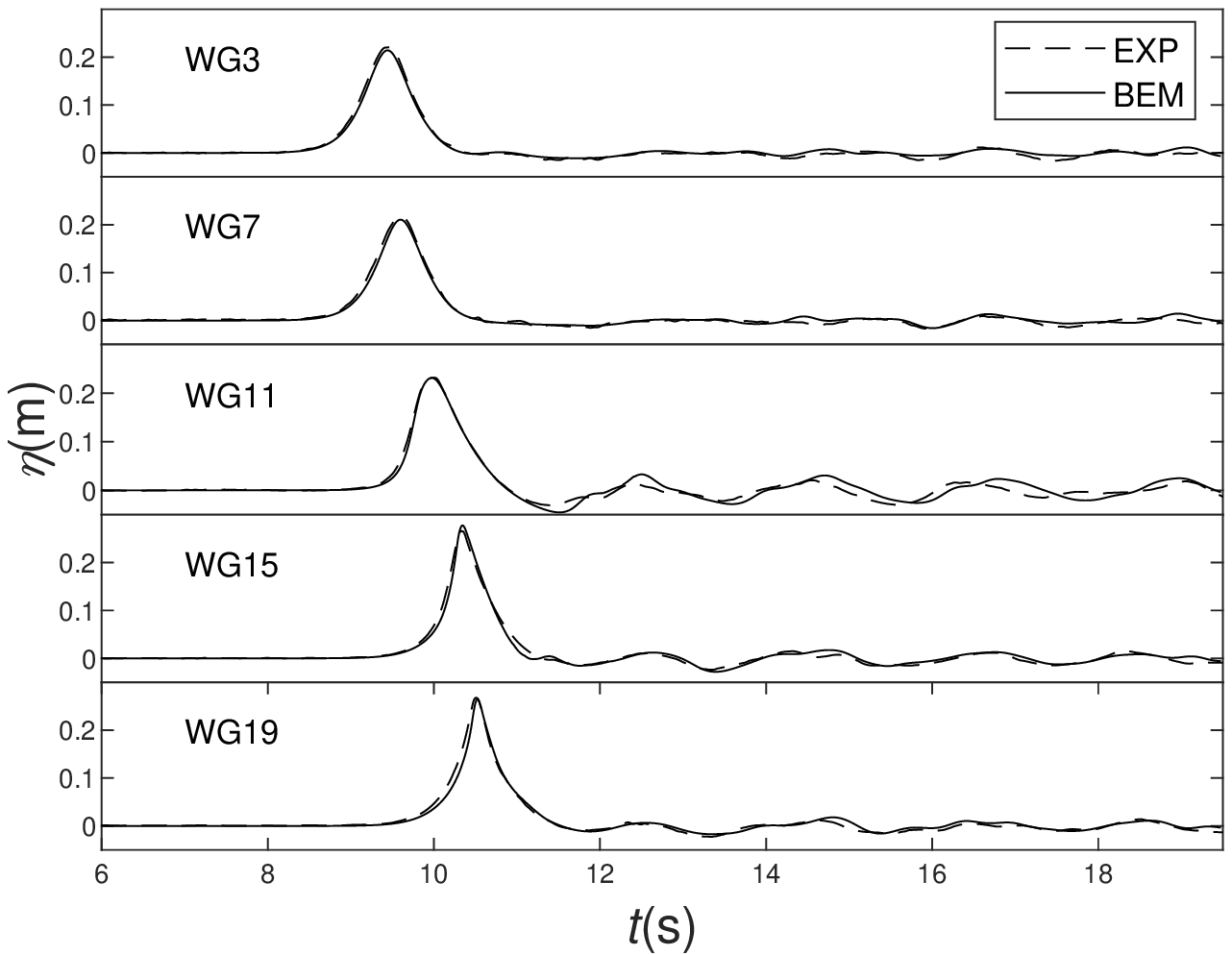}%
}
\caption{\label{fig_B30h50H05} Comparison of the free-surface elevation at 15 selected wave gauges when $B=30$ cm, $h=50$ cm, $H/h=0.5$ (solid line: numerical results, dashed line: experimental results).}
\end{figure*}

\begin{figure*}
\centering
    \includegraphics[width=0.65\columnwidth]{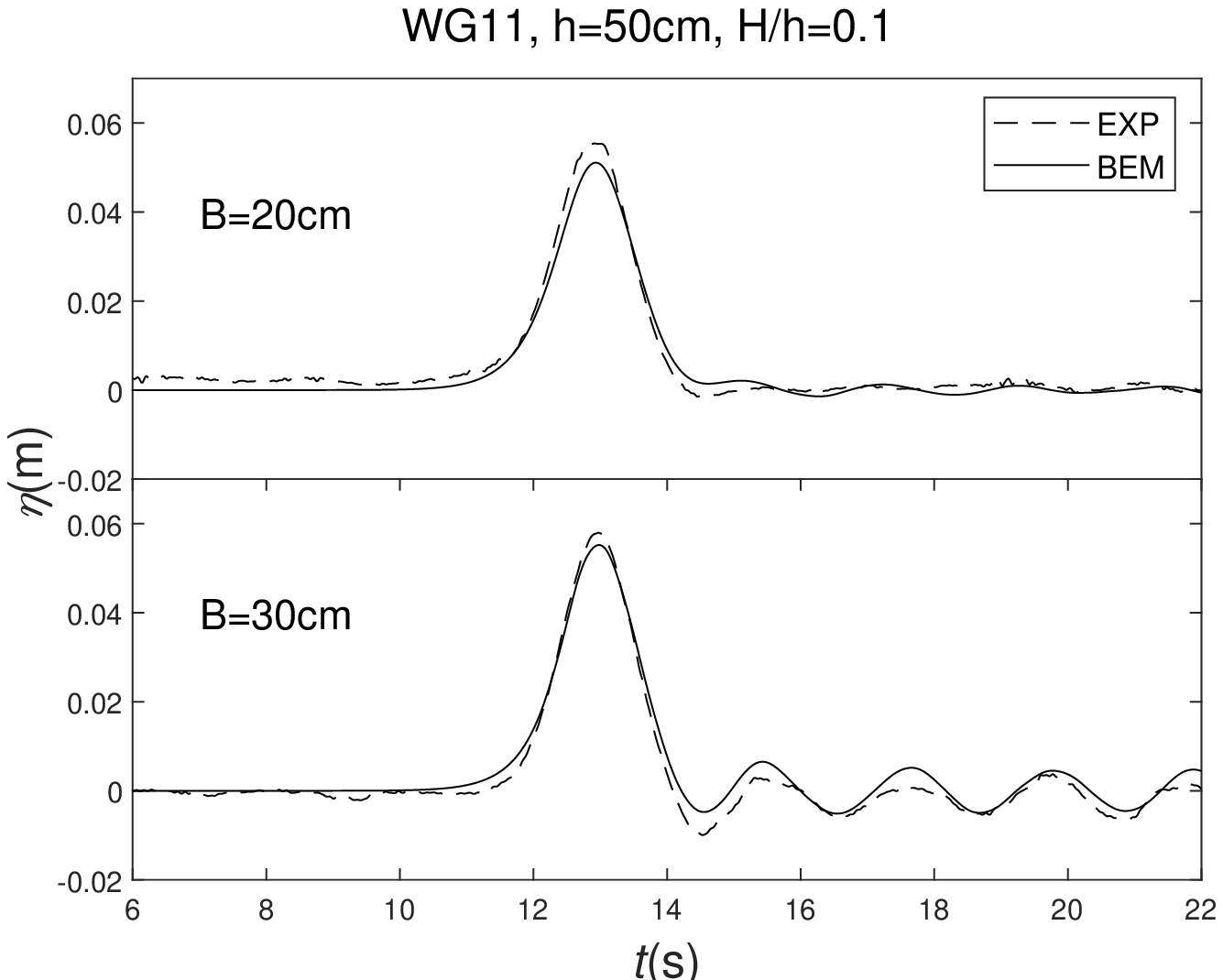}
    \includegraphics[width=0.65\columnwidth]{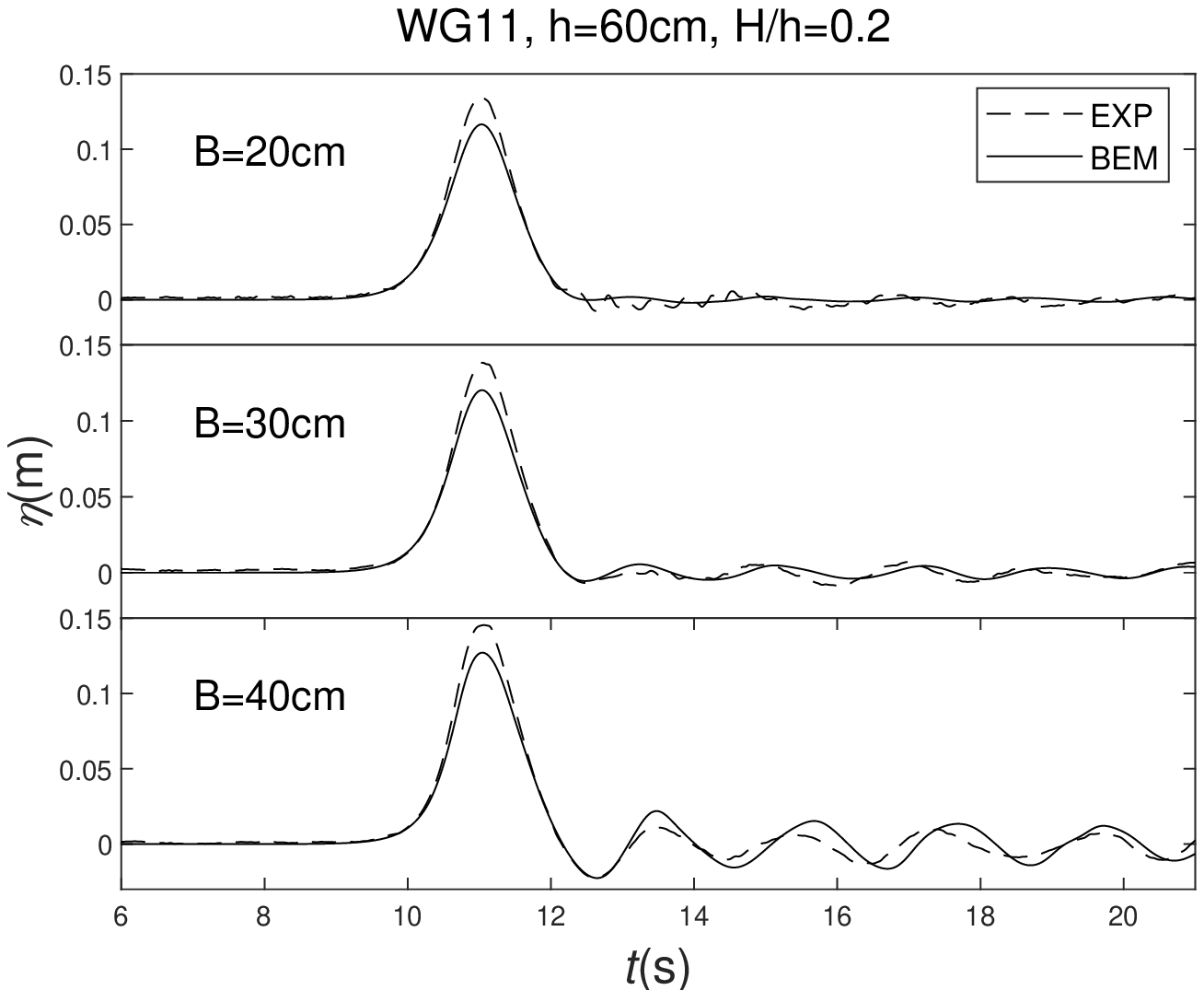}%
    \includegraphics[width=0.65\columnwidth]{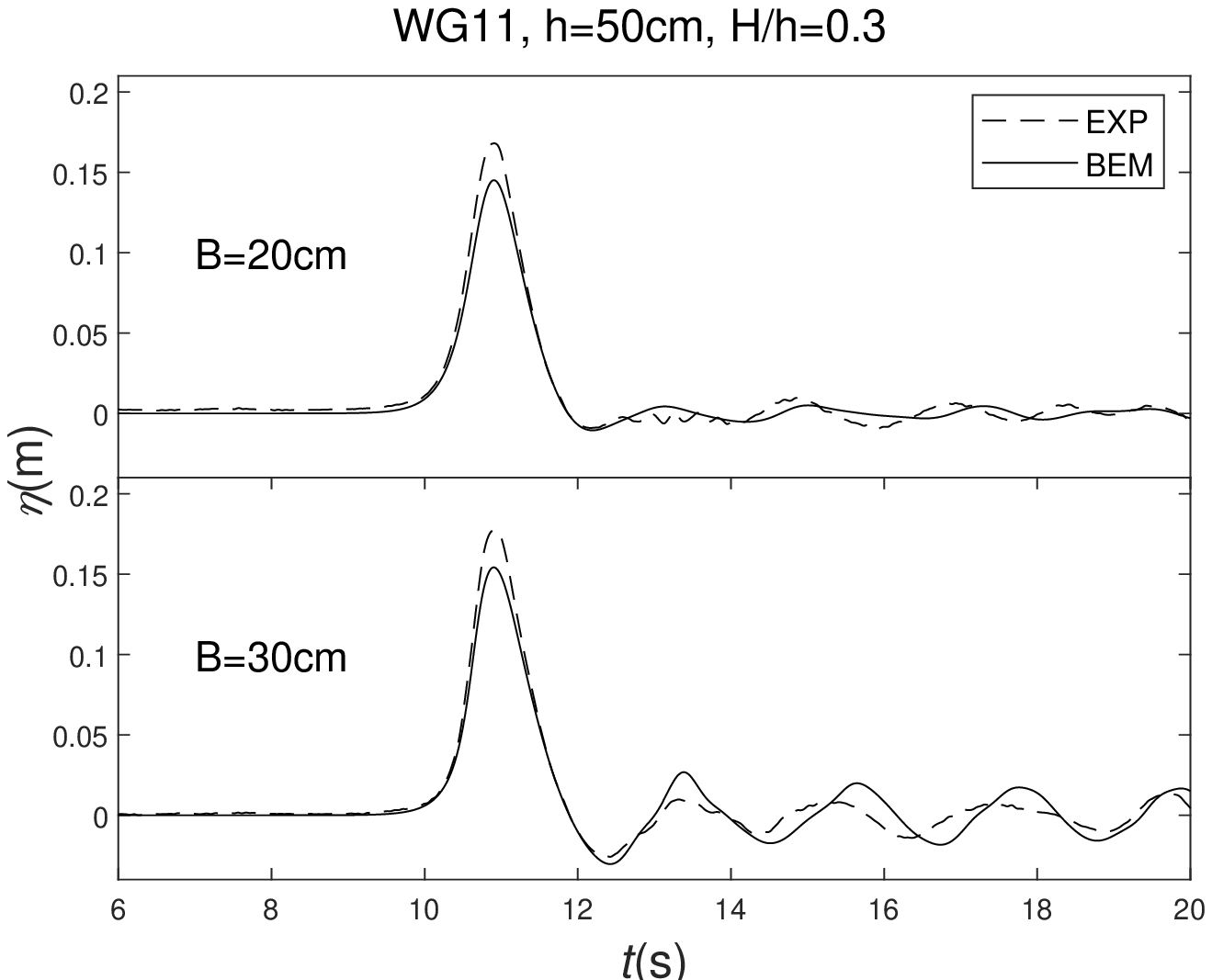} \\
    \includegraphics[width=0.65\columnwidth]{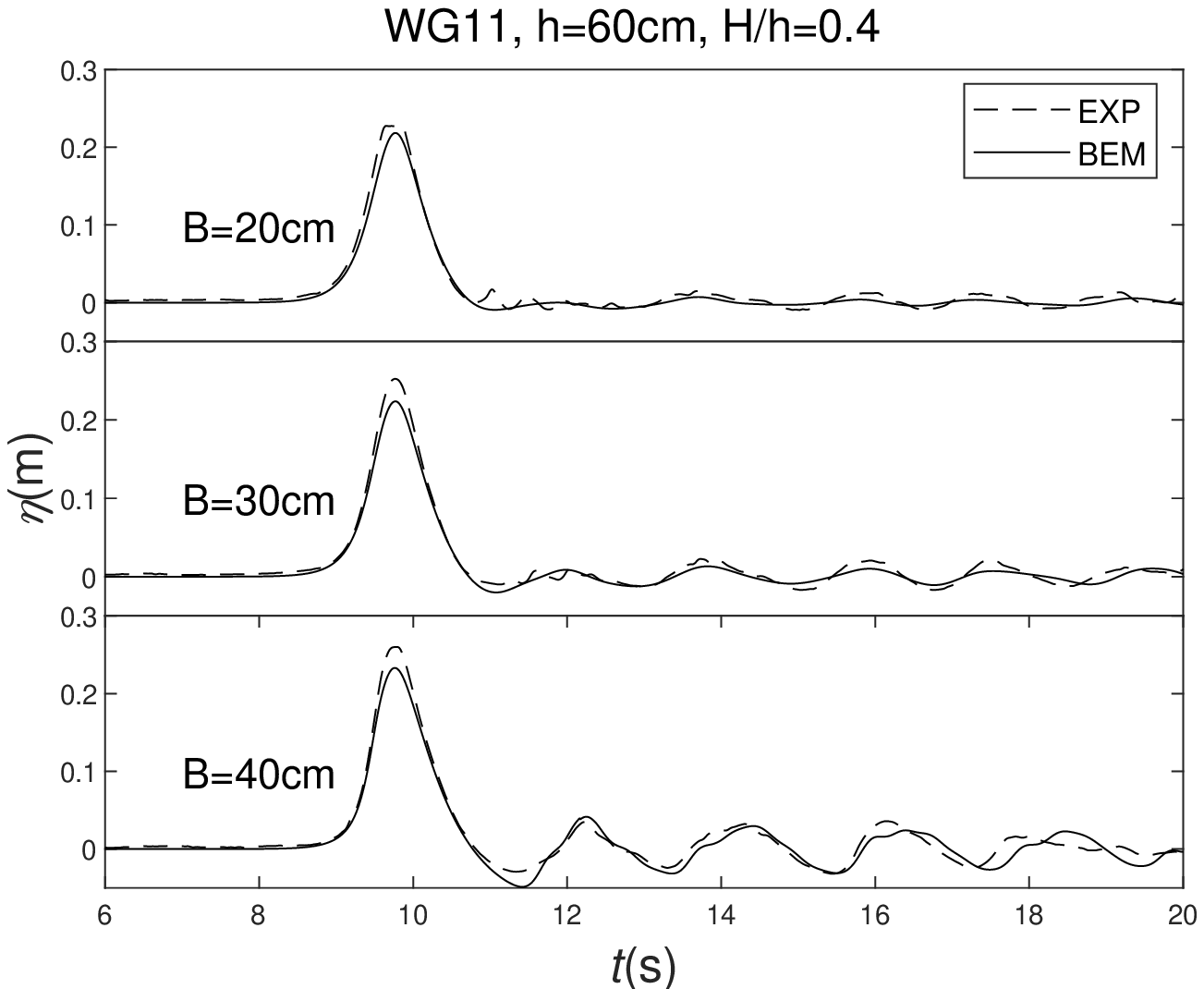}%
    \includegraphics[width=0.65\columnwidth]{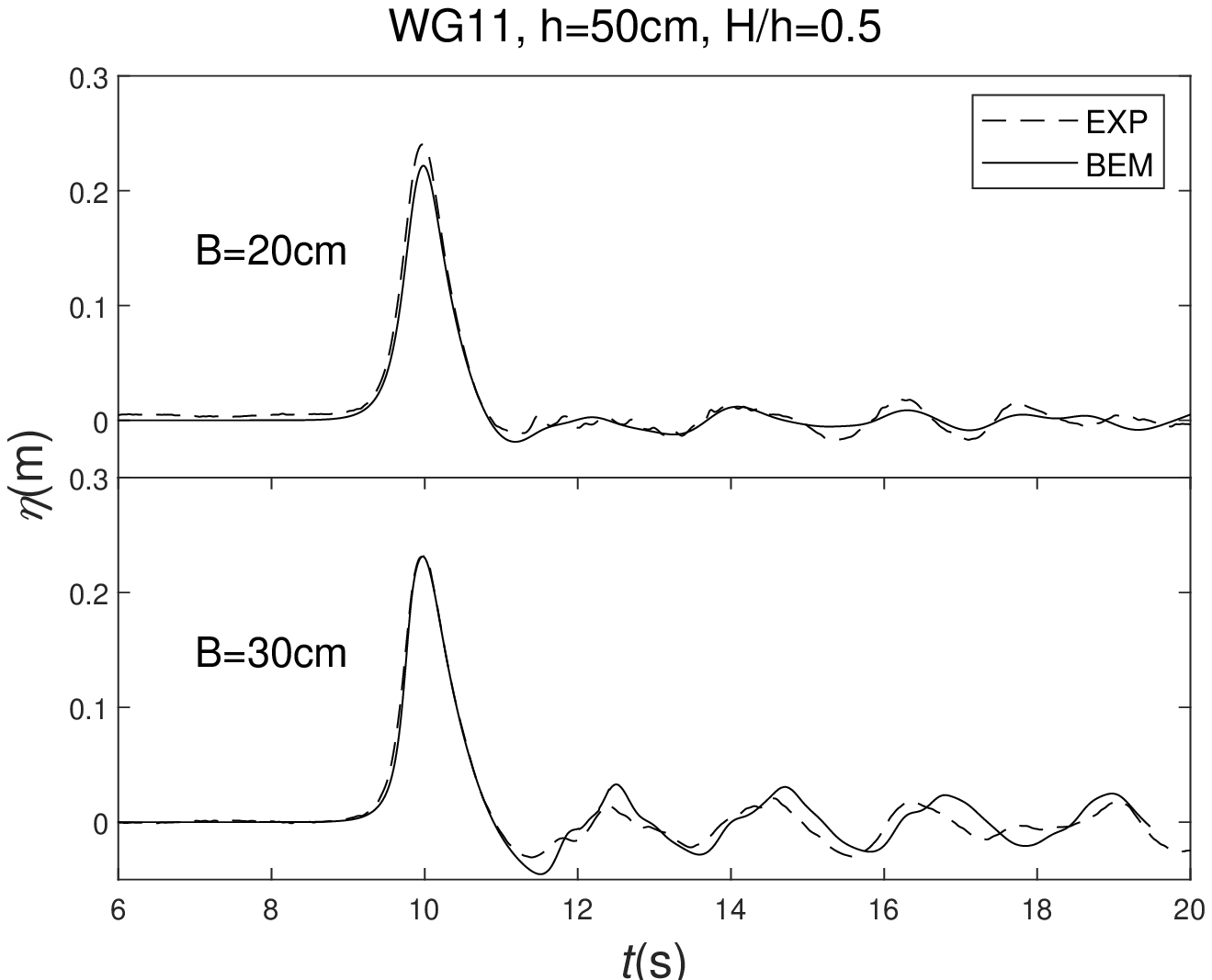}%
\caption{ Comparison of the free-surface elevation at wave gauge WG11 (solid line: numerical, dashed line: experimental).}
\label{fig_WG11}
\end{figure*}

To show all the cases at all the wave gauges would take too much space. Instead, we decided to show first the differences between the wave gauges for a given experimental setup -- see Fig \ref{fig_B30h50H05} -- and then the differences between the various cases at a given wave gauge -- see Fig \ref{fig_WG11}. Throughout the discussion, the numerical and experimental results are synchronized using the data from WG3, which is located on the middle line upstream of the plate. 

The experimental case we selected for the comparison at all wave gauges has the following characteristics: $B=30$ cm, $h=50$ cm, $H/h=0.5$. It corresponds to a large amplitude non-breaking wave ($H/h=0.5$) with a plate which is relatively close to the free surface ($G=10$ cm). Overall, the numerical results shown in Fig \ref{fig_B30h50H05} compare well with the experimental data, both for the peak and for the oscillations that follow (the oscillations are clearly visible at WG11 for example). The wave height increases from the lateral side to the middle line. The highest peak occurs at WG15 when the solitary wave leaves the plate. This is due to wave shoaling.

By looking at the first three wave gauges WG1, WG2 and WG3, we notice that the wave reflected by the leading edge of the plate is quite small compared to the large deformations at WG11 and WG15. This is different from what happens in 2D experiments \cite{Lo2013}, where the reflected wave is clearly observed. 

Checking the three wave gauges WG7, WG10 and WG15 that surround WG11 in Fig \ref{fig_B30h50H05}, it can be seen that the oscillations are noticeable at all three gauges. This indicates that the large deformation above the plate propagates in all directions.

Fig \ref{fig_WG11} shows the free surface elevation for all selected cases at WG11,  which is above the center of the plate. It can be seen that when the plate is closer to the free surface, the free surface deforms more. For example, after the solitary wave passes over the plate, the oscillations that appear are larger when $B=40$ cm than when $B = 20$ cm. 

Generally speaking, as the wave amplitude becomes larger, the agreement between experimental and numerical results becomes better. This is partly due to the electrical noise in the experiments. For large amplitude waves, the influence of electrical noise becomes negligible. For small amplitude waves, the numerical peak is smaller than the experimental peak. 

\begin{figure*}[ht]
\centering
\includegraphics[scale=0.15]{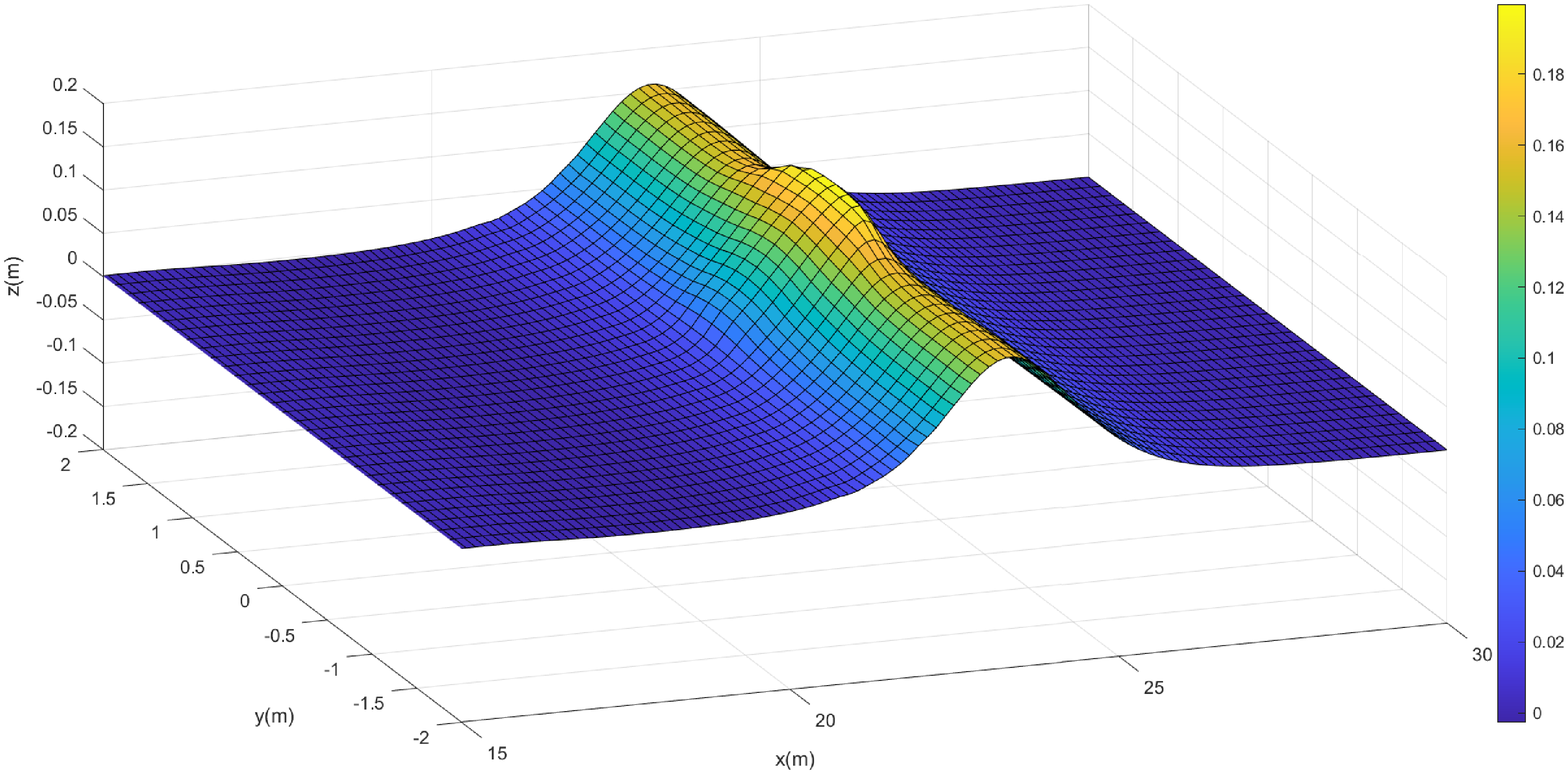}
\includegraphics[scale=0.15]{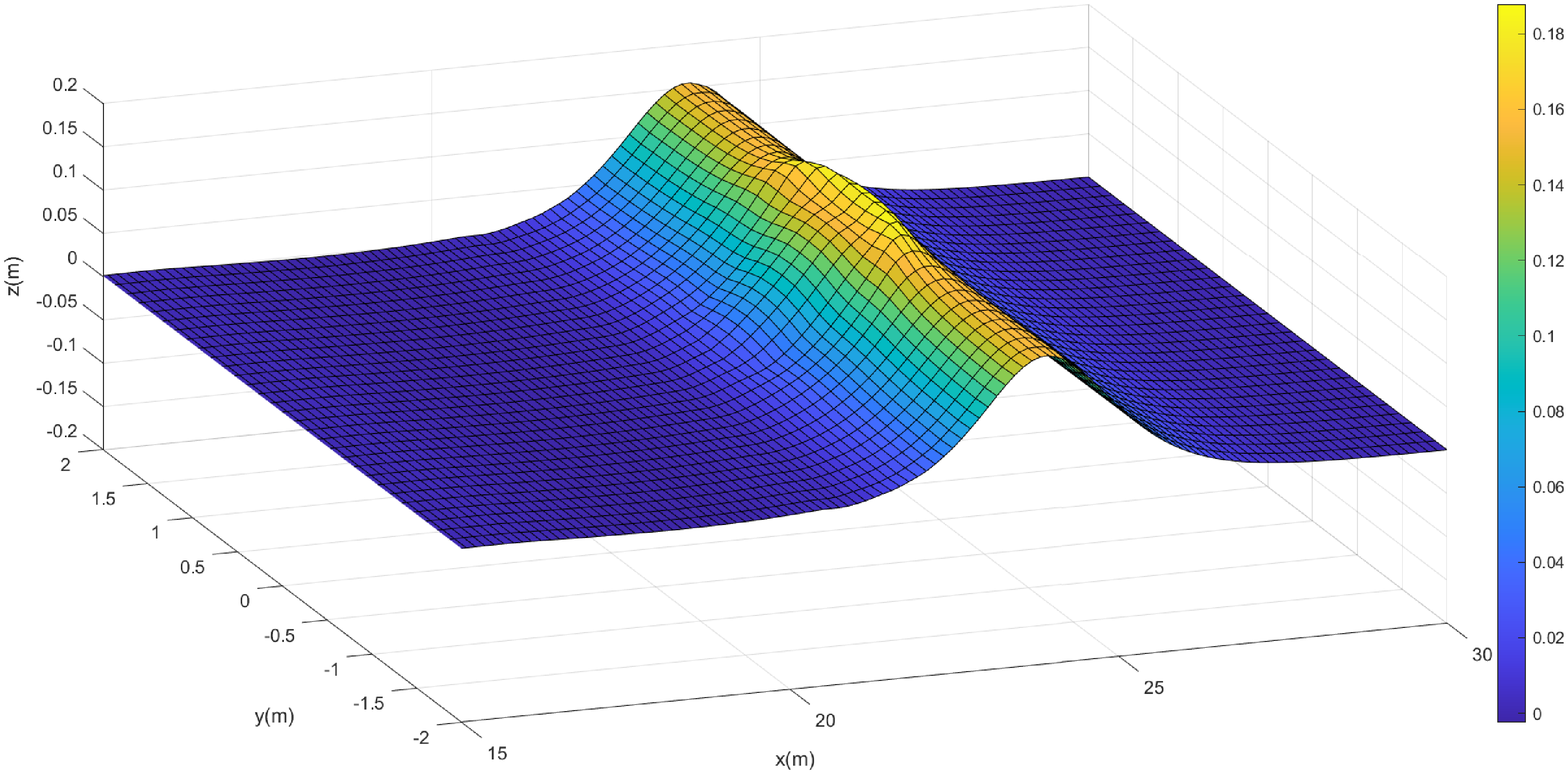}
\includegraphics[scale=0.15]{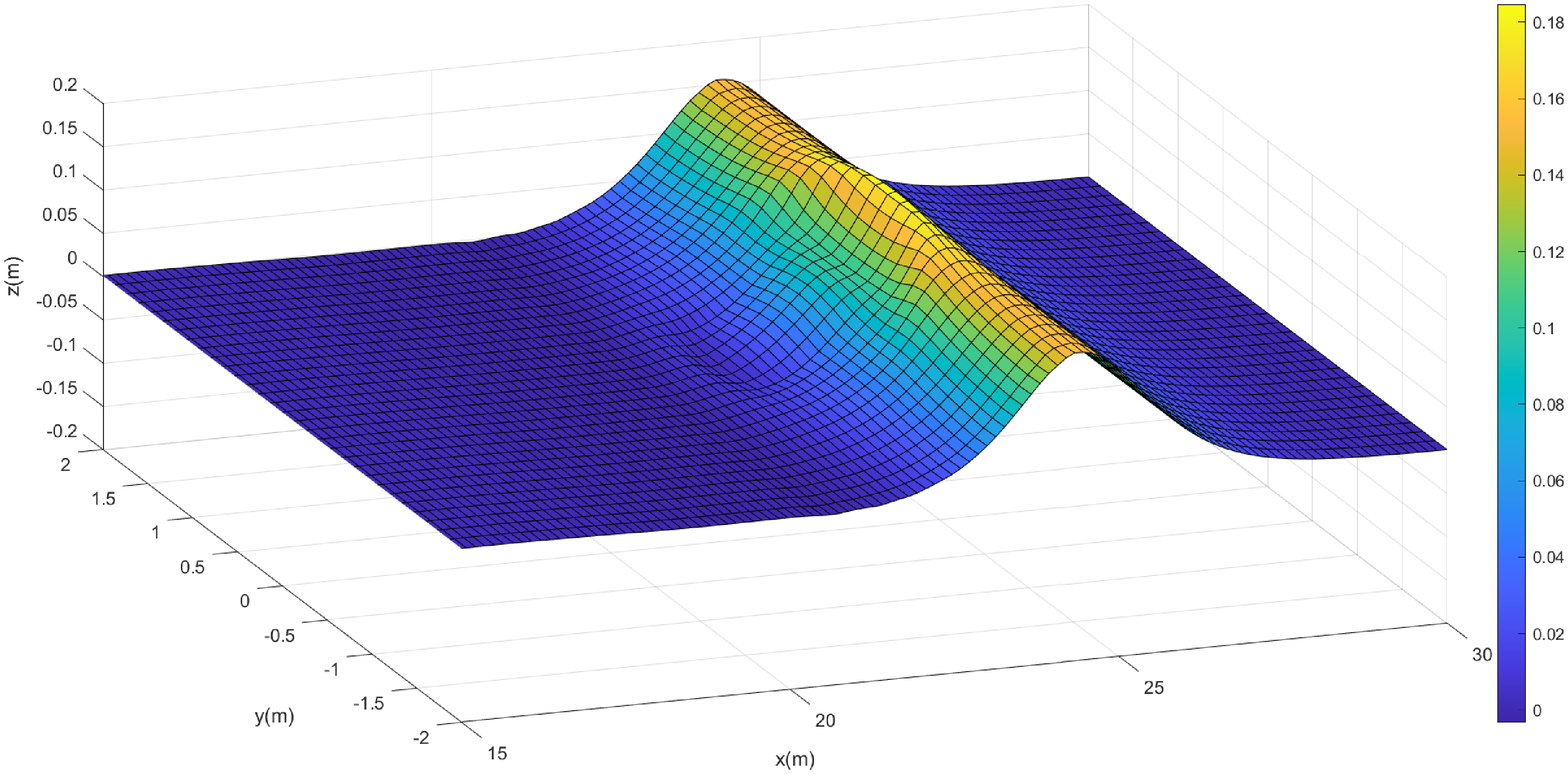}
\includegraphics[scale=0.15]{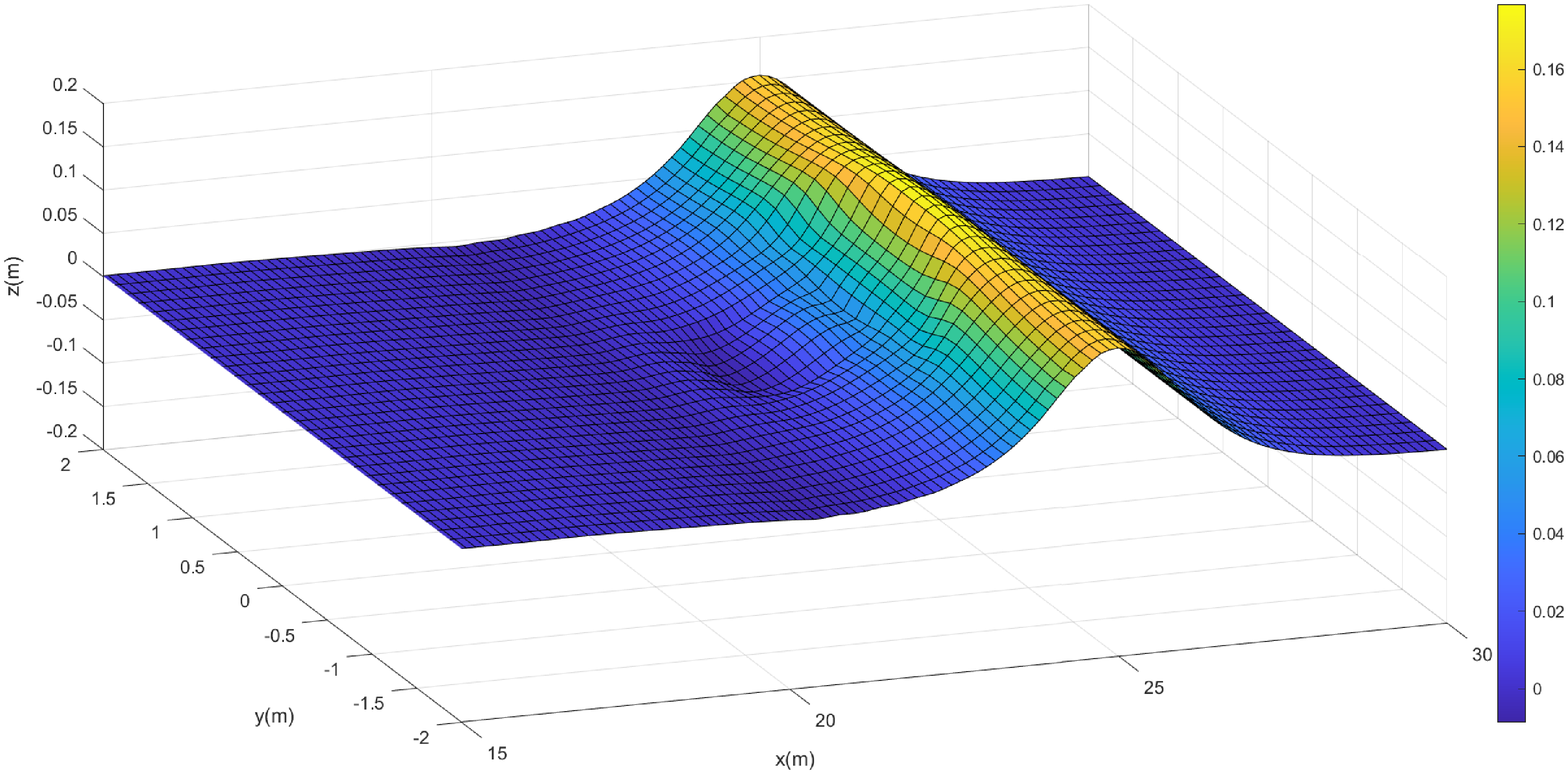}
\includegraphics[scale=0.15]{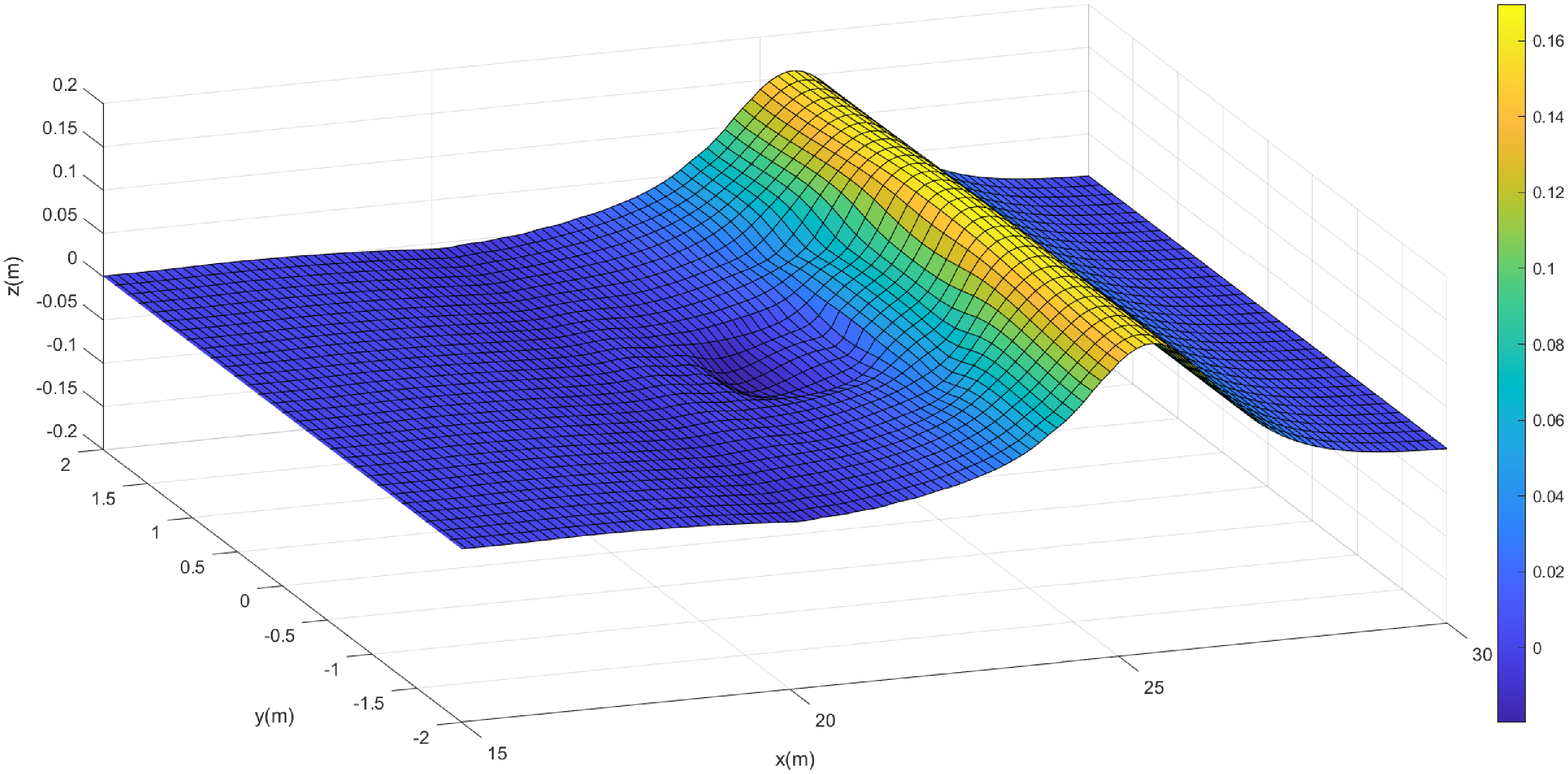}
\includegraphics[scale=0.15]{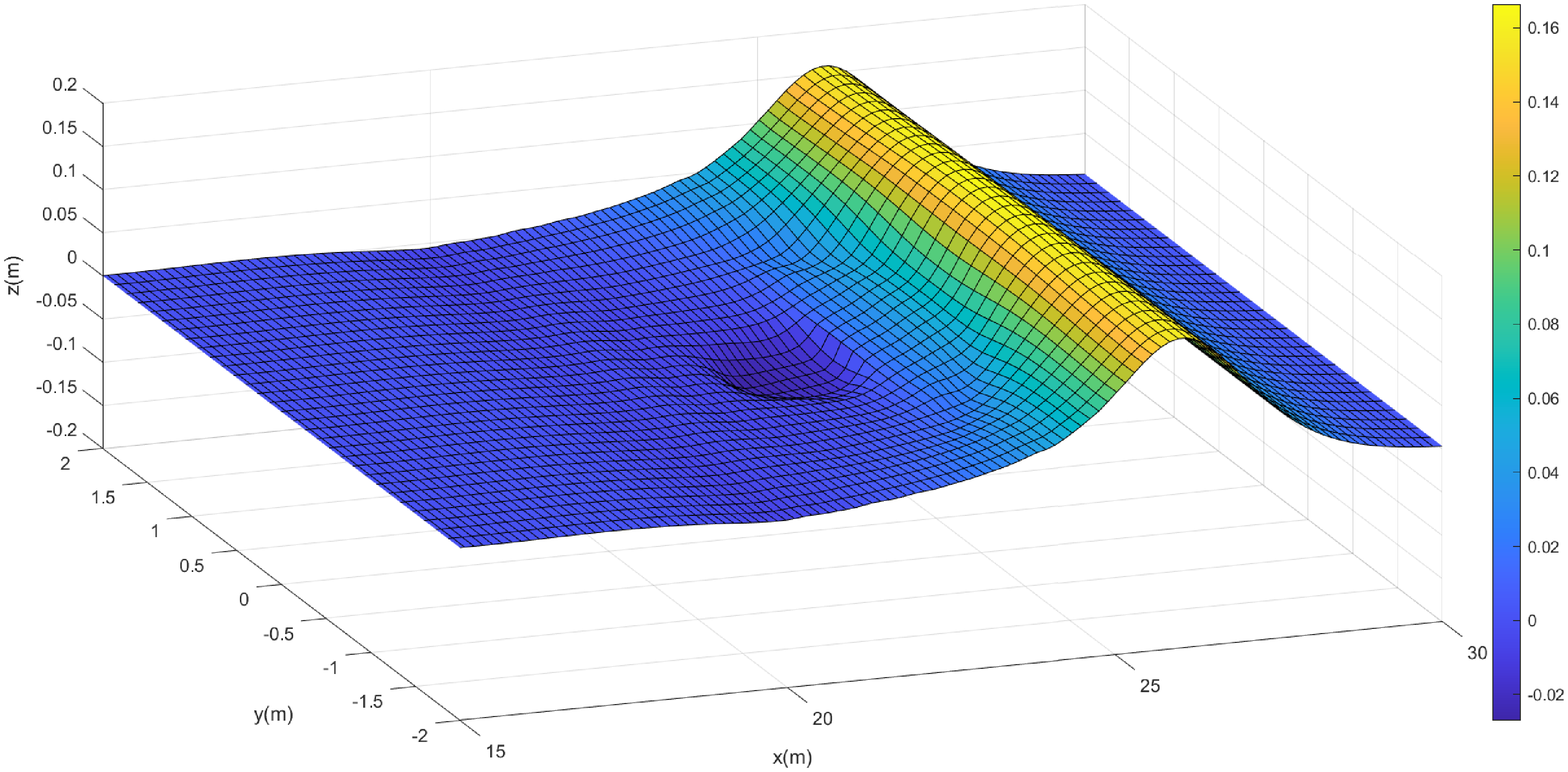}
\includegraphics[scale=0.15]{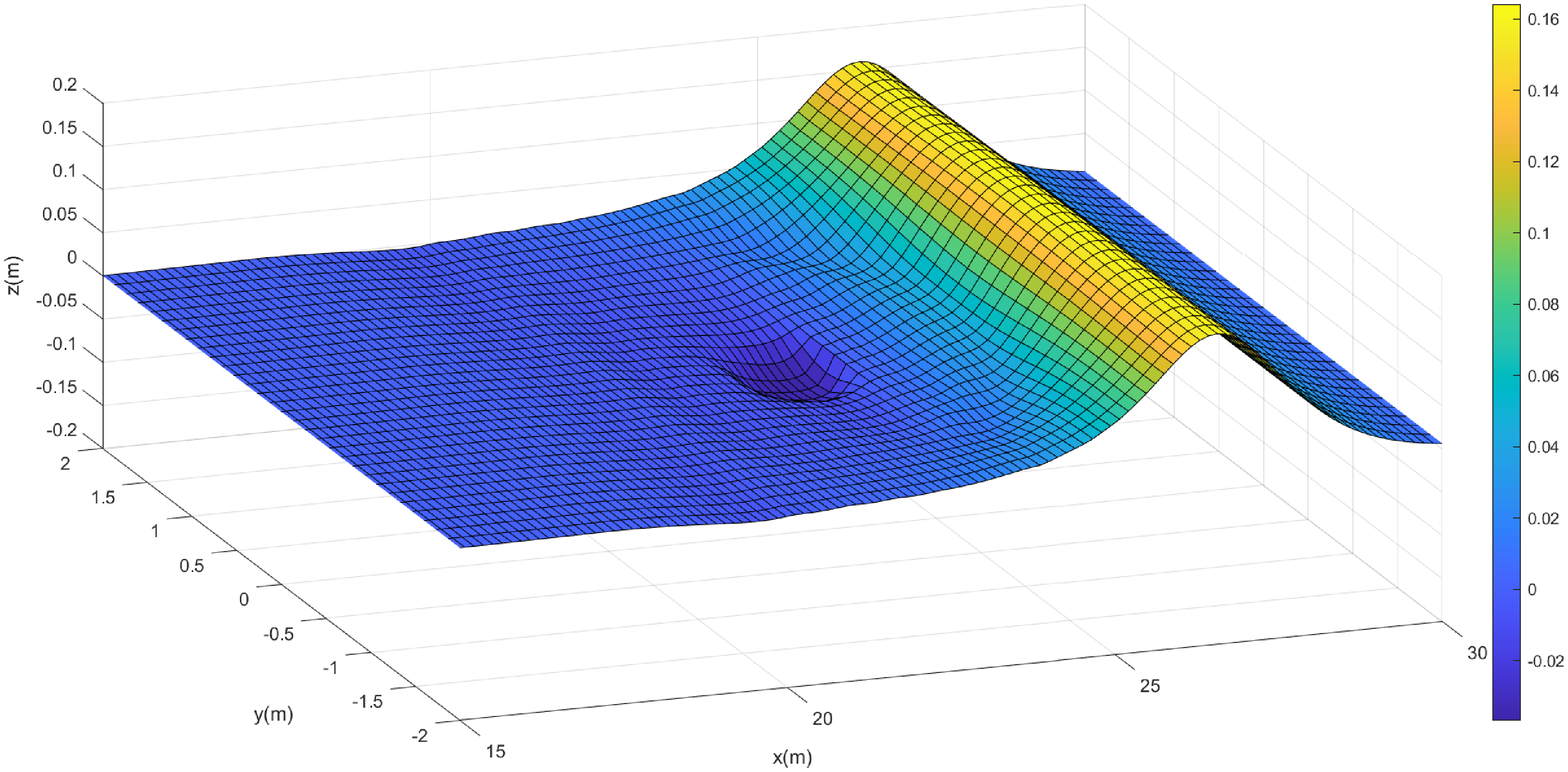}
\includegraphics[scale=0.15]{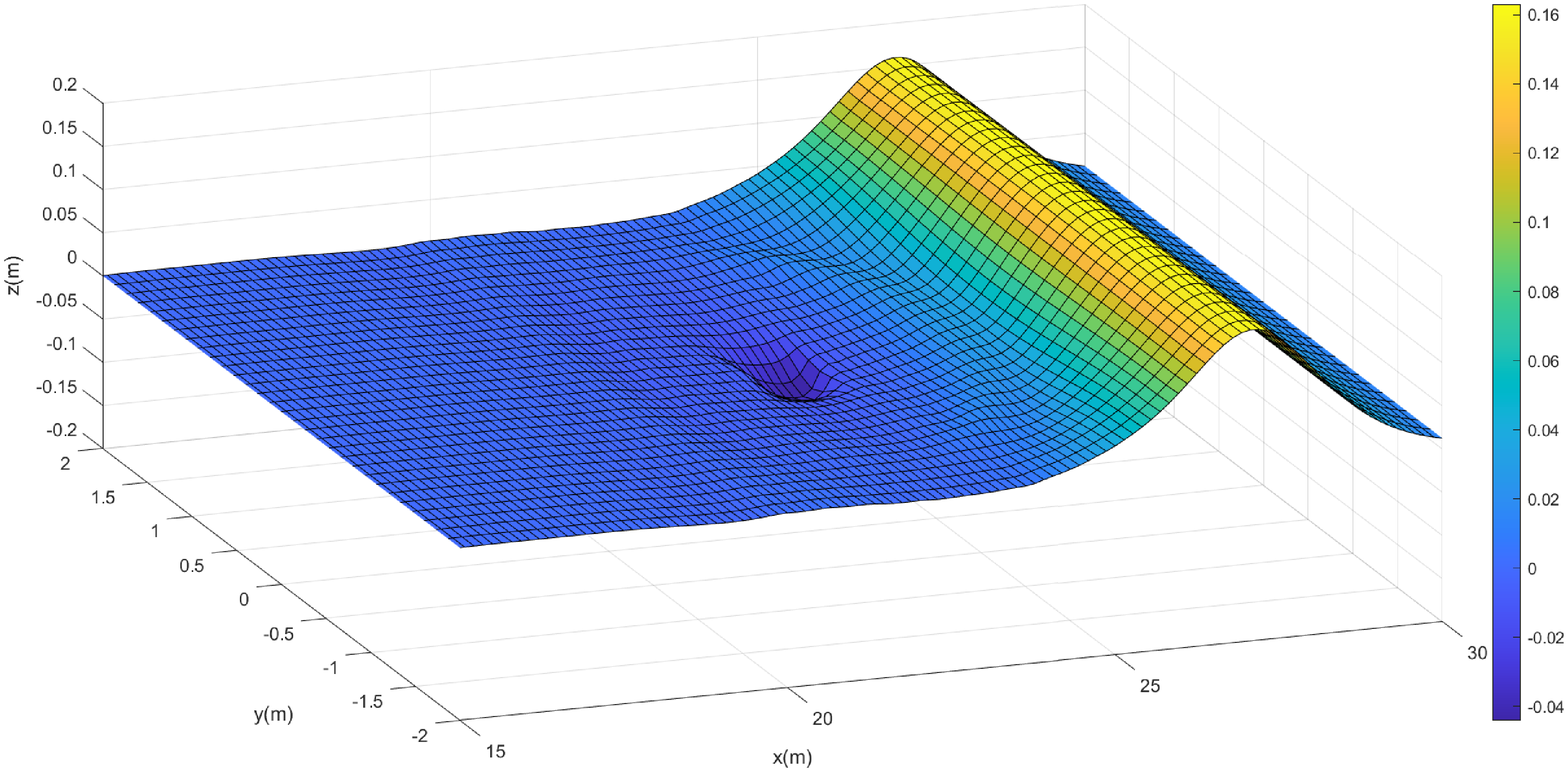}
\includegraphics[scale=0.15]{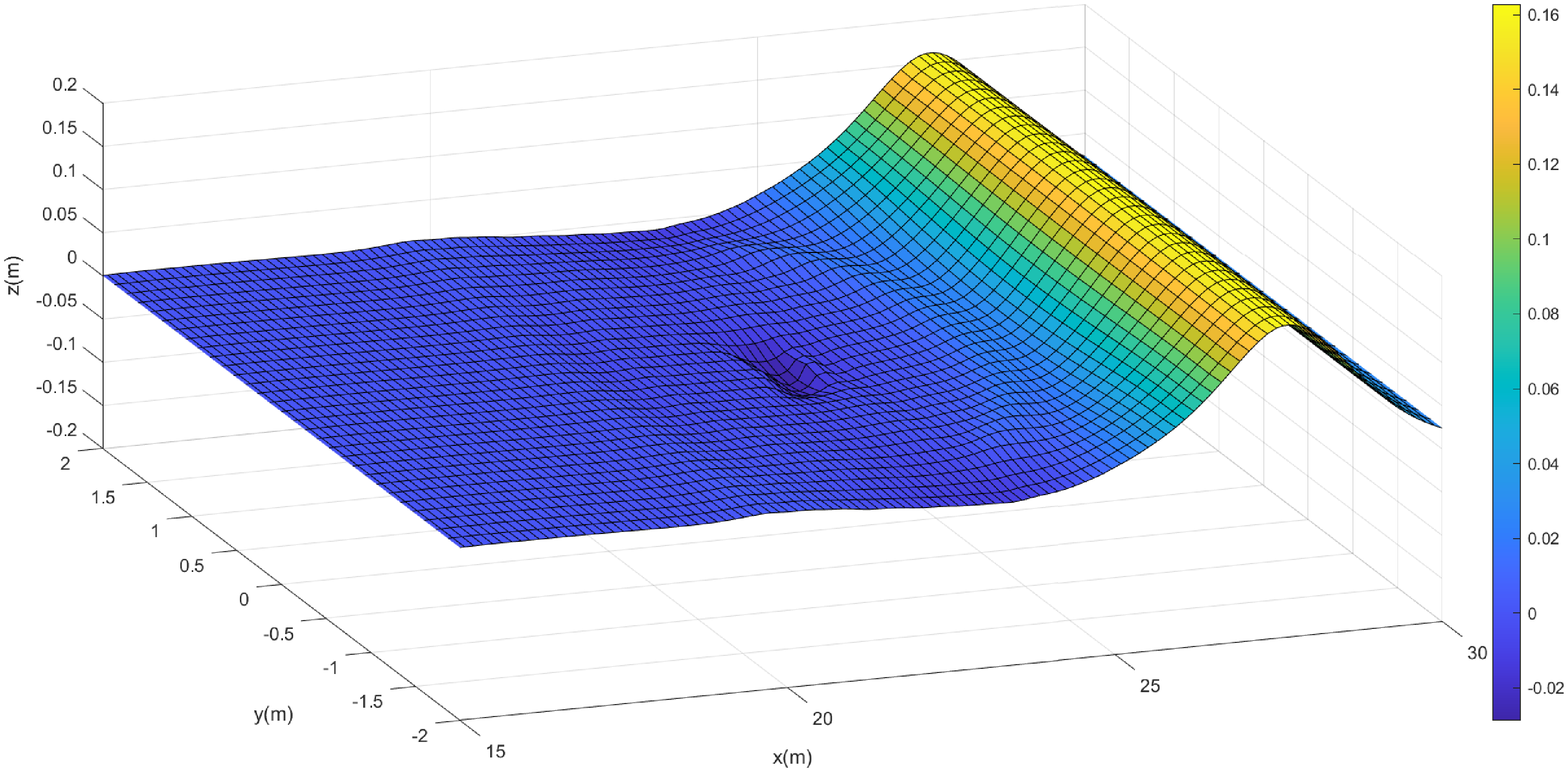}
\includegraphics[scale=0.15]{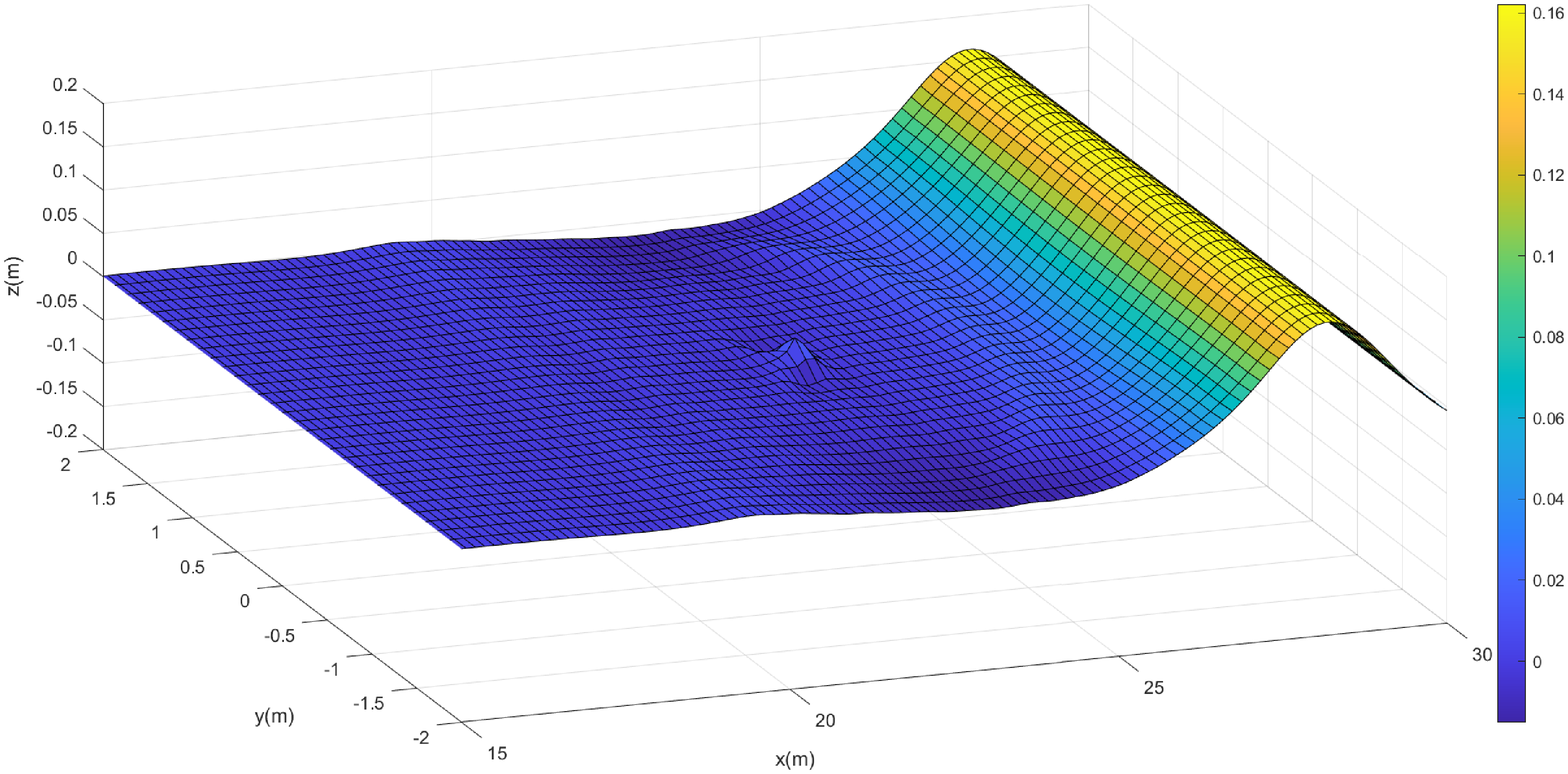}
\includegraphics[scale=0.15]{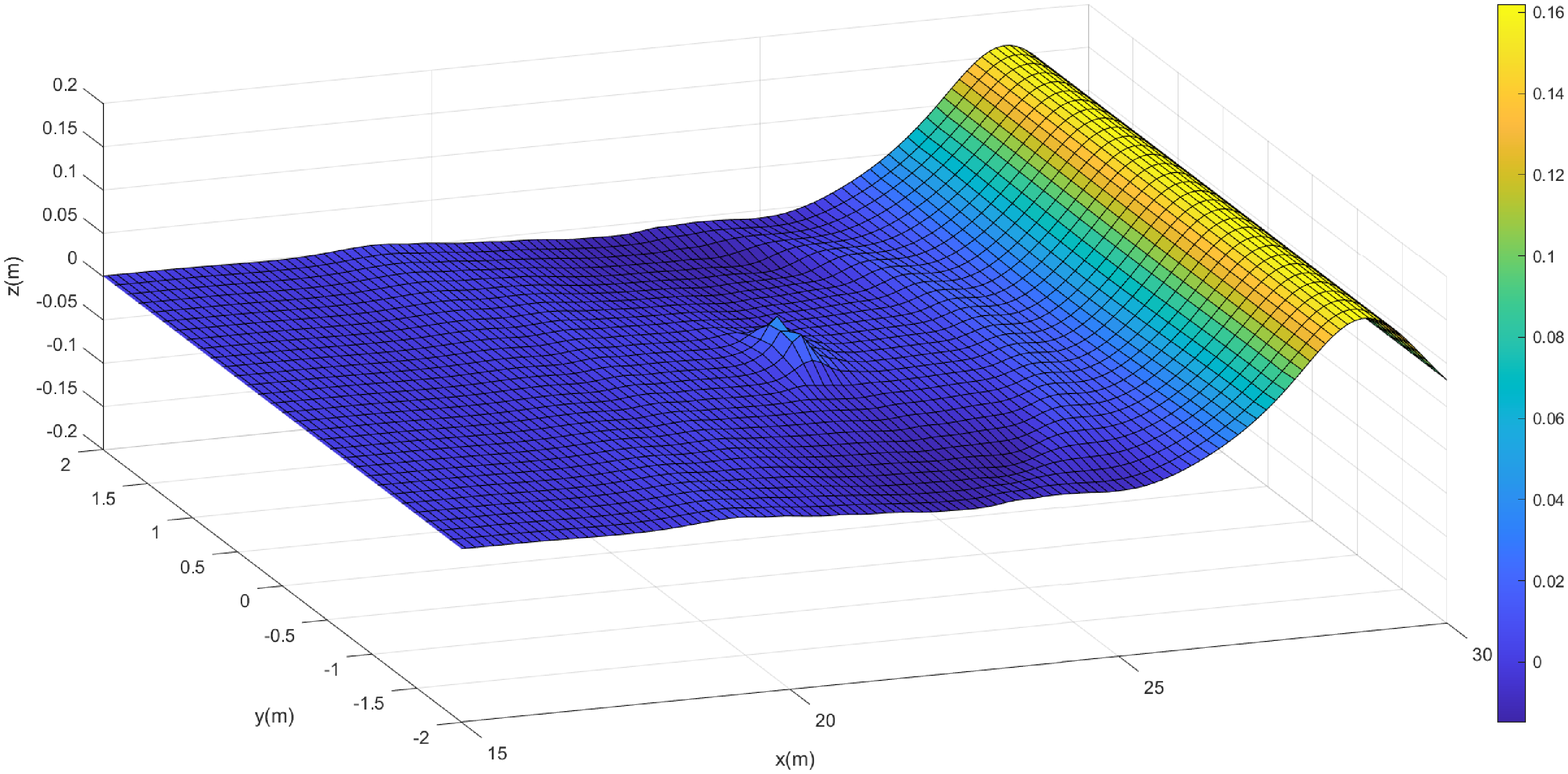}
\includegraphics[scale=0.15]{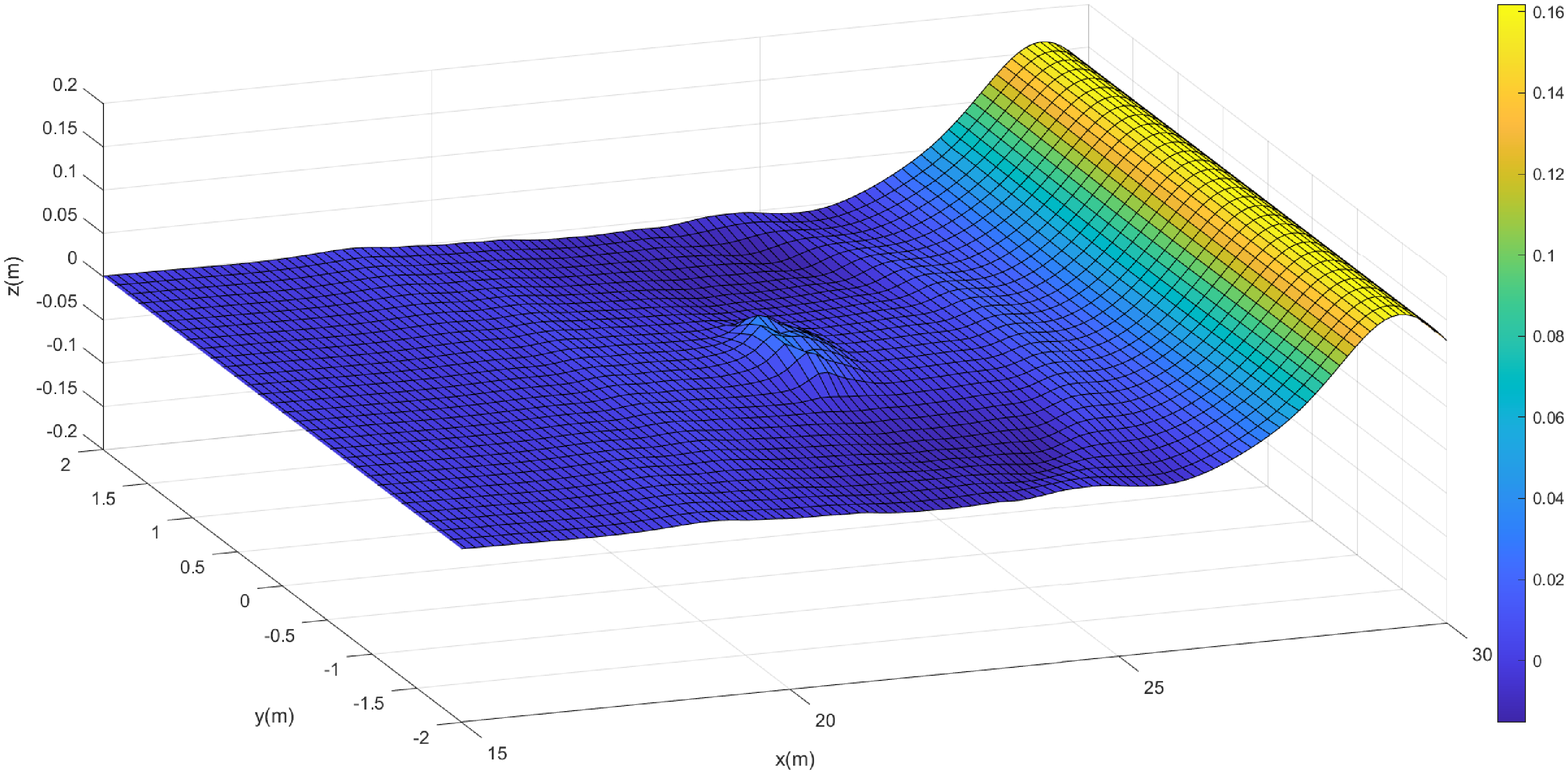}
\includegraphics[scale=0.15]{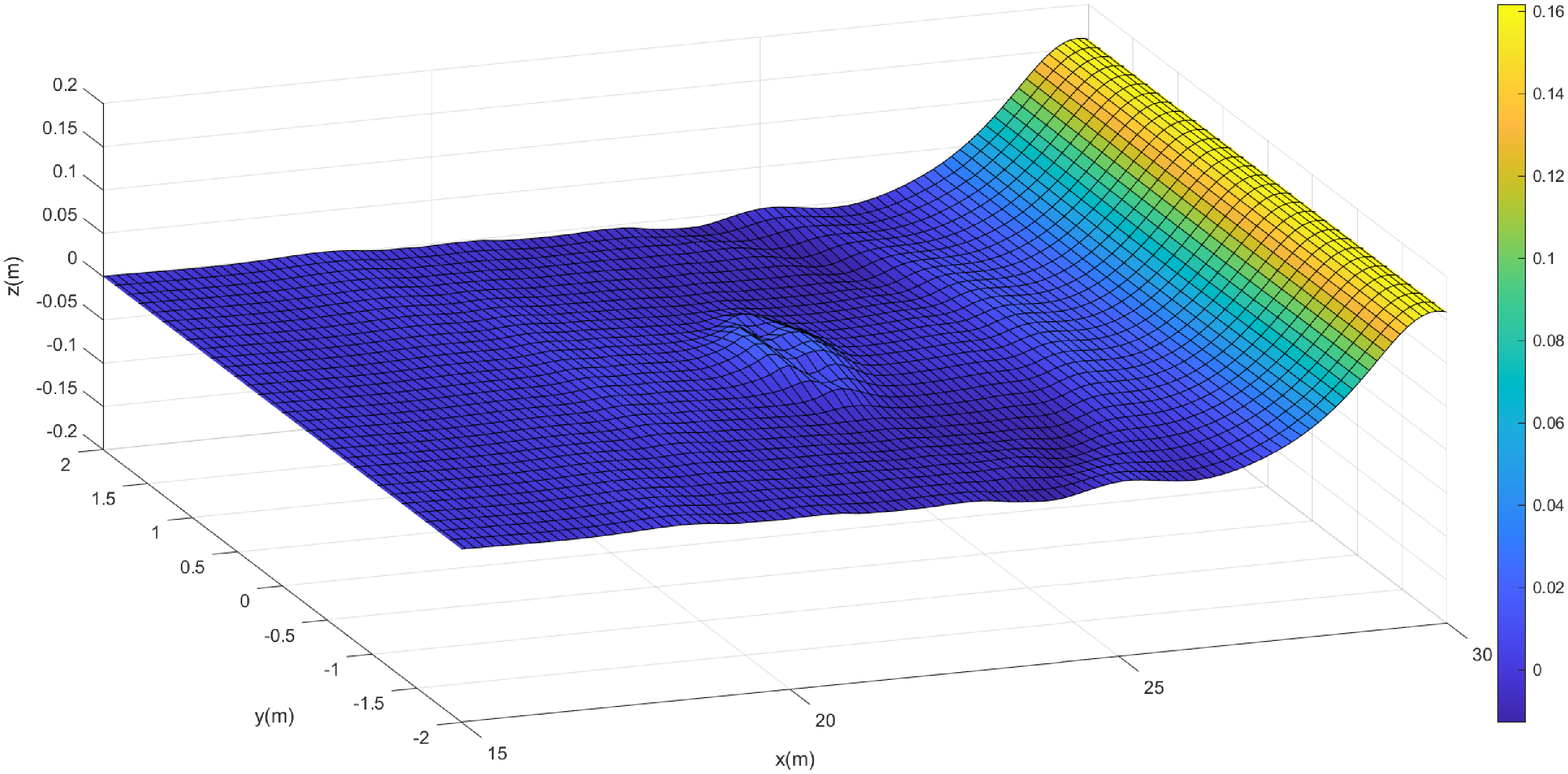}
\includegraphics[scale=0.15]{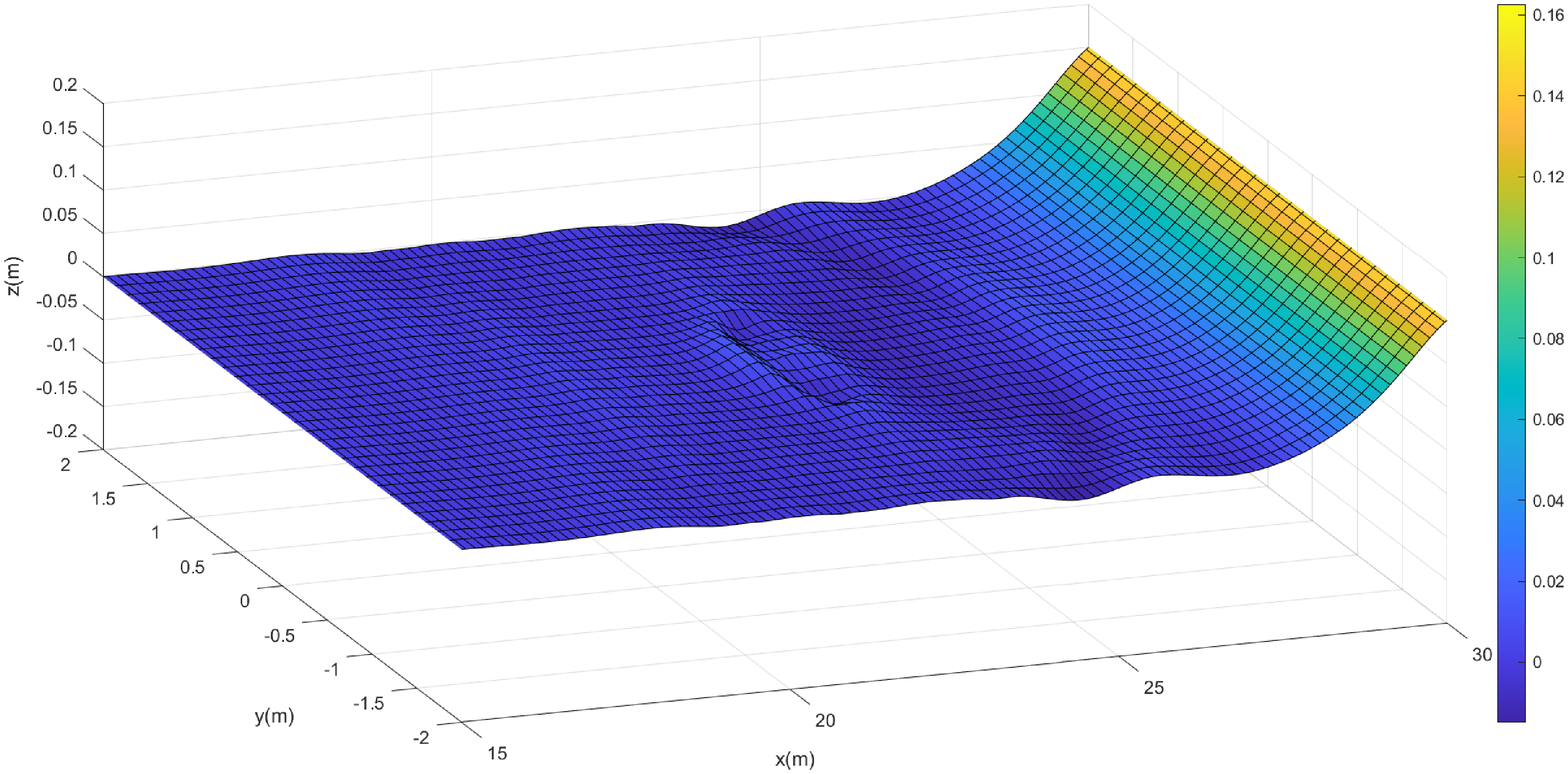}
\includegraphics[scale=0.15]{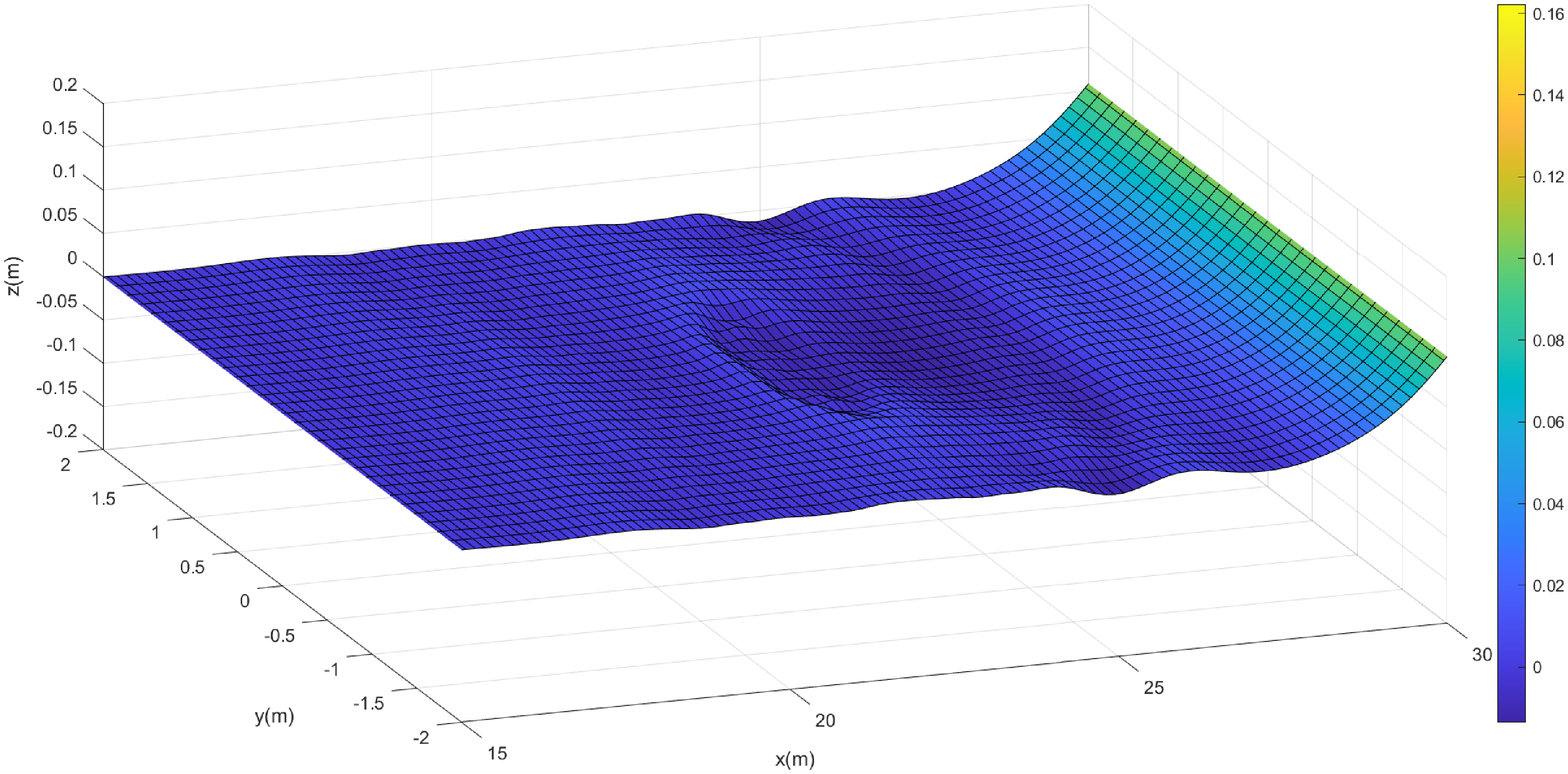}
\caption{Deformation of the free surface (from 10.6 s to 13.4 s -- the interval is 0.2 s -- left to right and top to bottom) when $B=40$ cm, $h=60$ cm and $H/h=0.3$.}
\label{fig_fsdeform}
\end{figure*}

Fig \ref{fig_fsdeform} shows the overall free-surface deformation above the plate for the case $B=40$ cm, $h=60$ cm and $H/h=0.3$. As the solitary wave passes above the plate, the wave amplitude becomes larger due to shoaling. Once the large amplitude wave reaches its maximum amplitude, it propagates in all directions, causing reflection from the trailing edge. Meanwhile, the reflection also occurs at the leading edge. These two effects compete simultaneously and leave a pitfall just above the plate. The large amplitude wave keeps radiating and overcomes the pitfall leaving a bulge above the plate. 

\subsection{Horizontal force, vertical force and moment}

As anticipated because of the symmetries involved in the experimental and numerical setups, it has been confirmed both by \cite{Wang2020} and in the present work that the lateral force $F_y$, the yaw moment $M_z$ and the roll moment $M_x$ are negligible. Therefore we concentrate on the horizontal force $F_x$, the vertical force $F_z$ and the (pitching) moment $M_y$ for comparisons between the various cases. The basic behavior of $F_x$, $F_z$ and $M_y$ was explained in \cite{Wang2020}, where a $6-$stage process was introduced to highlight the various peaks in the vertical force and moment. 

\begin{figure}
\centering
\includegraphics[width=0.45\columnwidth]{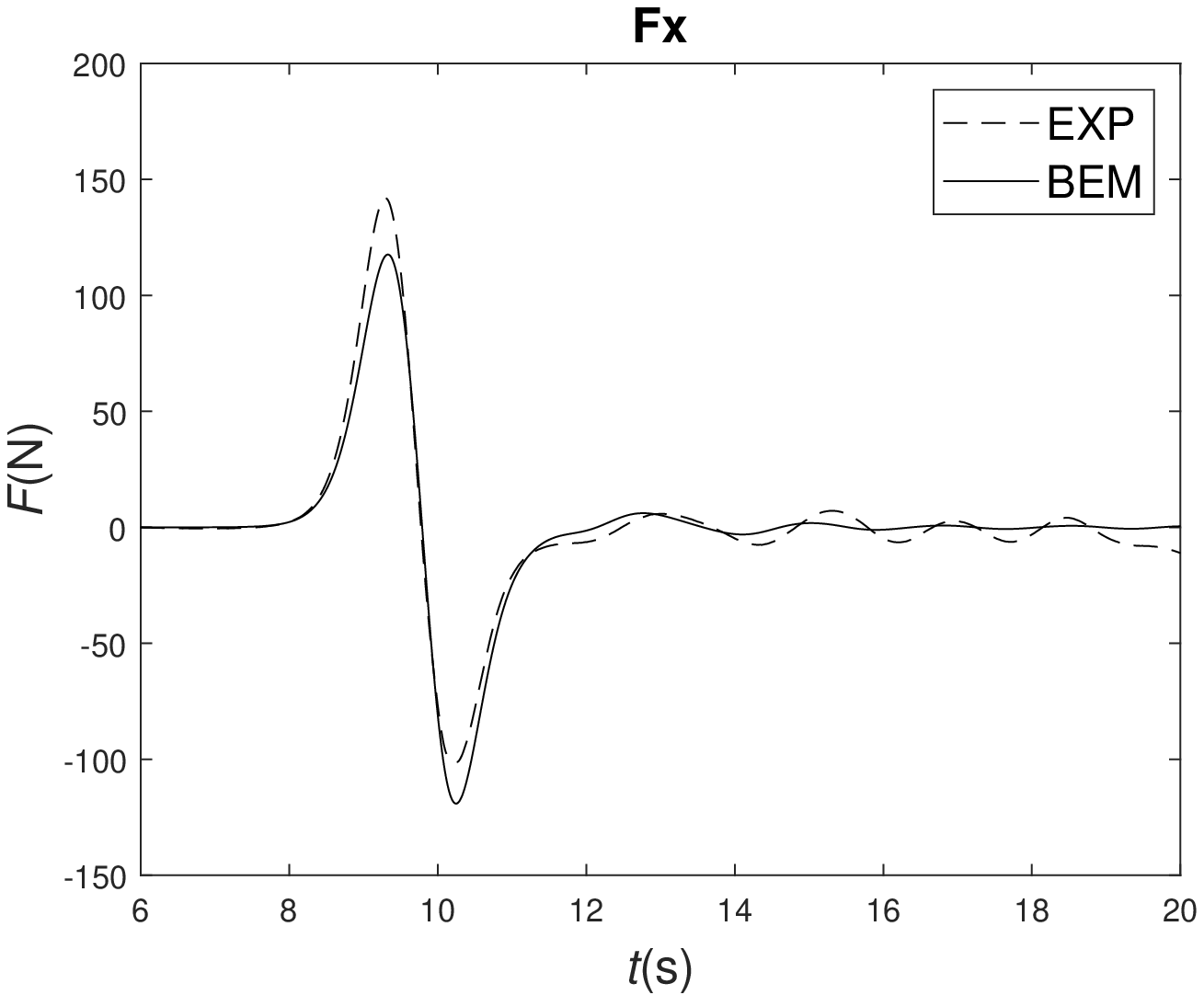}
\includegraphics[width=0.45\columnwidth]{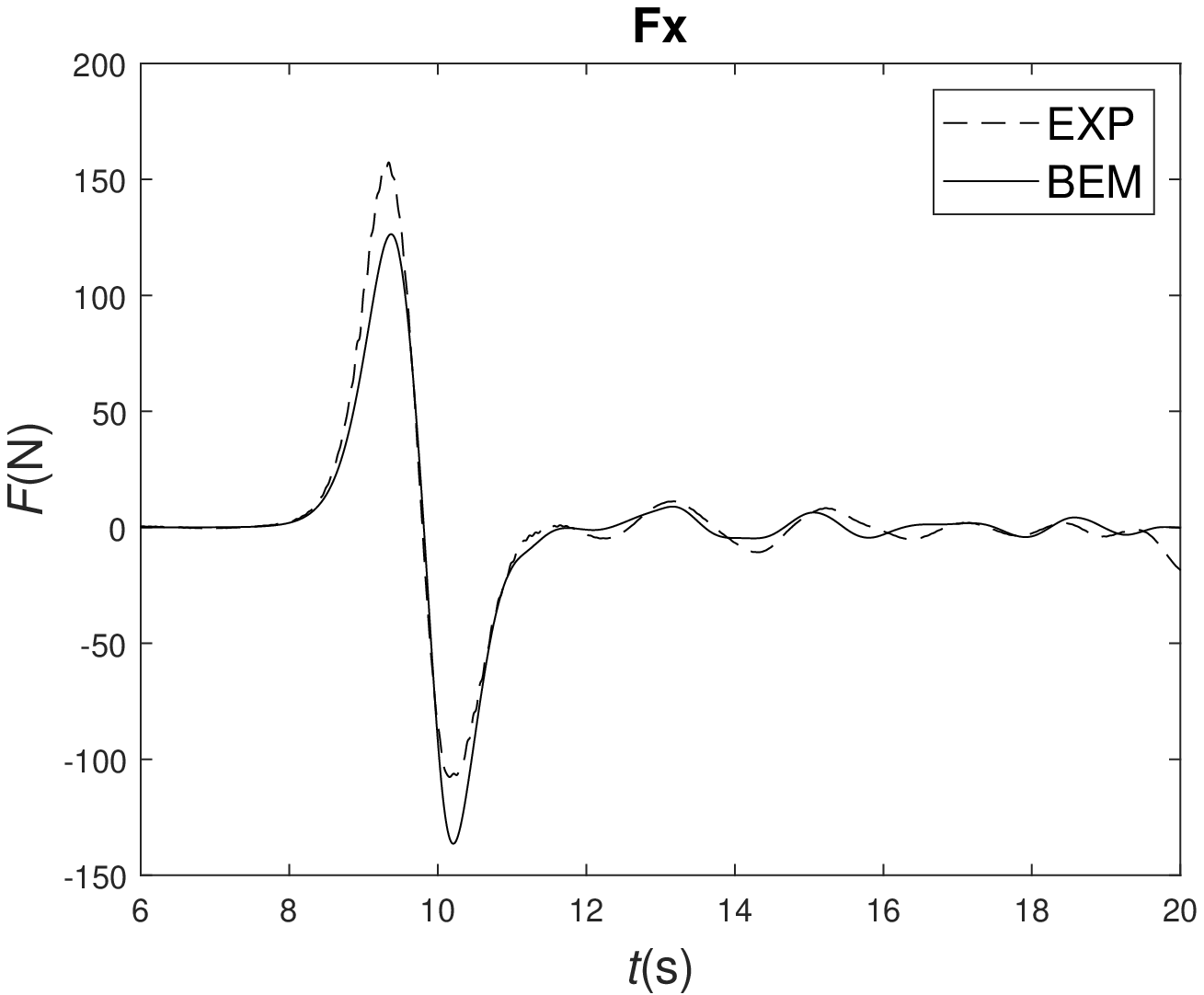} \\
\includegraphics[width=0.45\columnwidth]{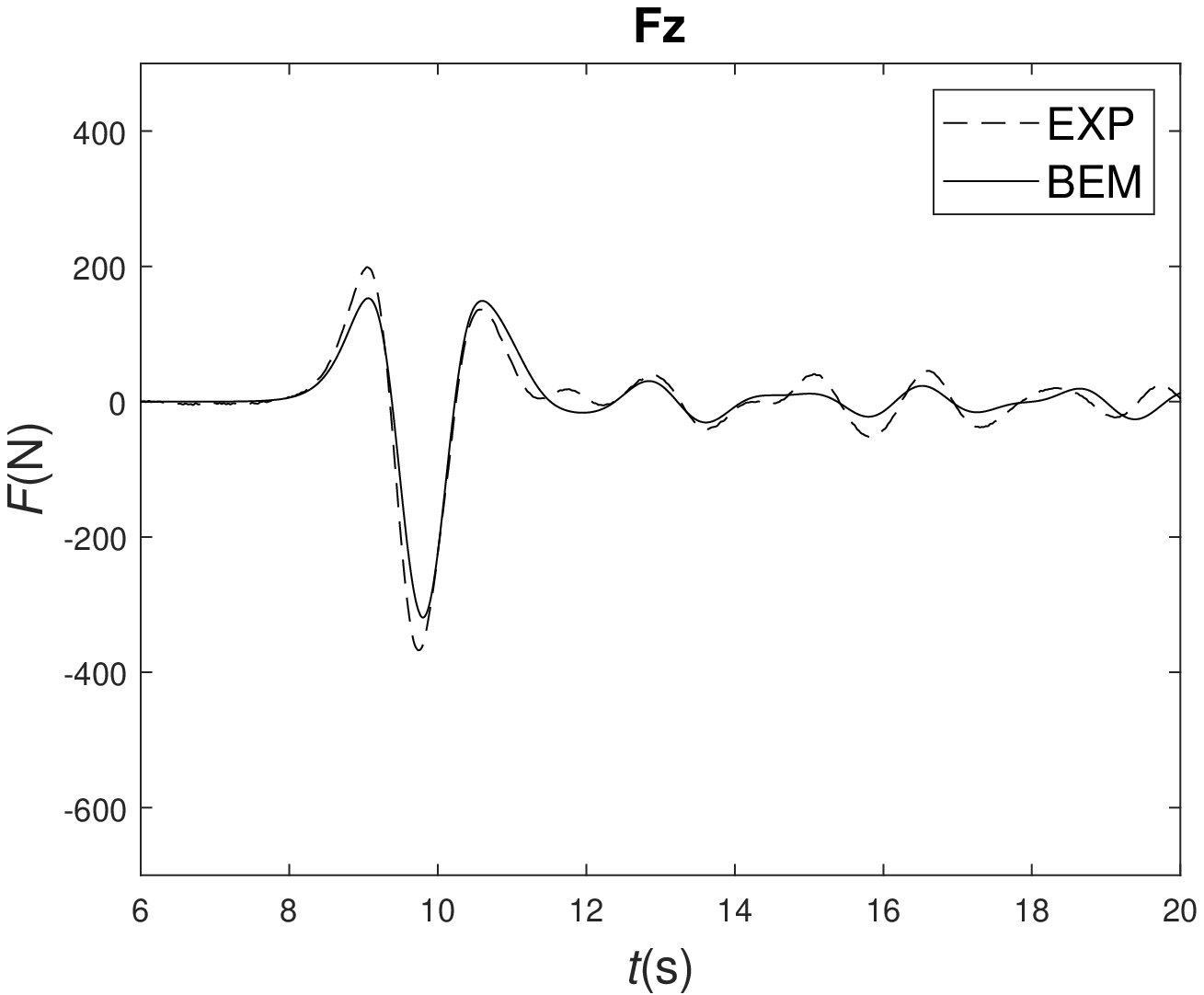}
\includegraphics[width=0.45\columnwidth]{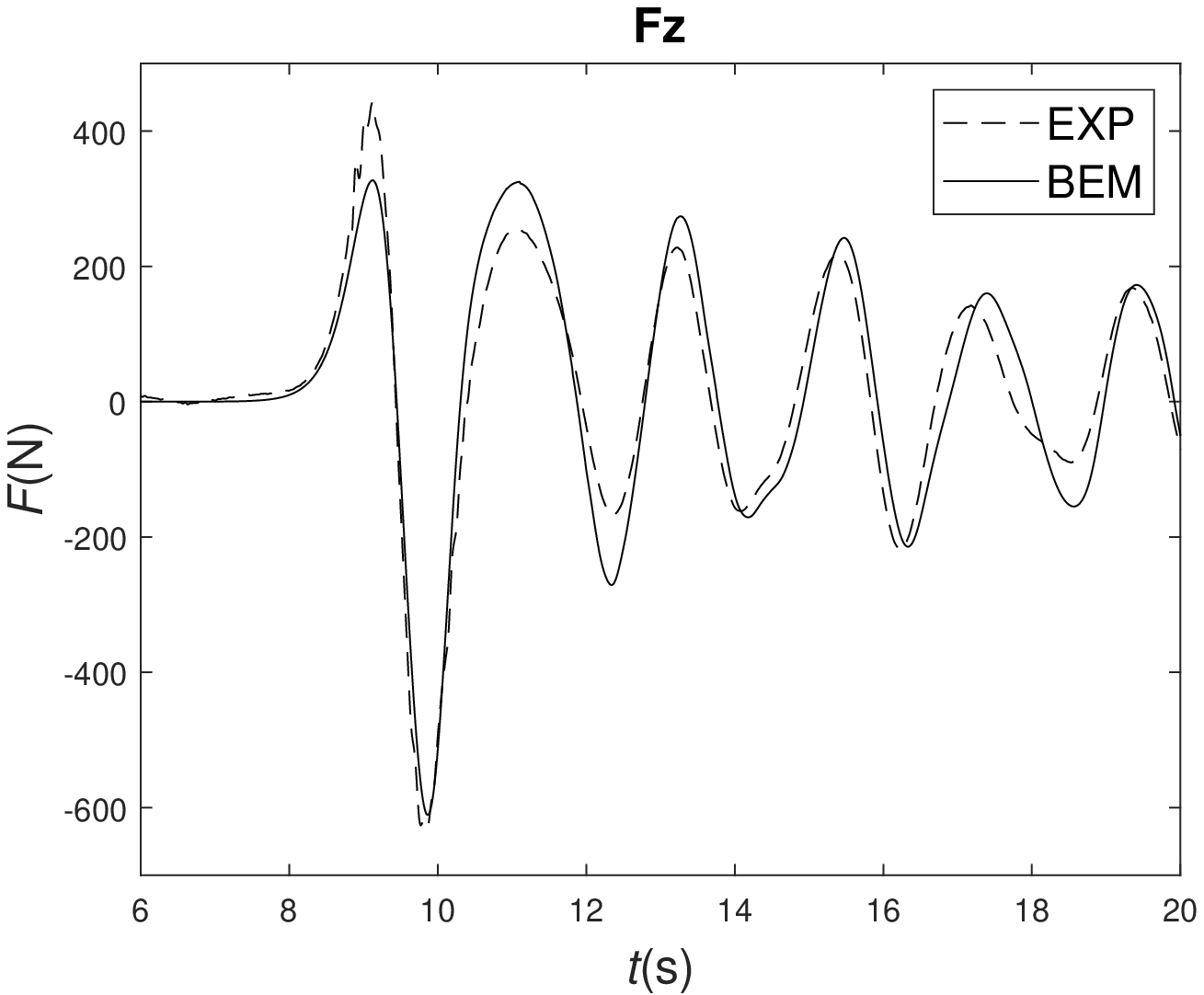} \\
\includegraphics[width=0.45\columnwidth]{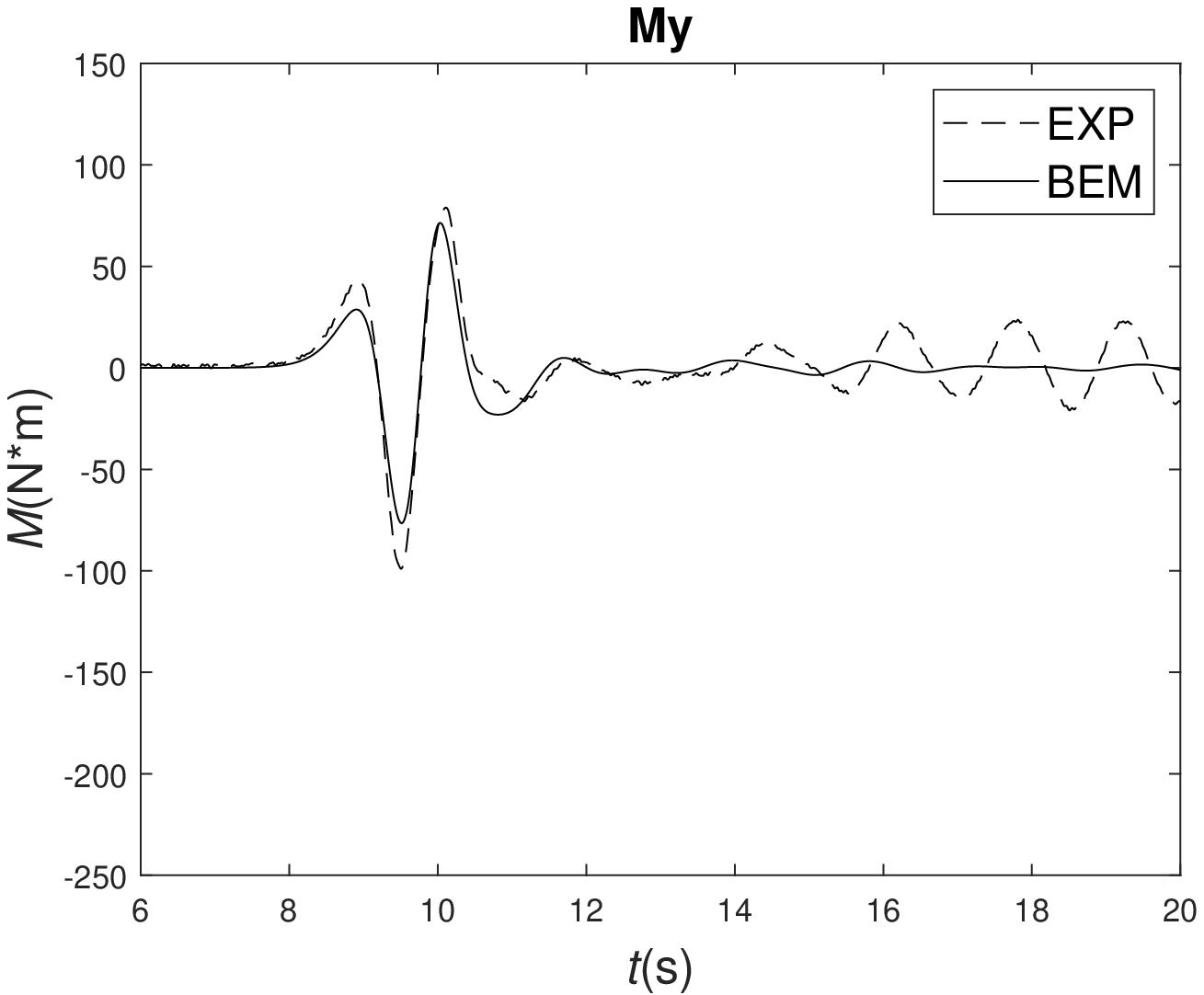}
\includegraphics[width=0.45\columnwidth]{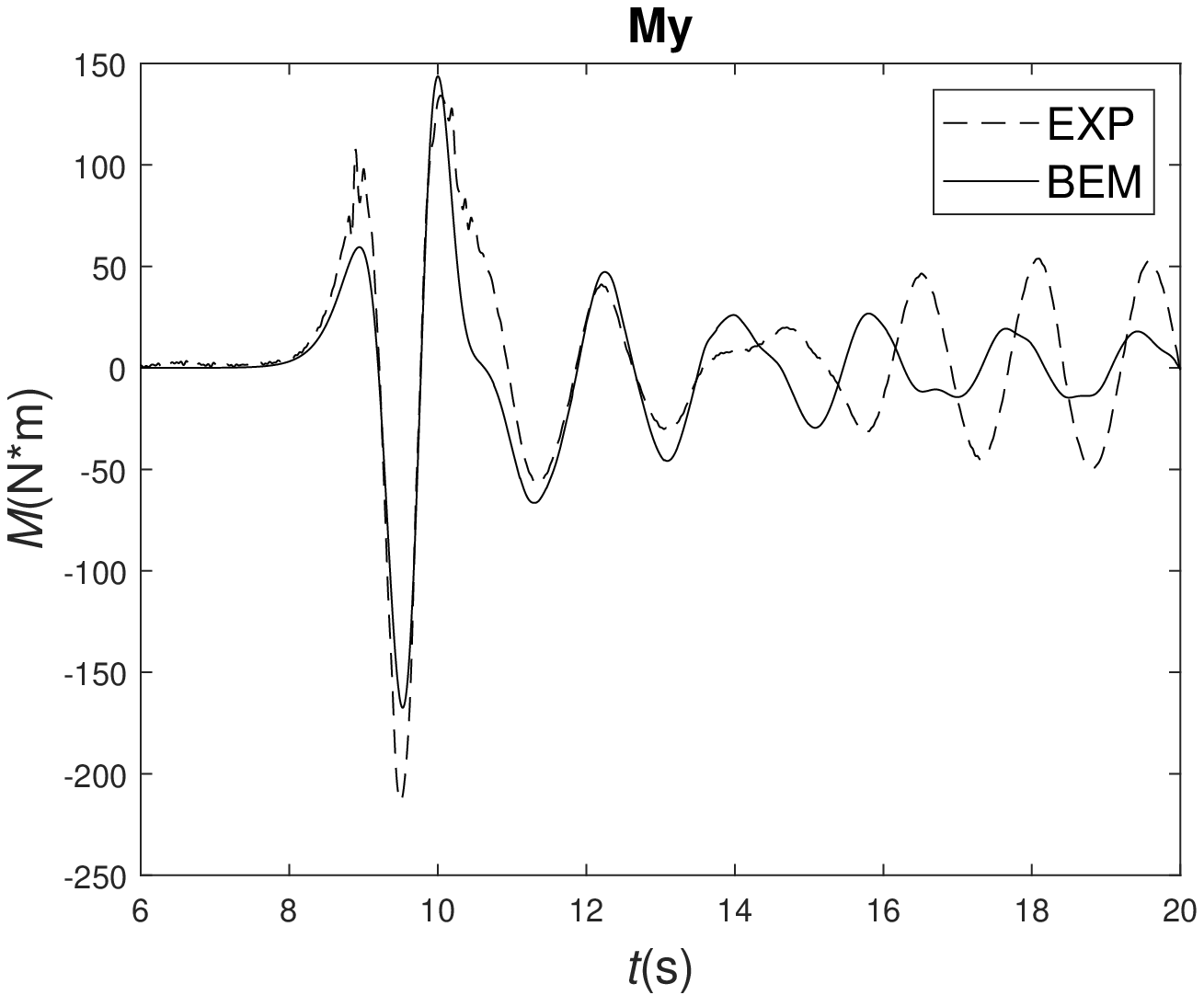}
\caption{Horizontal force $F_x$, vertical force $F_z$ and pitching moment $M_y$ for two different runs with $h=60$ cm, $H/h=0.4$; Left: $B=20$ cm, Right: $B=40$ cm.}
\label{fig_10}
\end{figure}

\begin{figure}[htbp]{
    \includegraphics[width=0.45\columnwidth]{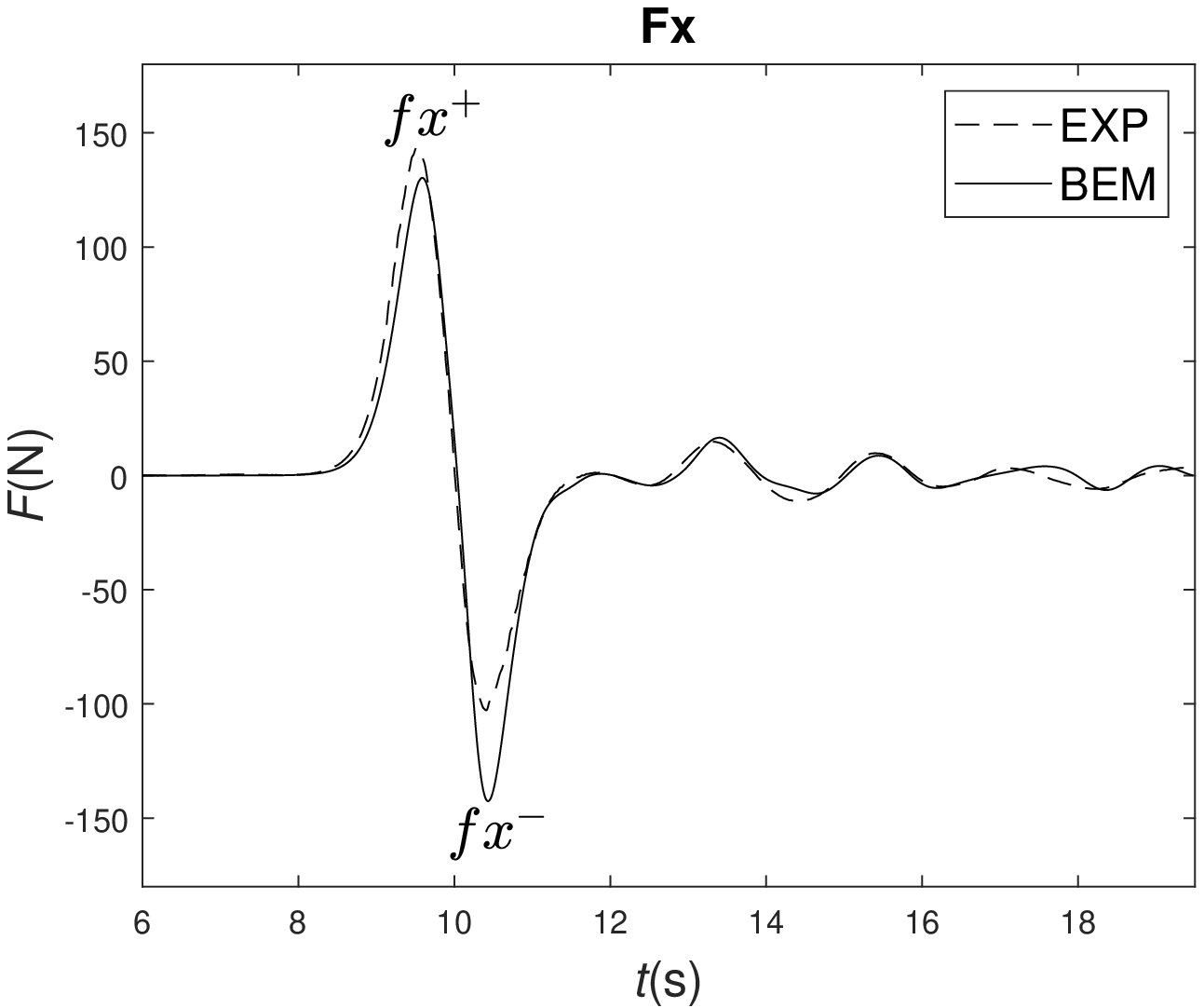} 
    \includegraphics[width=0.45\columnwidth]{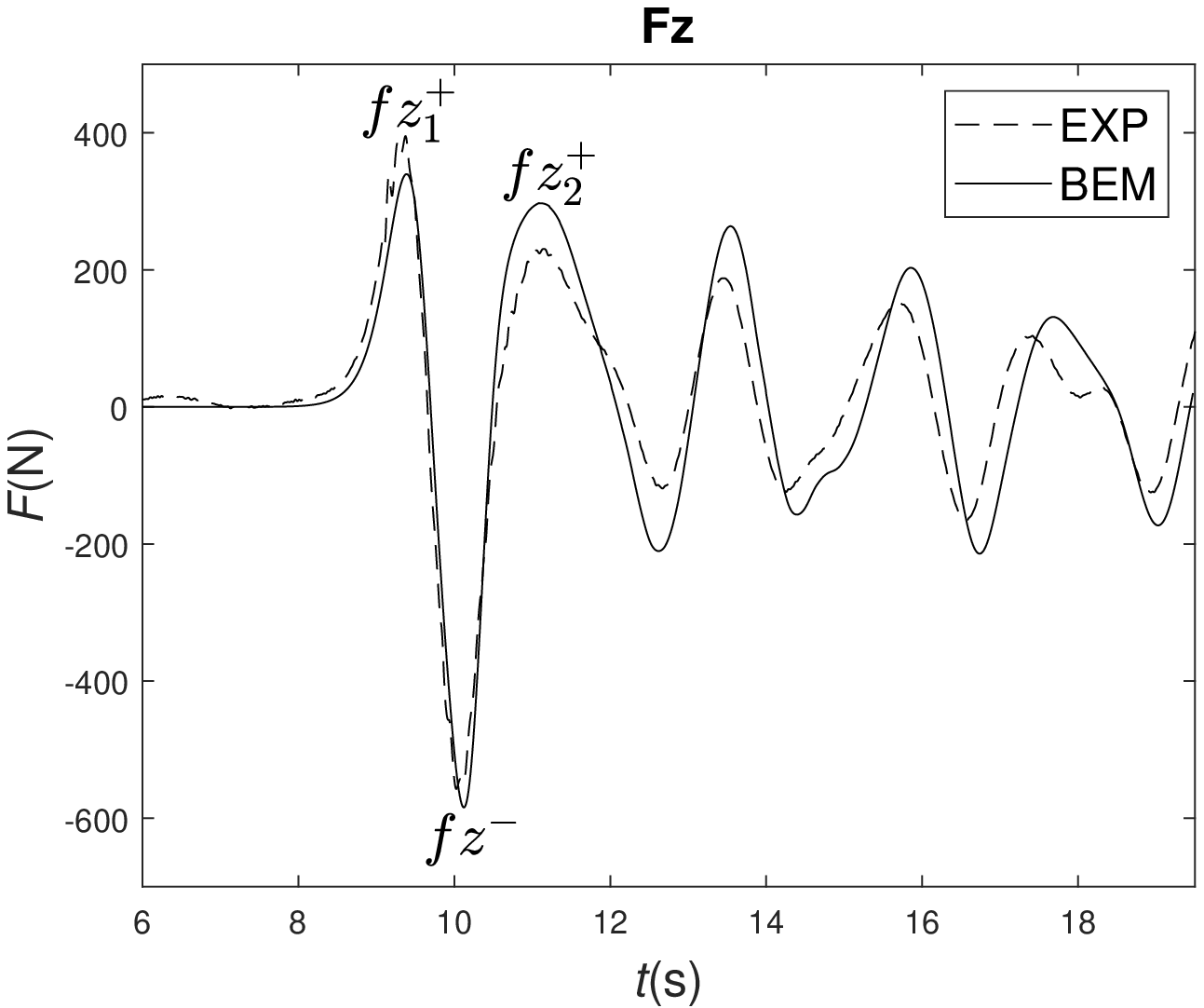} \\
    \includegraphics[width=0.45\columnwidth]{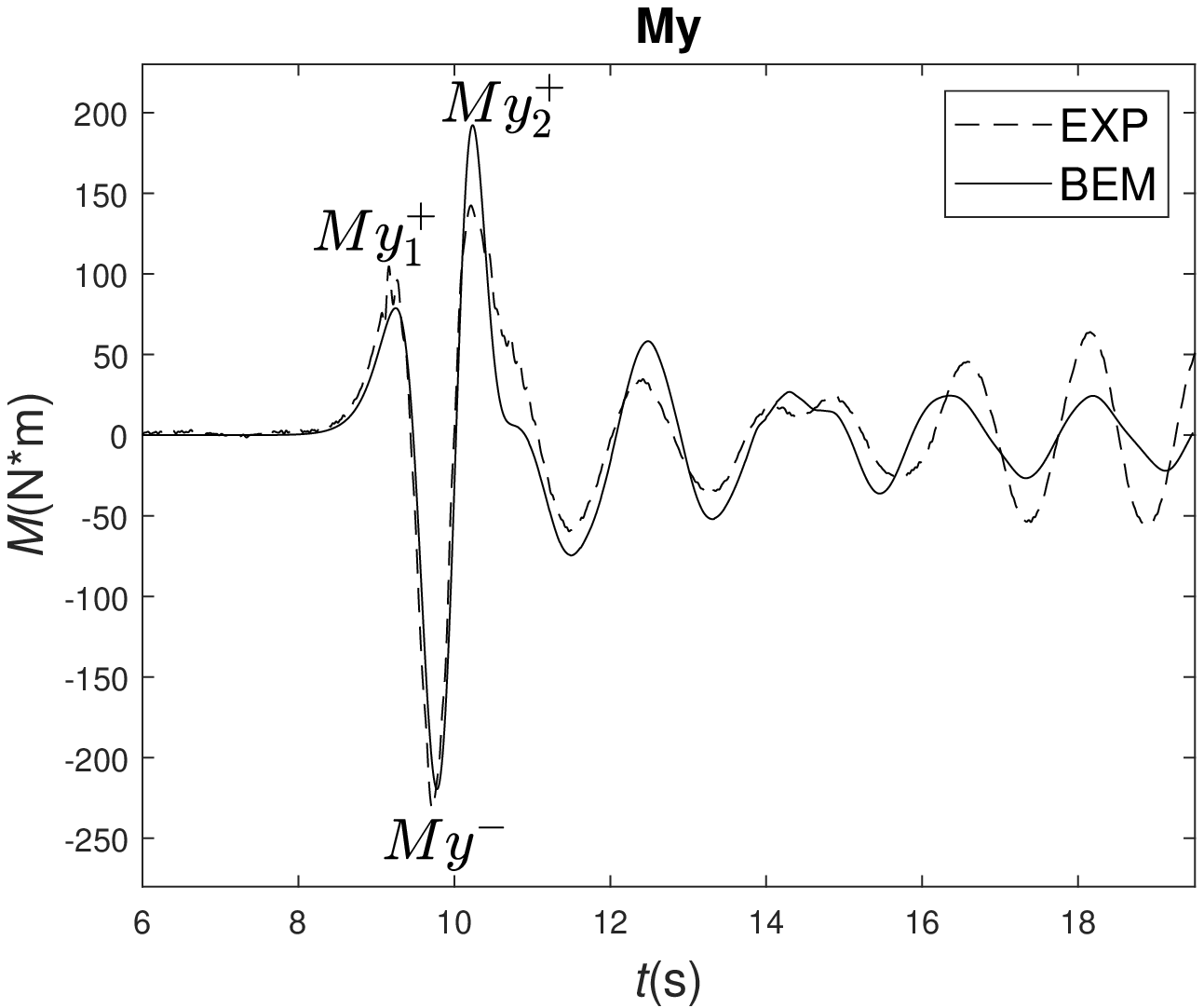}%
}
\caption{\label{fig_11} Extreme value of the hydrodynamic loads $F_x$, $F_z$ and $M_y$ when $h=50$ cm, $H/h=0.5$, $B=30$ cm.}
\end{figure}

As said above, we ran a lot of cases. In Fig \ref{fig_10}, we show the results for two cases which we believe are representative of all cases. In these two cases, all parameters are identical except the depth of submergence of the plate. A third case is shown in Fig \ref{fig_11}. In terms of horizontal and vertical forces, the numerical results agree well with the experimental data, even if the first peak is systematically under predicted by the numerical code. The agreement for the moment is not as good: after some time, the numerical values deviate from the experimental values. The greater the free-surface deformation, the longer the moment agrees (this is shown in the Supplementary Material). The moment is dominated by the pressure distribution on the upper and lower surfaces of the plate. What is intriguing in the experimental data is that the moment decreases first and then amplifies as seen for example in Fig \ref{fig_10}. In the subplot at the lower left, the experimental moment is very small for $t$ between 12 and 14 s but oscillations of increasing amplitude appear for $t > 14$ s. In the subplot at the lower right, the experimental moment develops oscillations of increasing amplitude for $t > 15$ s. The match between experimental and numerical values is poor and remains unexplained at this stage. For the horizontal force, as observed in both cases, the plate first experiences a positive force and then a negative force. This sequence occurs when the solitary wave hits the leading edge and then leaves the trailing edge. The basic structure of the vertical force has been explained in \cite{Lo2013}. It is of interest to explore the oscillations observed in the hydrodynamic loads as well as in the wave gauge data. Their connection will be explained in the next section.

\begin{figure}{
    \includegraphics[width=0.48\columnwidth]{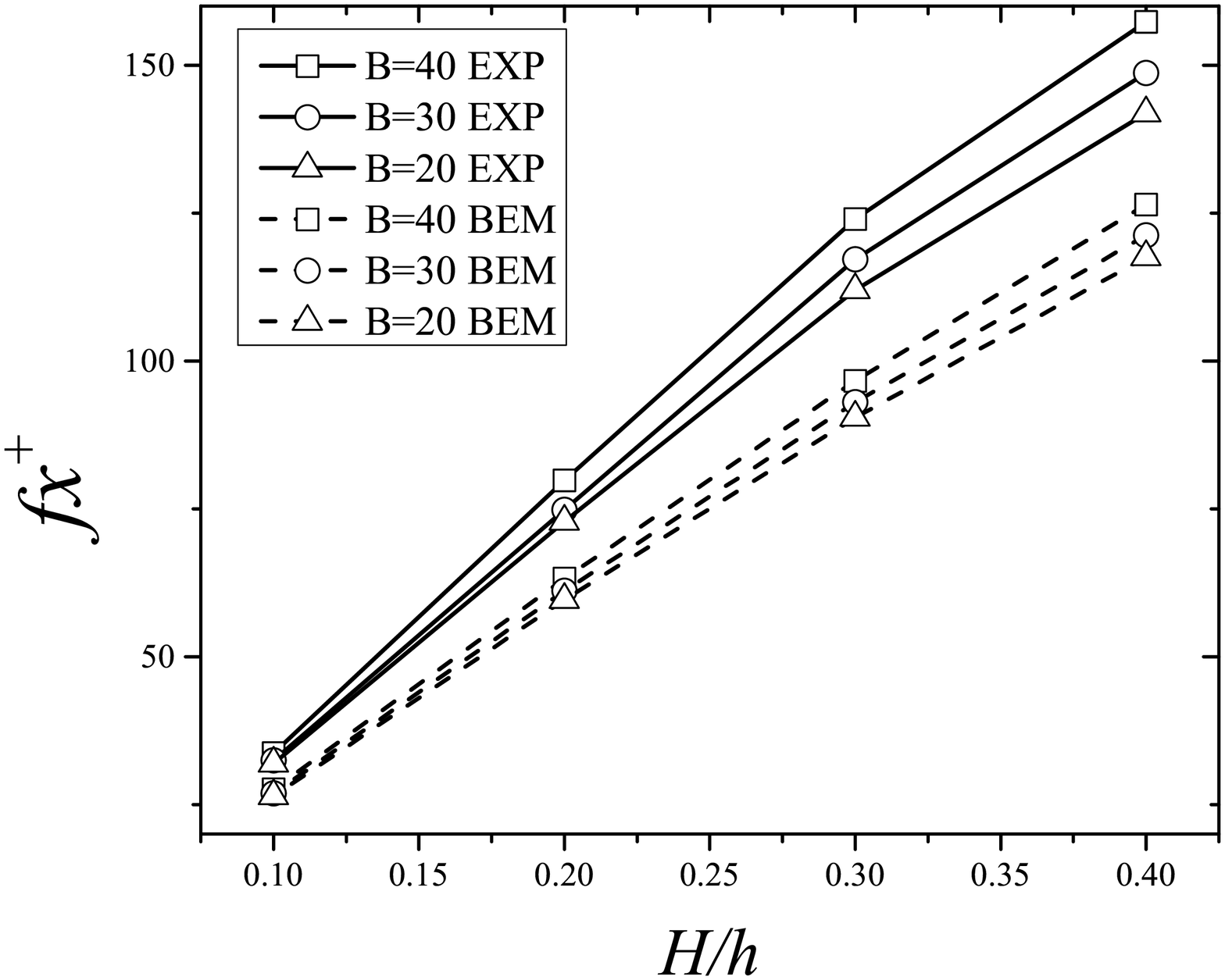} 
    \includegraphics[width=0.48\columnwidth]{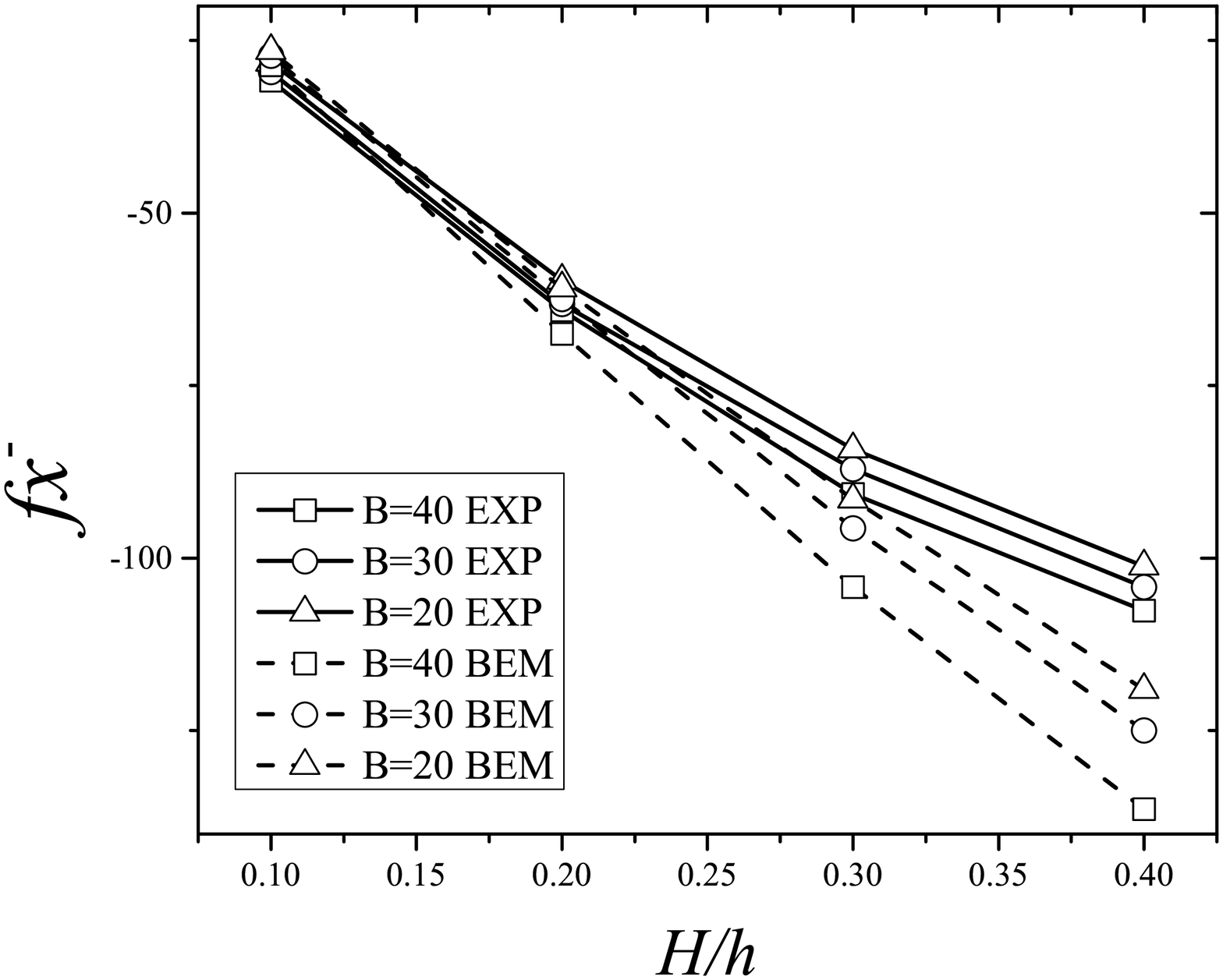}%
}
\caption{\label{fig_12} $h=60$ cm: Positive maximum and negative minimum horizontal force as a function of $H/h$ for various values of $B$.}
\end{figure}

\begin{figure}{
    \includegraphics[width=0.48\columnwidth]{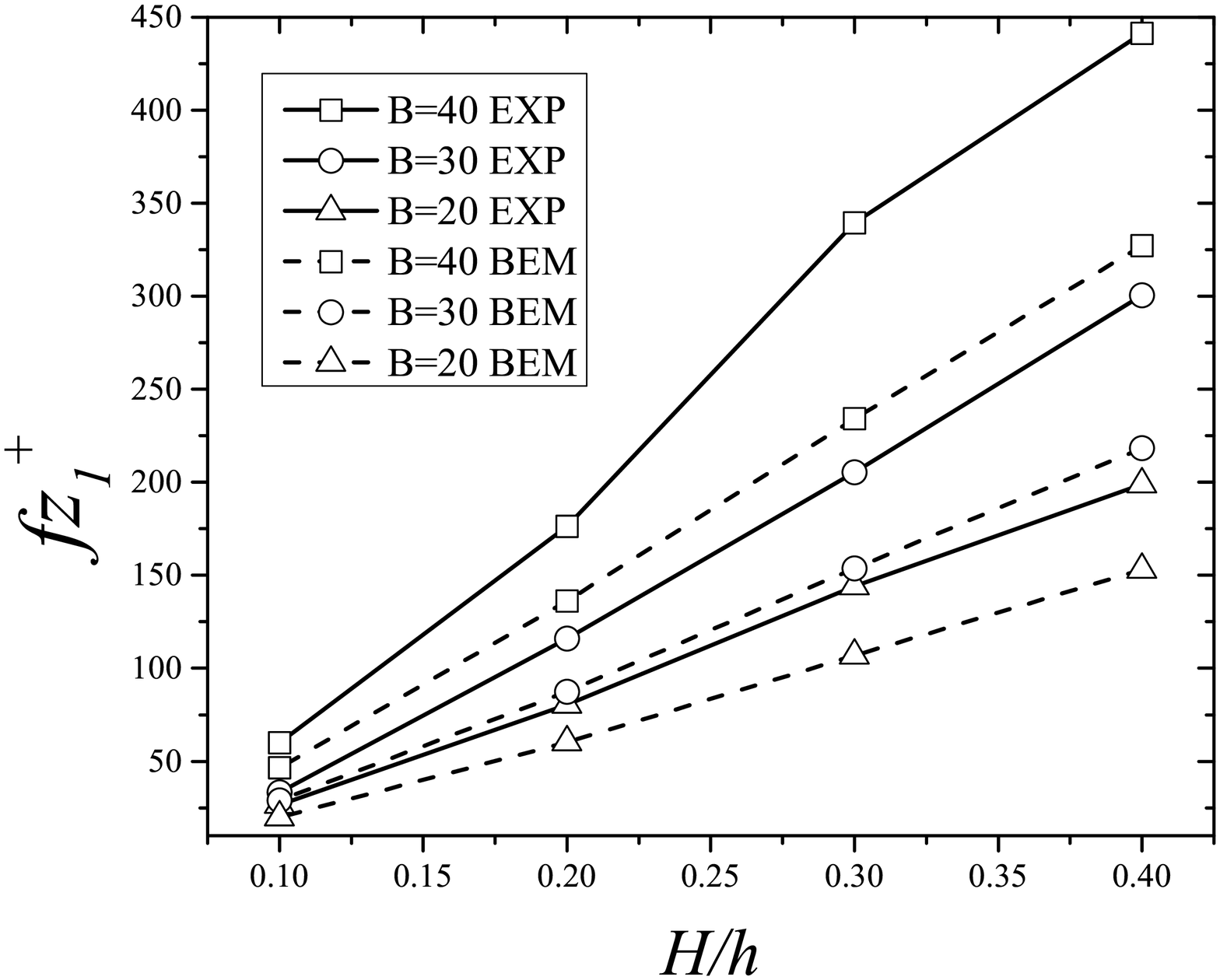} 
    \includegraphics[width=0.48\columnwidth]{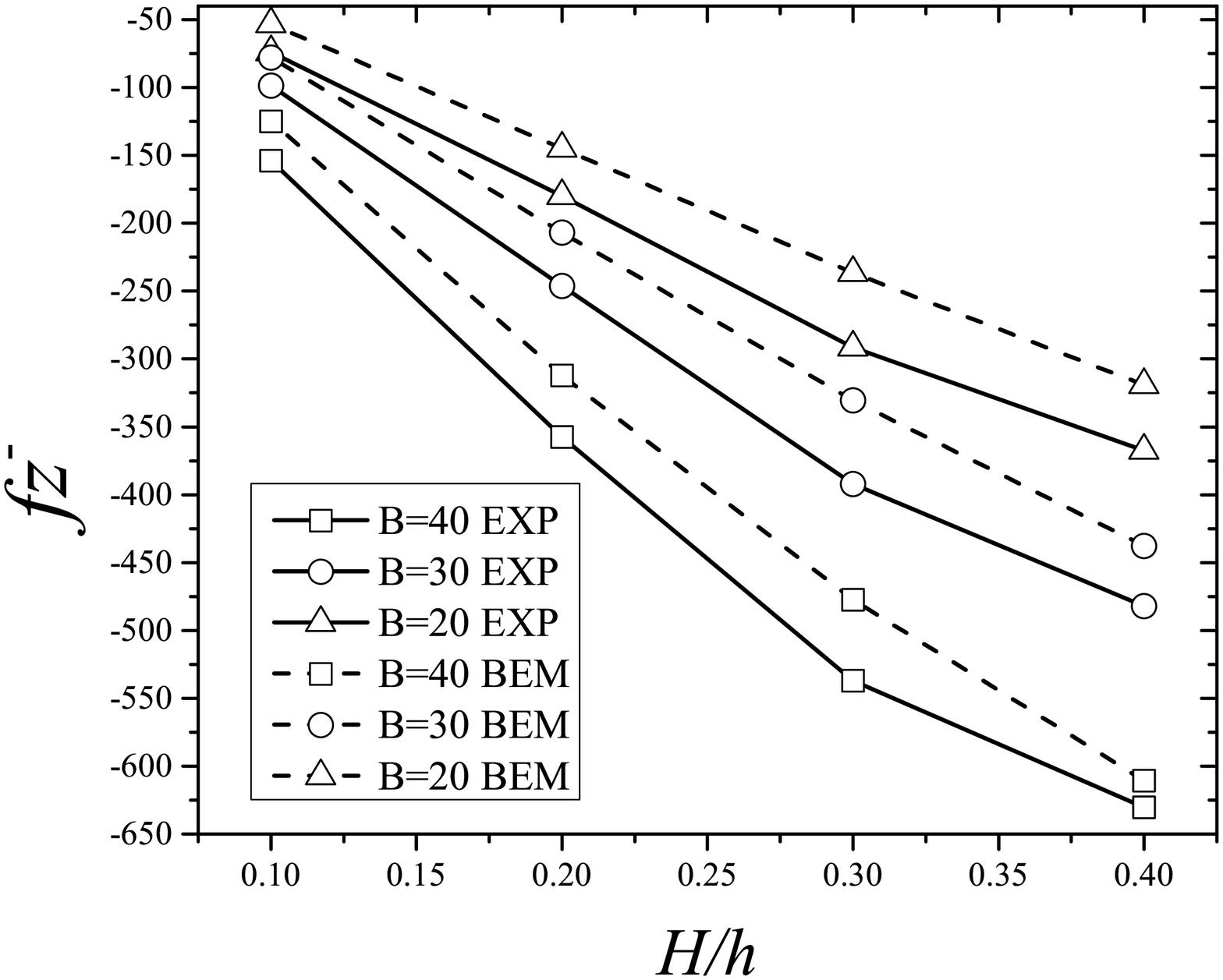} \\
    \includegraphics[width=0.48\columnwidth]{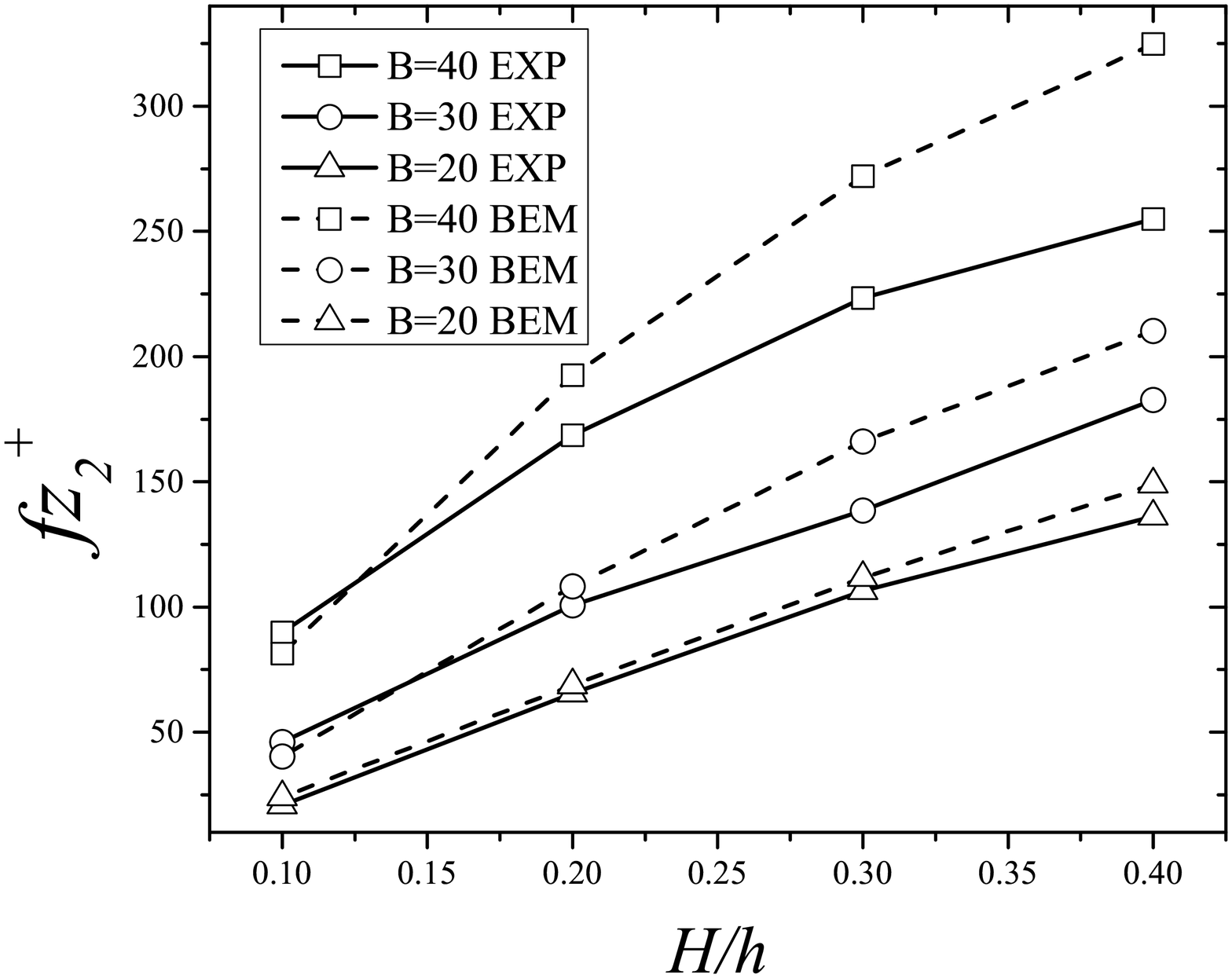}%
}
\caption{\label{fig_13} $h=60$ cm: Positive maximum and negative minimum vertical force as a function of $H/h$ for various values of $B$.}
\end{figure}

\begin{figure}{
    \includegraphics[width=0.48\columnwidth]{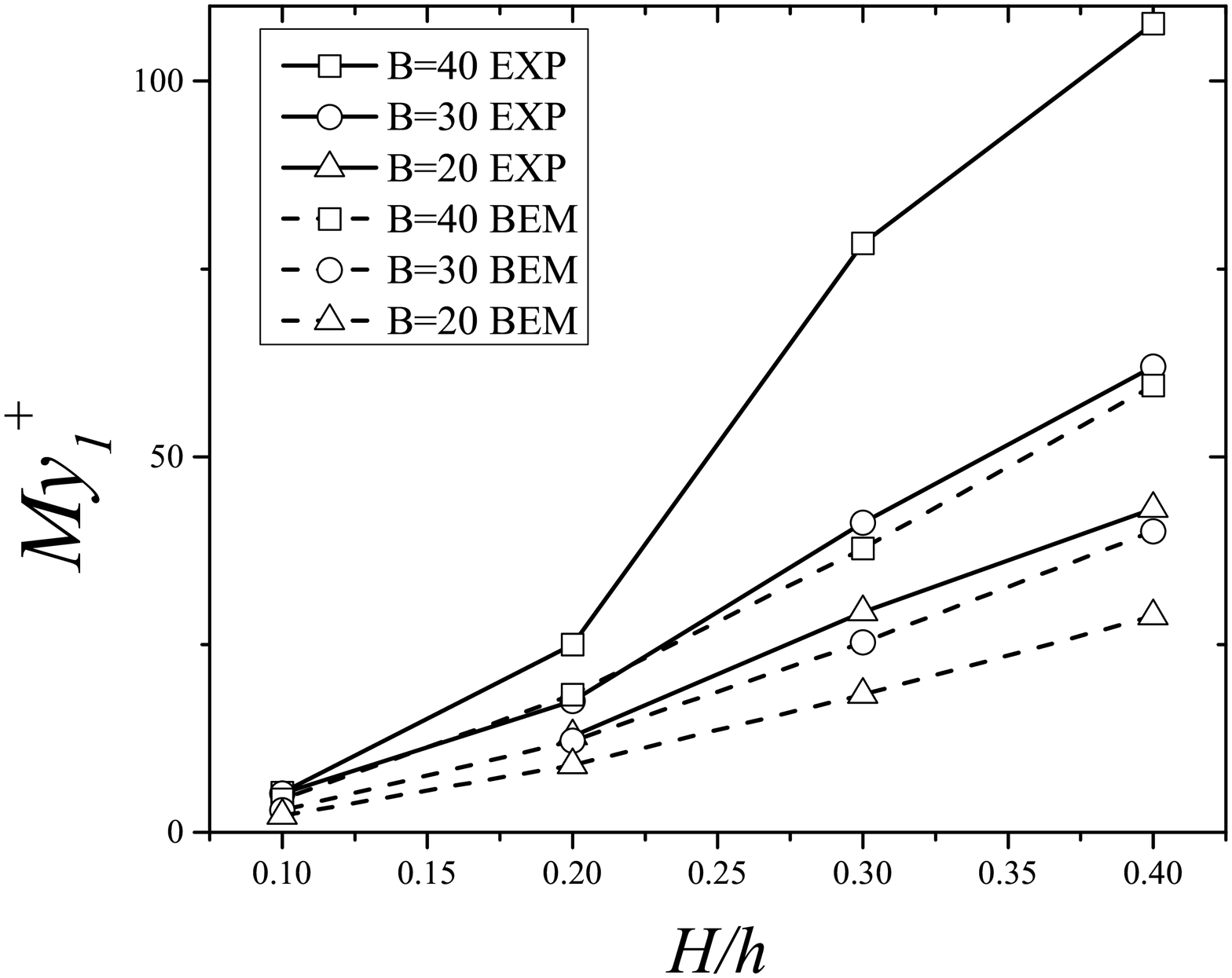} 
    \includegraphics[width=0.48\columnwidth]{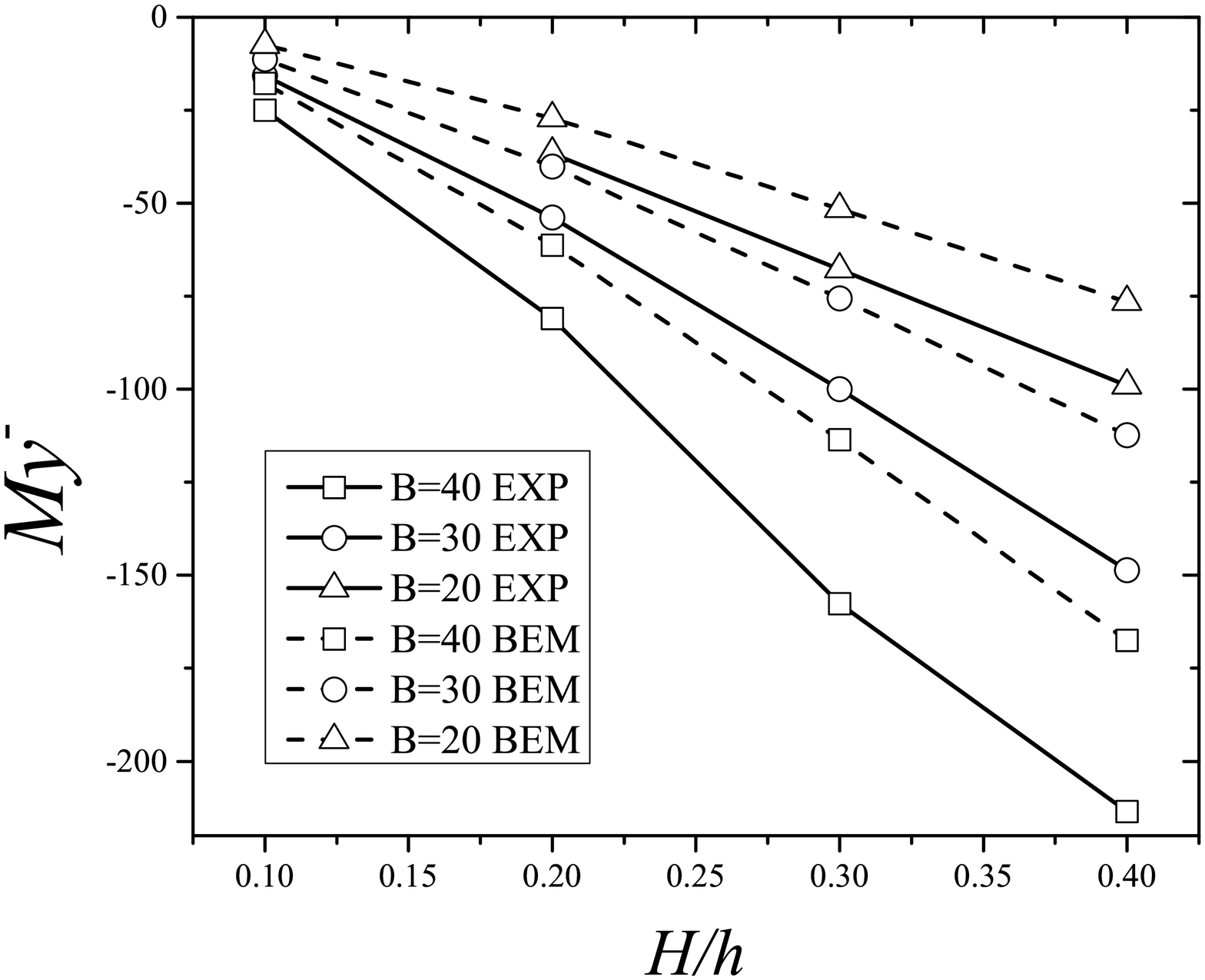} \\
    \includegraphics[width=0.48\columnwidth]{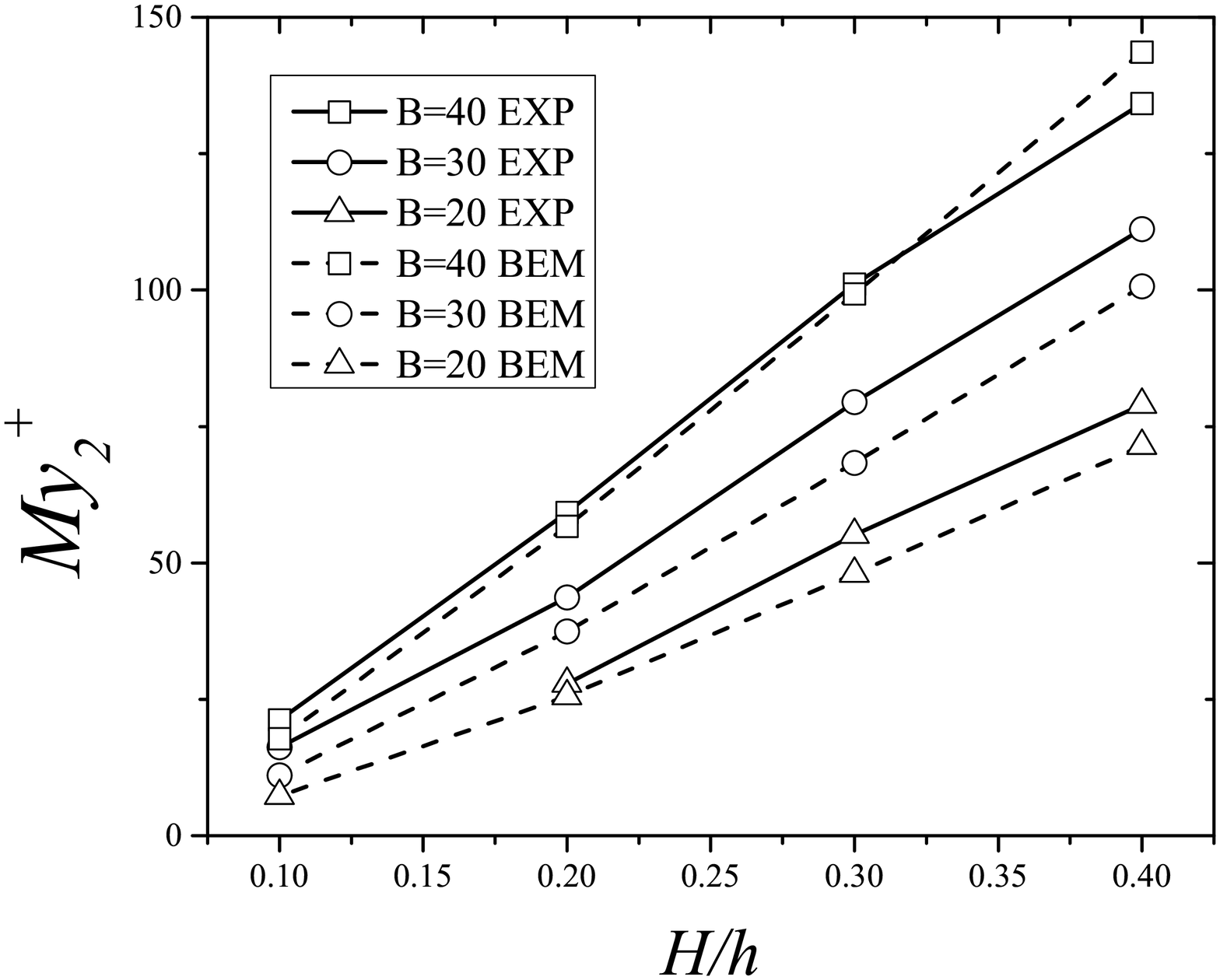}%
}
\caption{\label{fig_14} $h=60$ cm: Positive maximum and negative moment as a function of $H/h$ for various values of $B$.}
\end{figure}

As shown in Fig \ref{fig_11}, where results for a different set of parameter values are presented, we can define the positive maximum horizontal force as $fx^+$ and the negative minimum force as $fx^-$. The first peak values of the vertical force and moment are defined as $fz^+_1$ and $My^+_1$. The second peaks are defined as $fz^+_2$ and $My^+_2$. The negative minima are defined as $fz^-$ and $My^-$. As can be seen in Figs \ref{fig_12}, \ref{fig_13} and \ref{fig_14}, the discrepancy between experimental and numerical values increases as the wave amplitude increases. As explained in \cite{Lo2013,Poupardin2012}, the discrepancy is most likely due to the presence of a boundary layer and vortex shedding along the plate, which alter the pressure distribution on the plate. In general, the numerical results capture the trend of these extreme values. Note that the purpose of the experiments of \cite{Wang2020} was not to detect boundary layer effects. The flow near the four edges is obviously more complicated than the rest of the flow. Strong vortices can form near these edges. 

In the following figures (Figs \ref{fzandfs} and \ref{fig2Dfs}), we show the 2D free-surface elevation at four selected times. Fig \ref{fzandfs}, which corresponds to the case $h=60$ cm, $H/h=0.3$, $B=40$ cm, shows the time evolution of the vertical force $F_z$ and of the free-surface elevation at the center of the plate. Four times labelled A, B, C and D have been selected for the snapshots of the 2D free surface shown in Fig \ref{fig2Dfs}. The oscillations of the vertical force are strongly linked with the free-surface deformation at the center of the plate. When the free surface reaches a crest, the vertical force appears to reach a trough, and vice versa. Figs \ref{figup} and \ref{figlp} show the pressure distribution on the upper and lower surface of the plate respectively at the four selected times A, B, C and D. As the solitary wave approaches the front edge of the plate, at time A, the dynamic pressure near the front edge increases both on the upper and lower surfaces. The pressure on the lower surface is greater than that on the upper surface due to the channel flow effect between the lower surface of the plate and the bottom, and leads to the first positive vertical force. The solitary wave keeps propagating and a bulge forms above the plate due to shoaling, which generates great dynamic pressure on the upper surface at time B, and leads to the first negative vertical force. Then at time C the solitary wave just passes over the plate. The pressure at the trailing edge is larger than that at the front edge. The large negative dynamic pressure on the upper surface leads to the second positive vertical force. Finally, at time D, the second bulge forms above the plate and generates positive dynamic pressure on the upper surface. Though the dynamic pressure is now much smaller than that at time B, it still leads to the second negative vertical force because the dynamic pressure on the lower surface becomes negligible. A discussion on the pressure distribution on a submerged circular plate due to a solitary wave can be found in Wu \textit{et al}. \cite{Wu2021}.

\begin{figure}
\centering
\includegraphics[width=0.65\columnwidth]{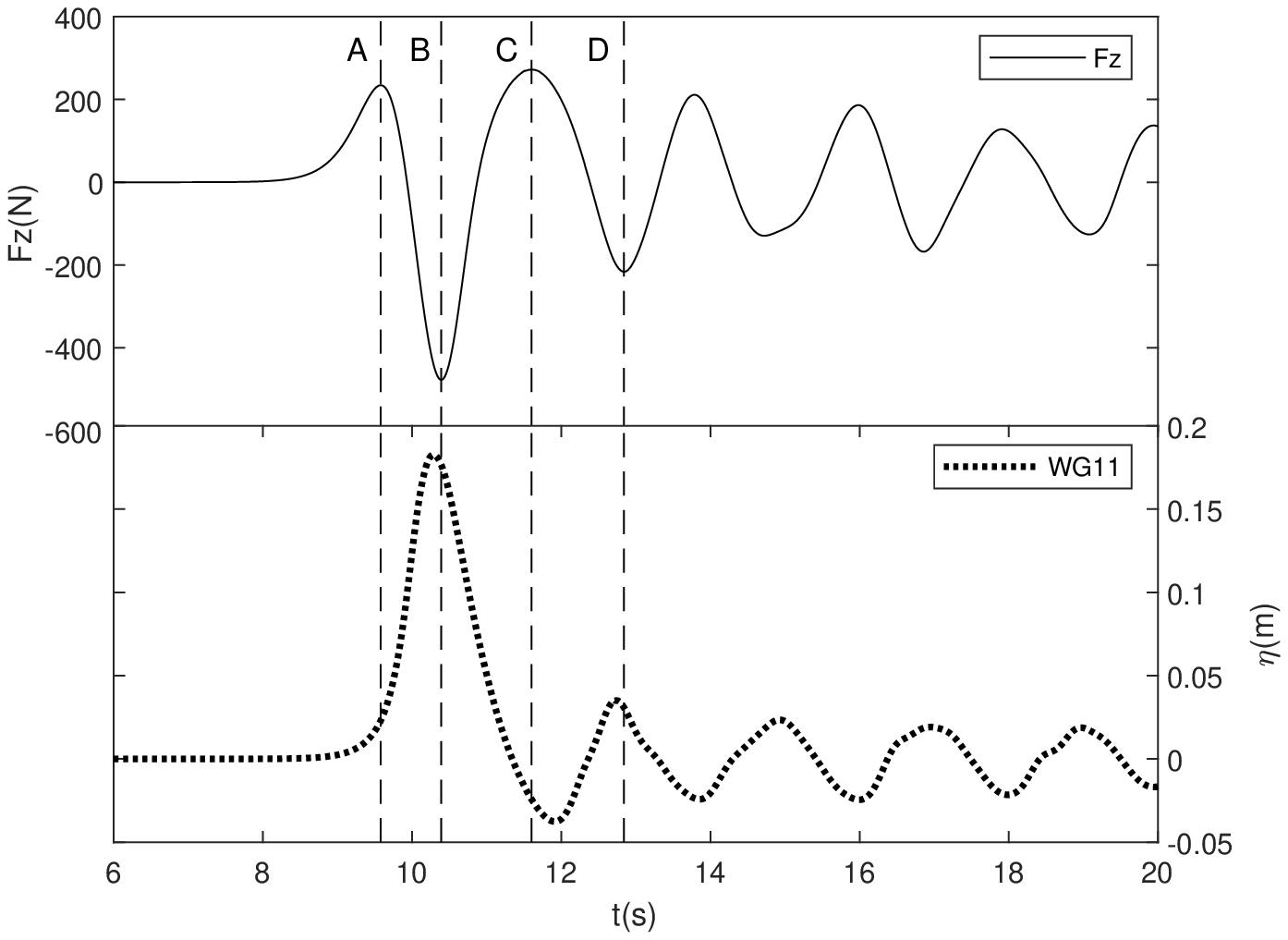}
\caption{Case $h=60$ cm, $H/h=0.3$, $B=40$ cm: Time history of the vertical force and of the free-surface elevation at WG11. Four times A, B, C and D have been selected, corresponding, respectively, to the first peak $fz_1^+$, first trough $fz_2^-$, second peak $fz_3^+$ and second trough $fz_4^-$ of the vertical force.}
\label{fzandfs}
\end{figure}

\begin{figure}
\centering
\includegraphics[width=0.45\columnwidth]{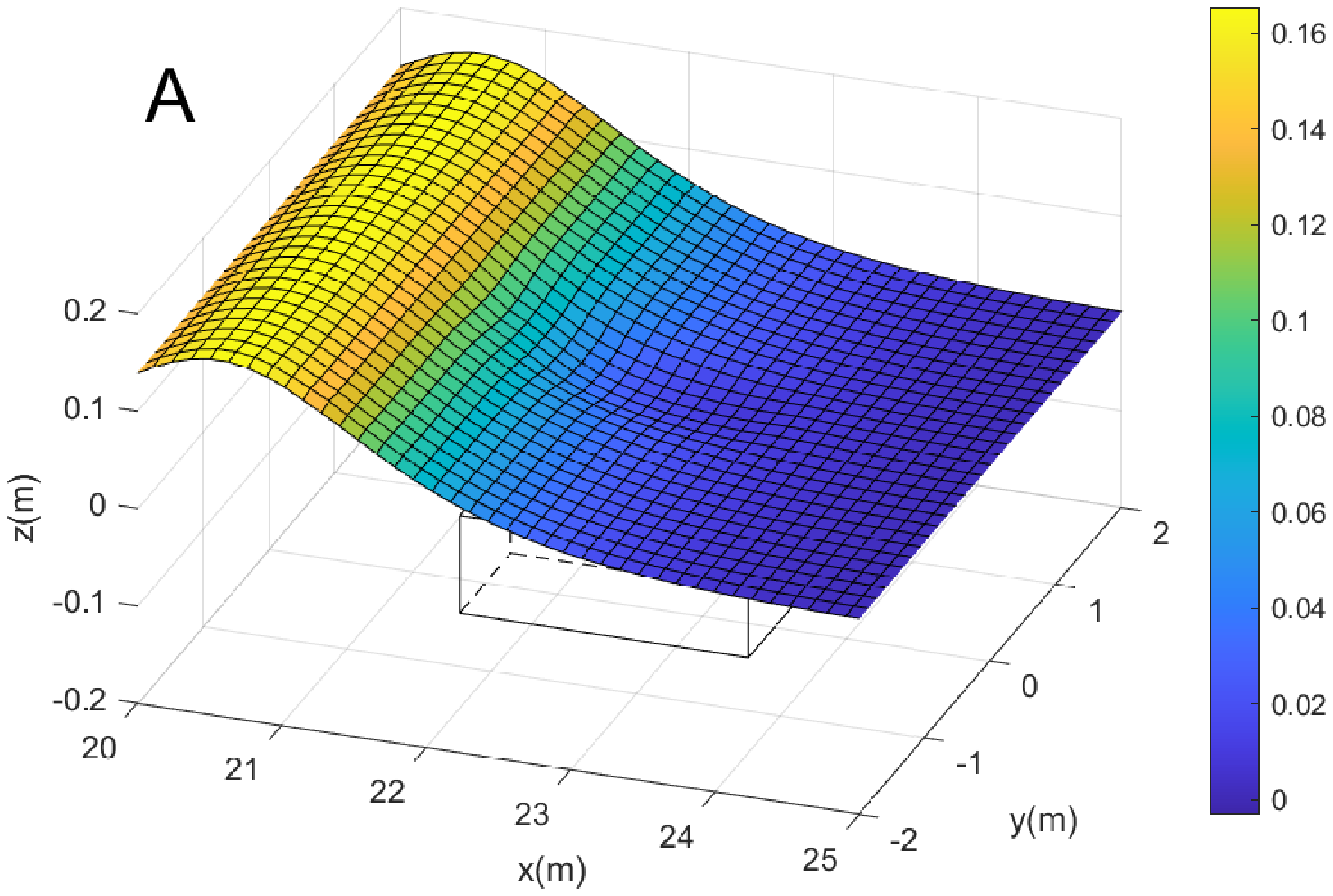}
\includegraphics[width=0.45\columnwidth]{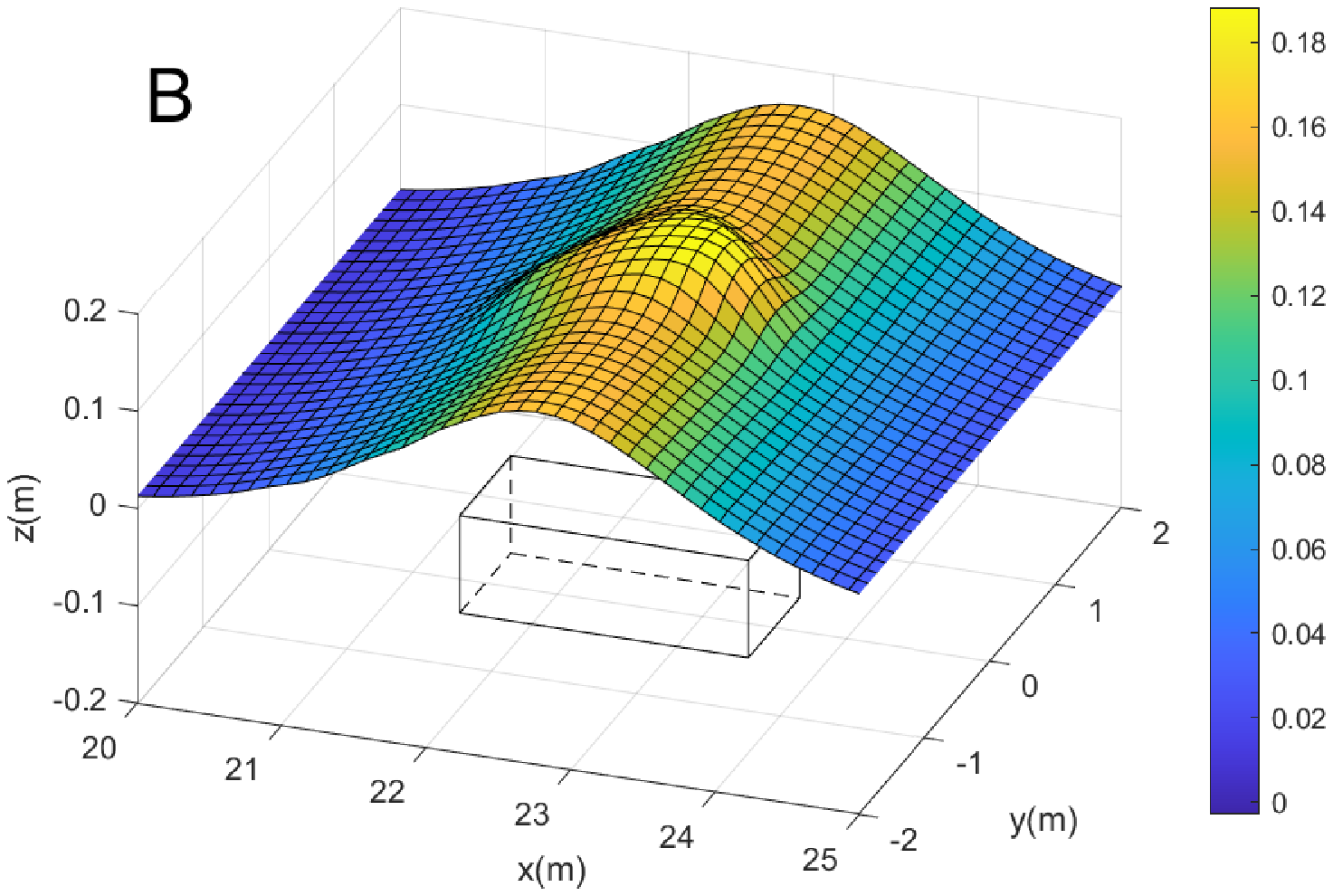} \\
\includegraphics[width=0.45\columnwidth]{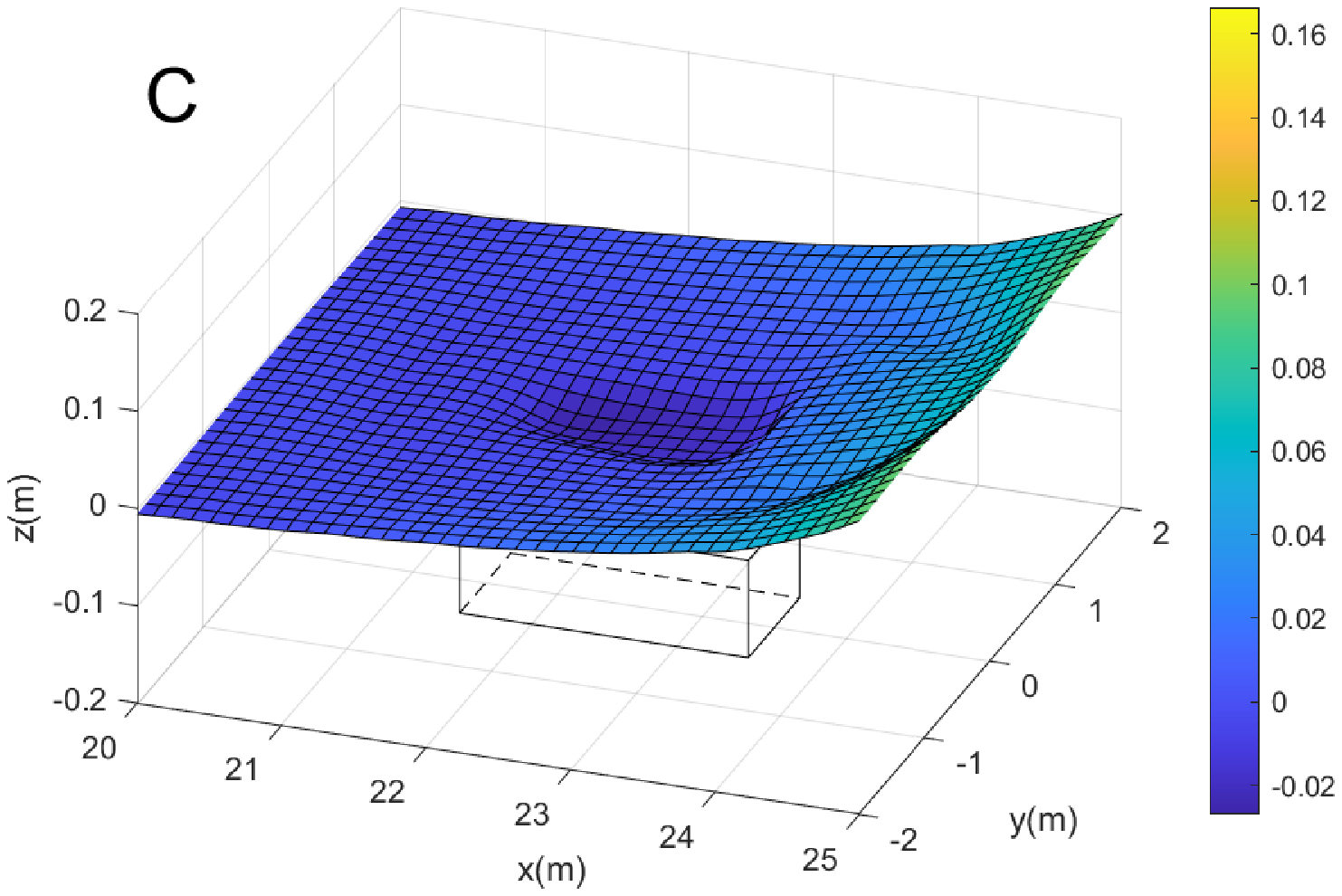}
\includegraphics[width=0.45\columnwidth]{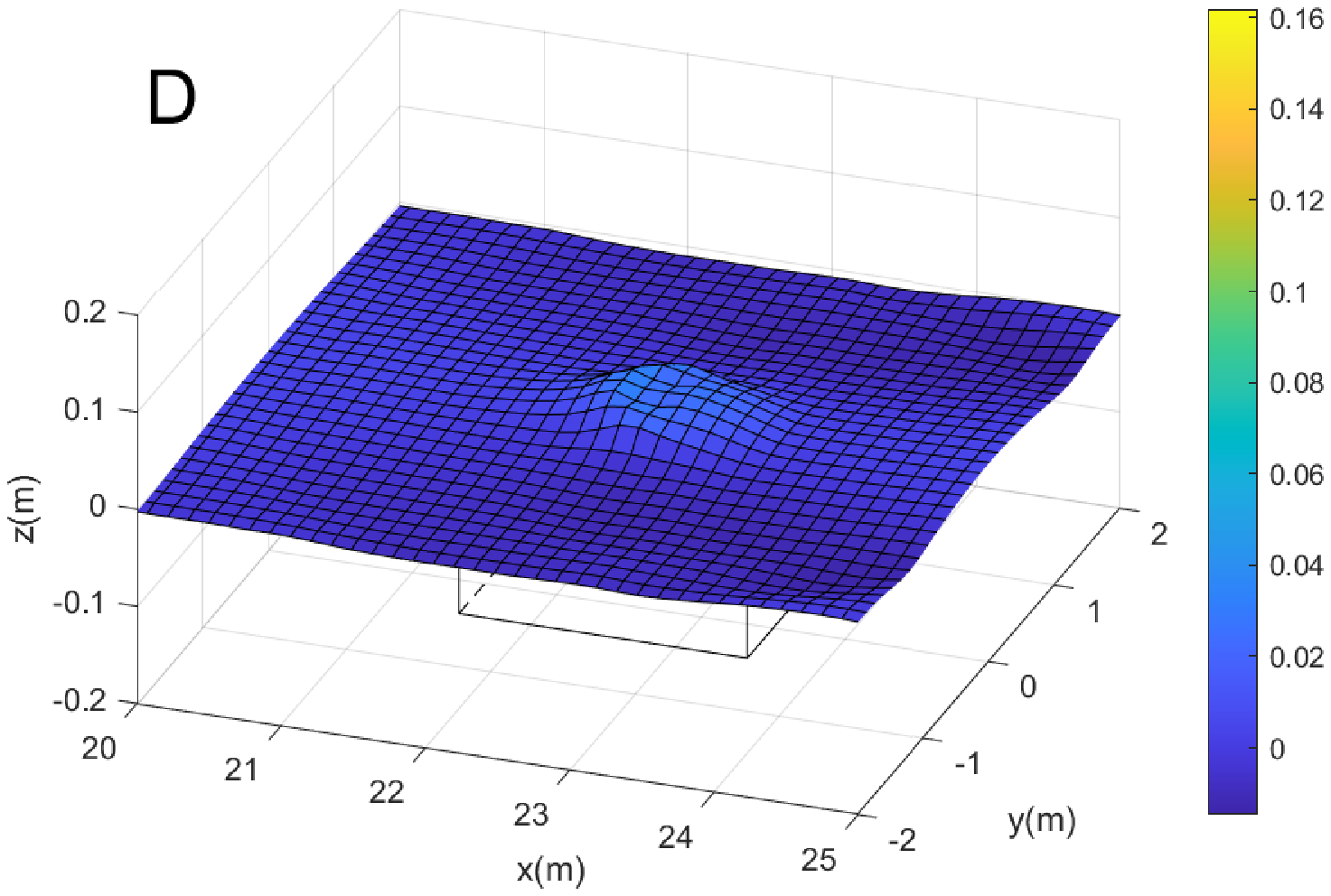}
\caption{Case $h=60$ cm, $H/h=0.3$, $B=40$ cm: Two-dimensional free-surface elevation at the four times A, B, C and D selected in Fig \ref{fzandfs}.}
\label{fig2Dfs}
\end{figure}

\begin{figure}
\centering
\includegraphics[width=0.45\columnwidth]{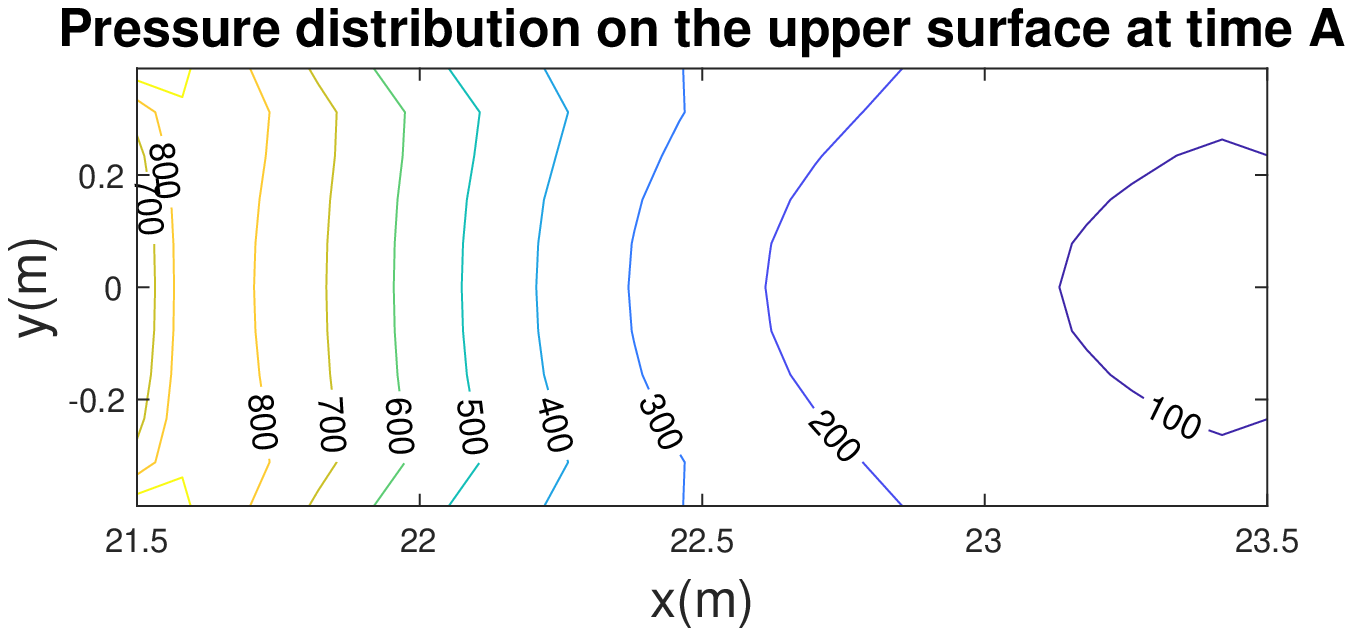}
\includegraphics[width=0.45\columnwidth]{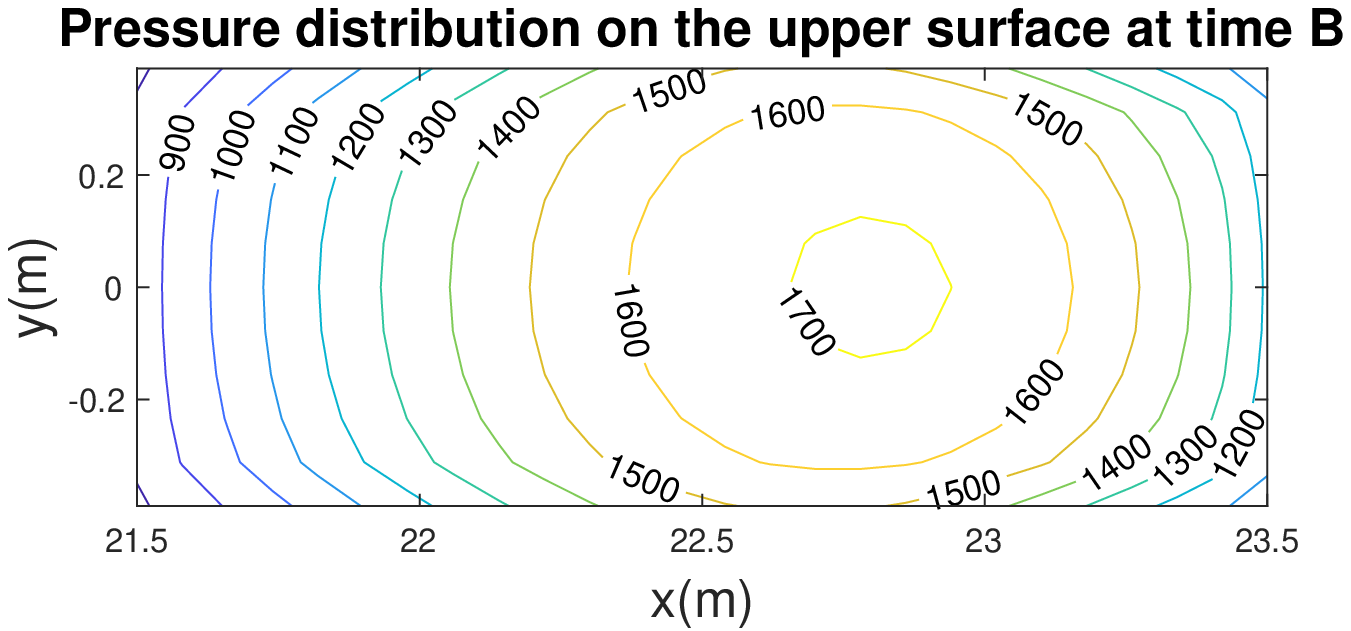} \\
\includegraphics[width=0.45\columnwidth]{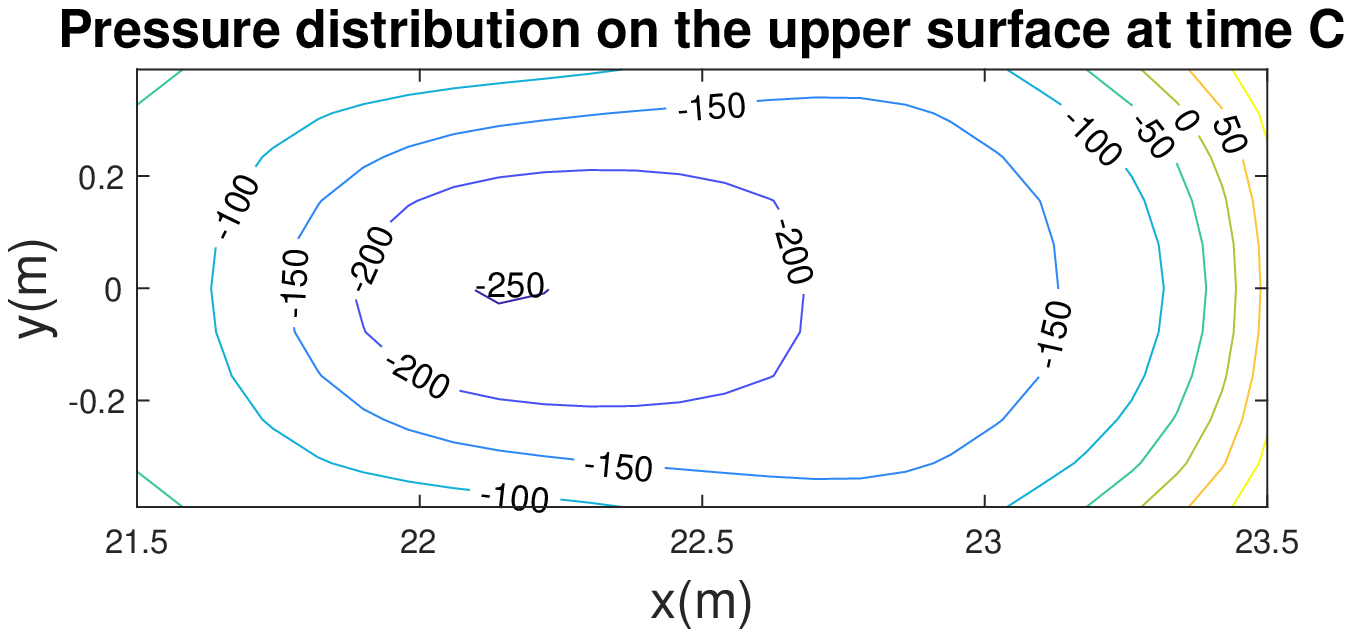}
\includegraphics[width=0.45\columnwidth]{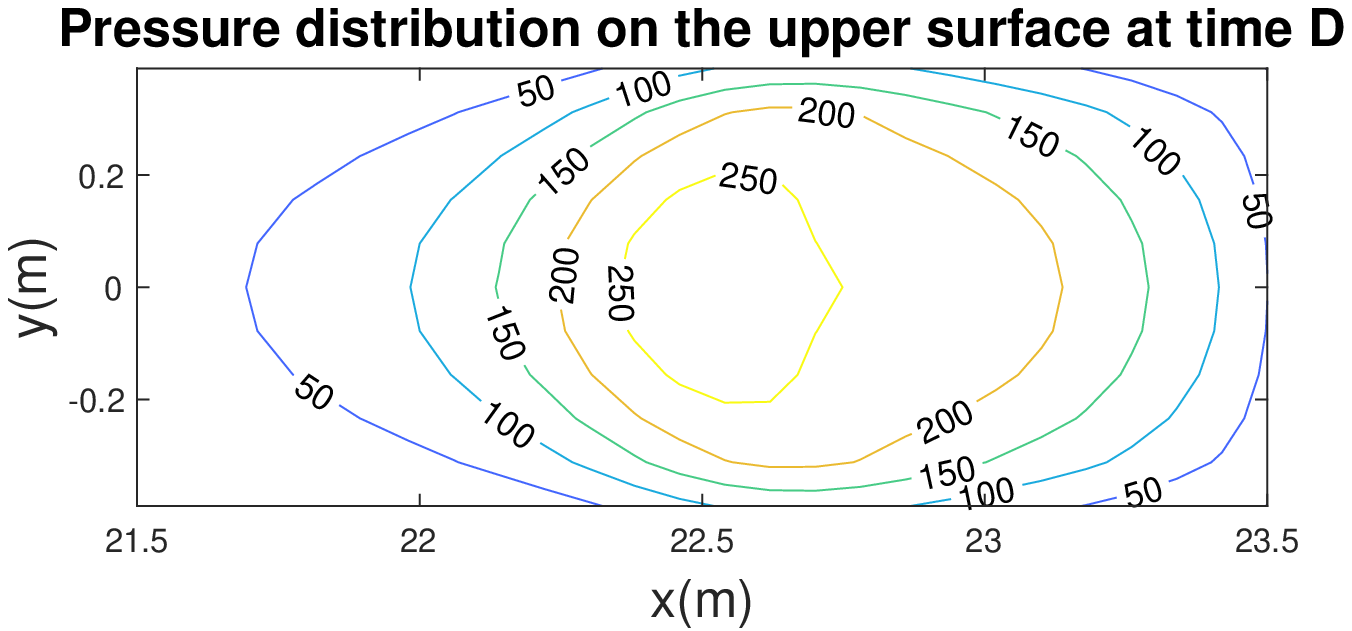}
\caption{Case $h=60$ cm, $H/h=0.3$, $B=40$ cm: Spatial distributions of the hydrodynamic pressure (isobars in Pa) on the upper surface of the plate at the four times A, B, C and D selected in Fig \ref{fzandfs}.}
\label{figup}
\end{figure}

\begin{figure}
\centering
\includegraphics[width=0.45\columnwidth]{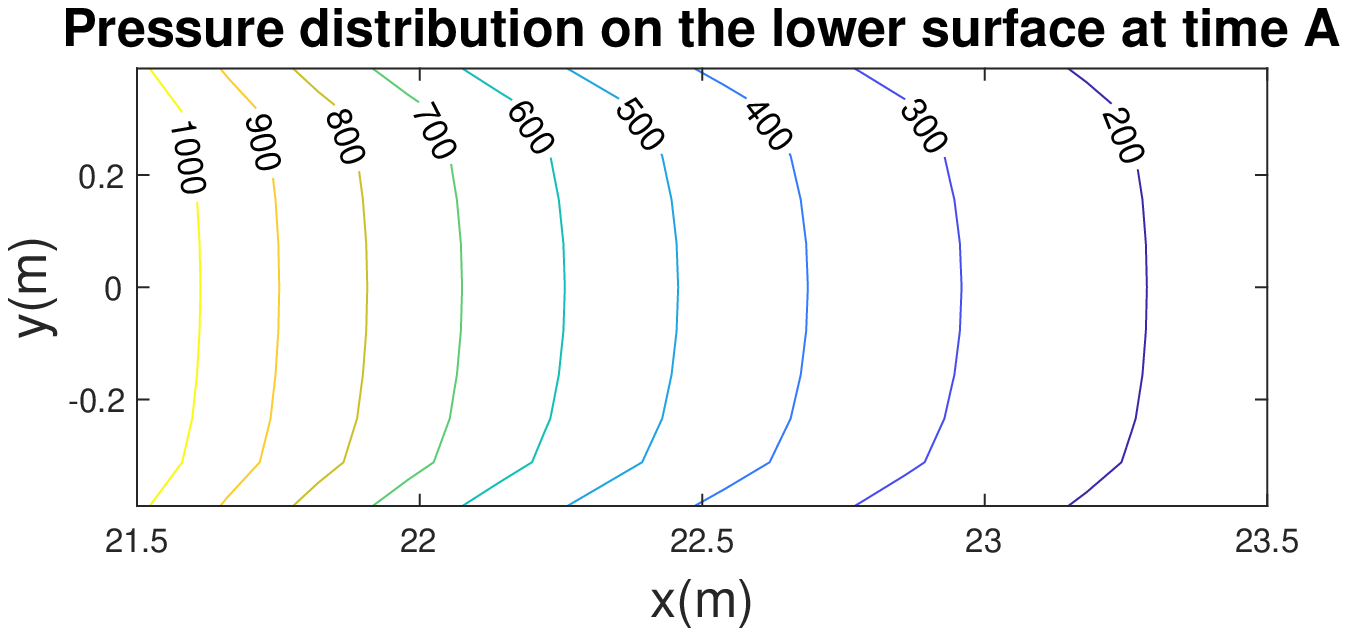}
\includegraphics[width=0.45\columnwidth]{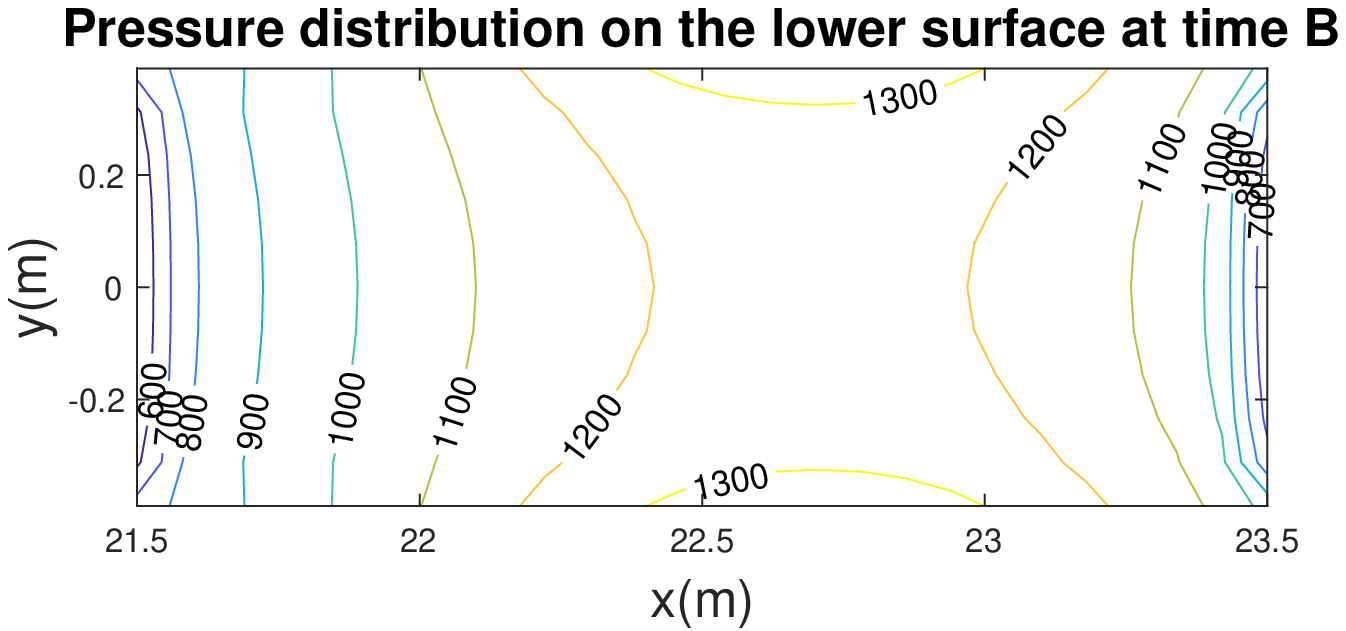} \\
\includegraphics[width=0.45\columnwidth]{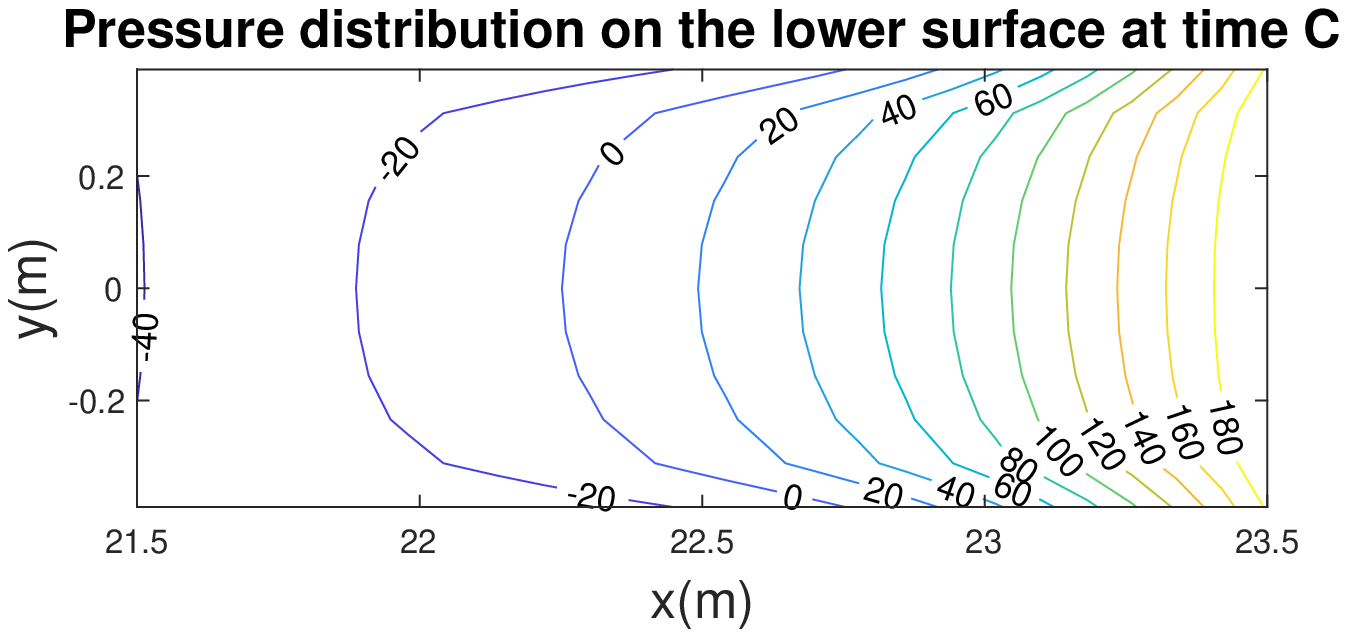}
\includegraphics[width=0.45\columnwidth]{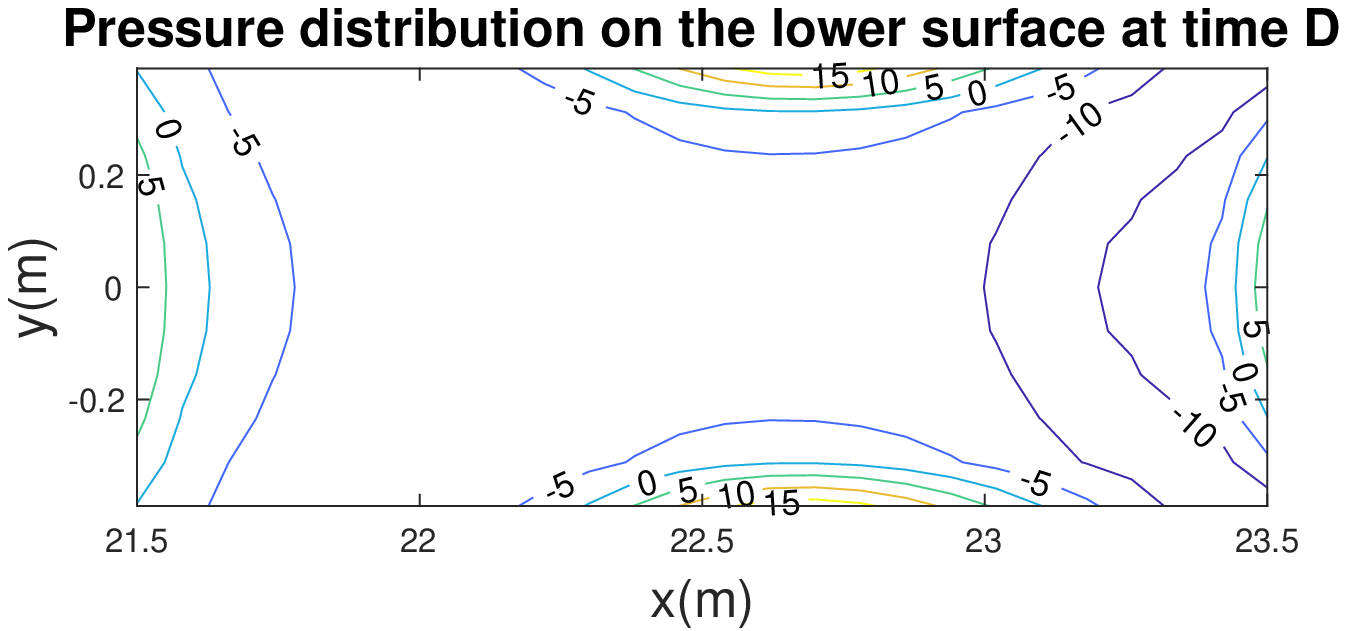}
\caption{Case $h=60$ cm, $H/h=0.3$, $B=40$ cm: Spatial distributions of the hydrodynamic pressure (isobars in Pa) on the lower surface of the plate at the four times A, B, C and D selected in Fig \ref{fzandfs}.}
\label{figlp}
\end{figure}

\begin{figure}
\centering
\includegraphics[width=0.65\columnwidth]{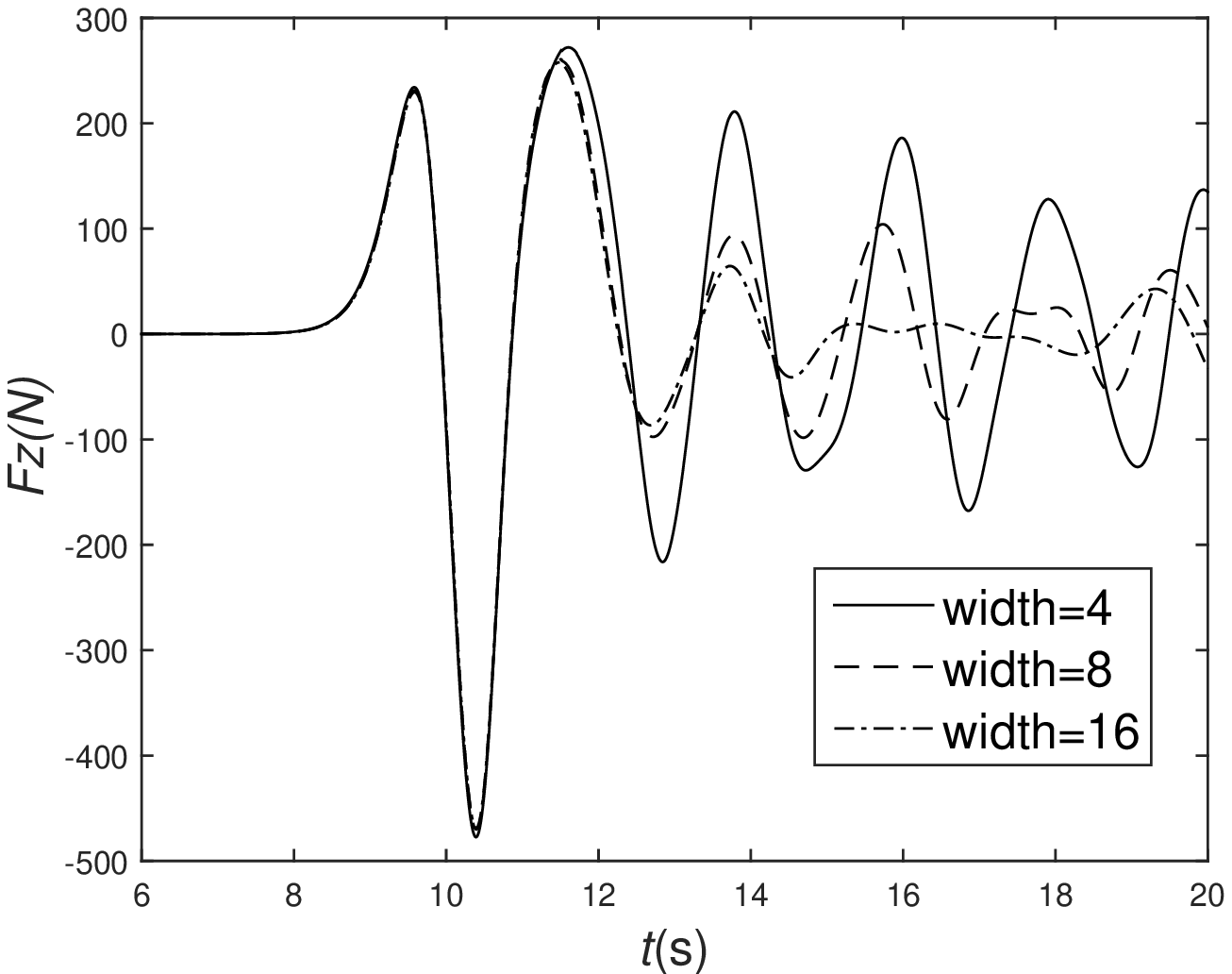}
\caption{Time evolution of the vertical force $F_z$ on the plate with different widths of the wave tank. The other parameters are $h=60$ cm, $H/h=0.3$, $B=40$ cm.}
\label{fig_width}
\end{figure}

We have partially explained the process of oscillations in section 4.1 that covers the above figures. 
Another important factor for these sustained oscillations is the reflection from the lateral walls. To explore that, we changed the width of the wave tank. The results for the vertical force are shown in Fig \ref{fig_width}. When the flume width is multiplied by 2, the oscillations have a smaller amplitude. It is surprising that the frequency does not change much. With four times the width, a similar structure persists up to the third peak of the vertical force $fz_5^+$. It looks like the magnitude of $fz_5^+$ is not affected by an even larger width, since with doubling the width from 8 to 16, the value only decreases a little bit.

\subsection{Velocity field}

\begin{figure*}
\centering
\includegraphics[scale=0.32]{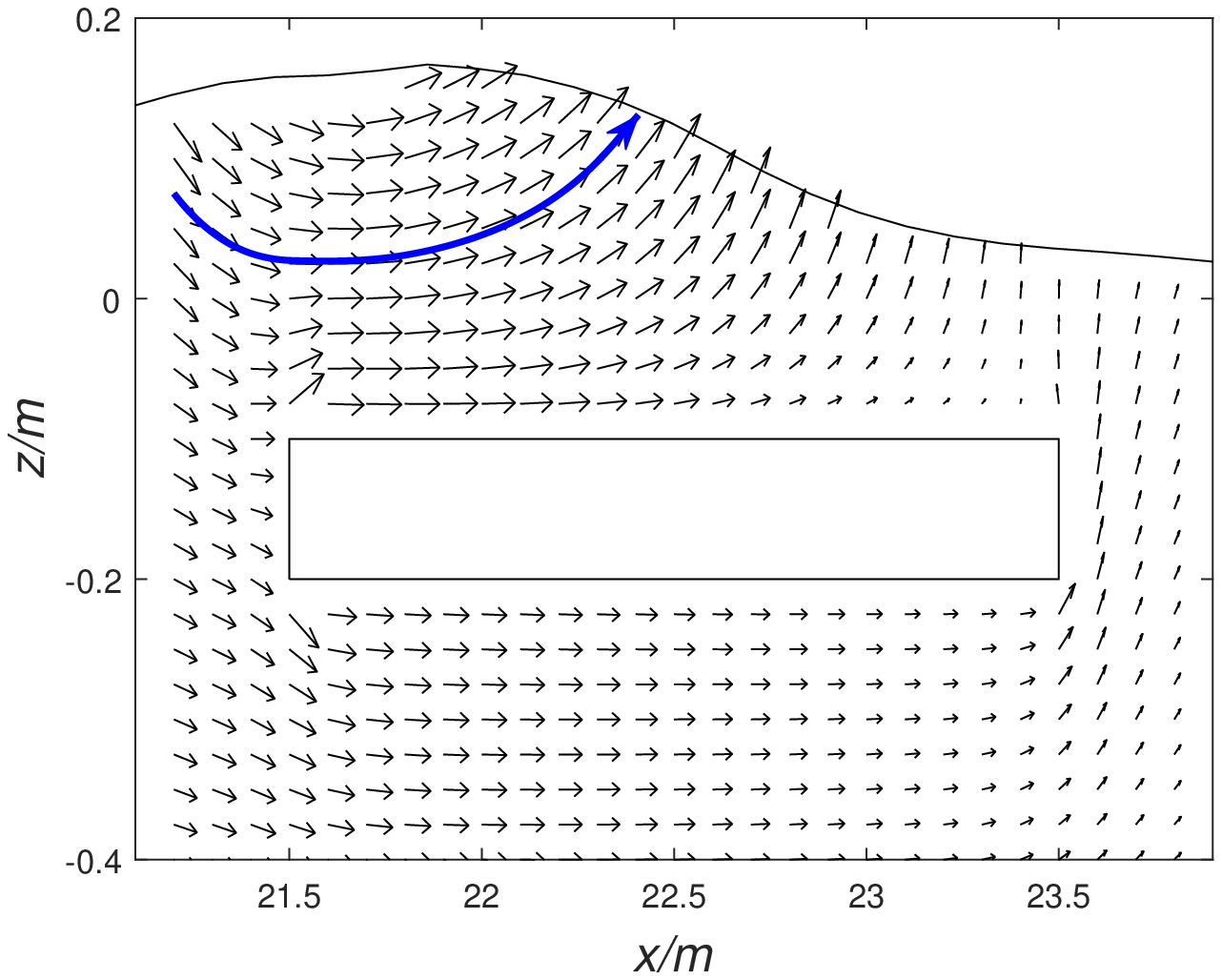}
\includegraphics[scale=0.32]{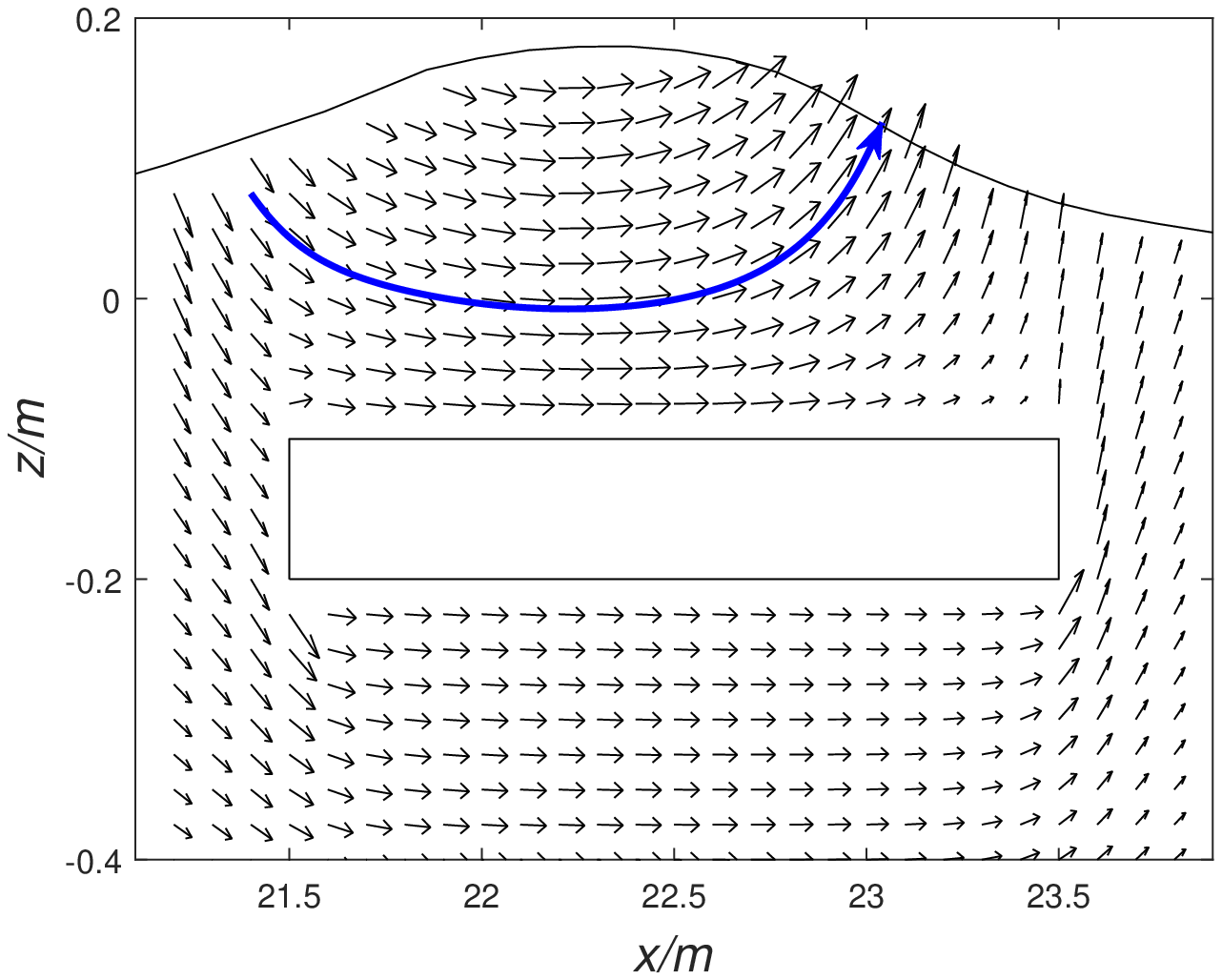}
\includegraphics[scale=0.32]{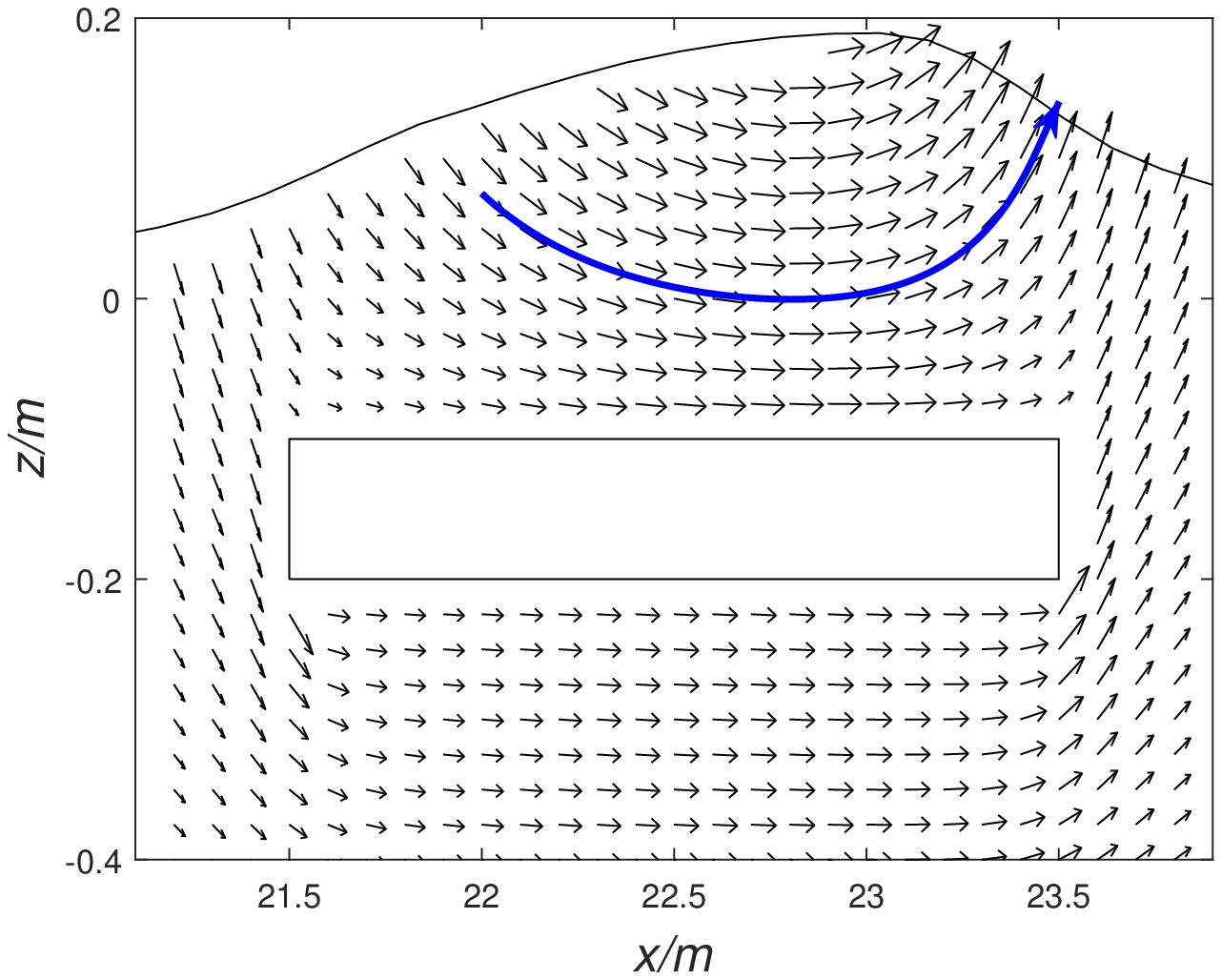}
\includegraphics[scale=0.32]{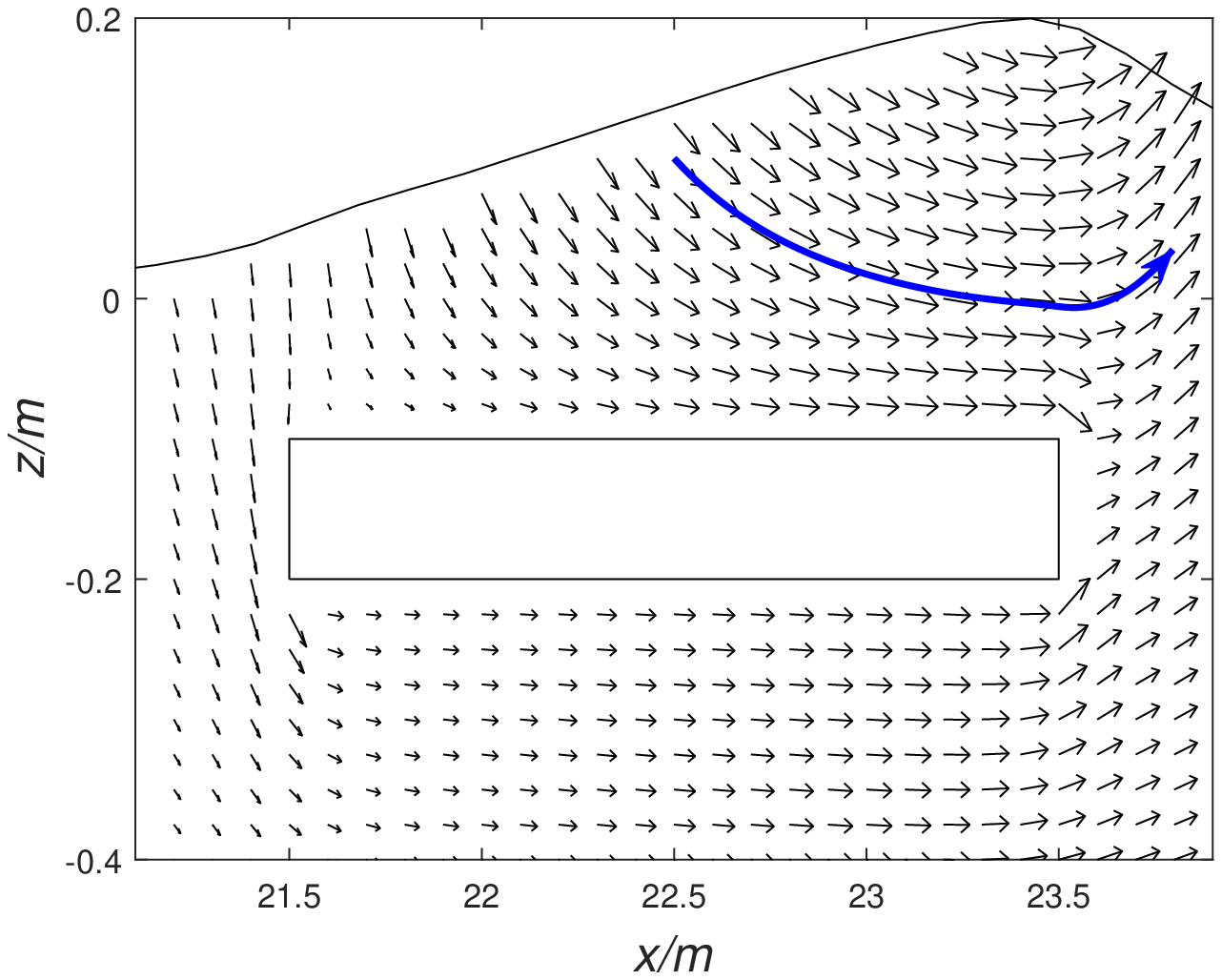}
\includegraphics[scale=0.32]{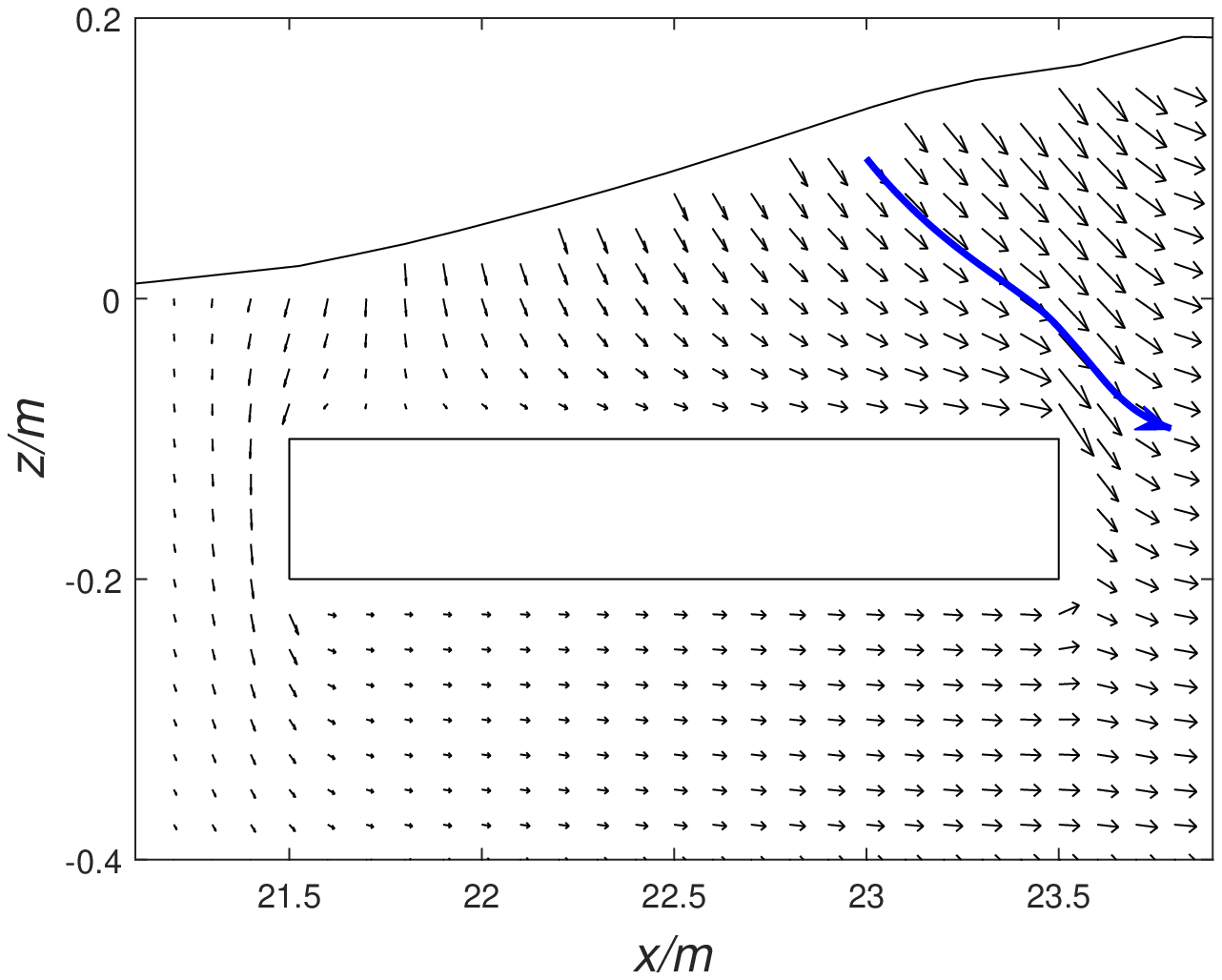}
\includegraphics[scale=0.32]{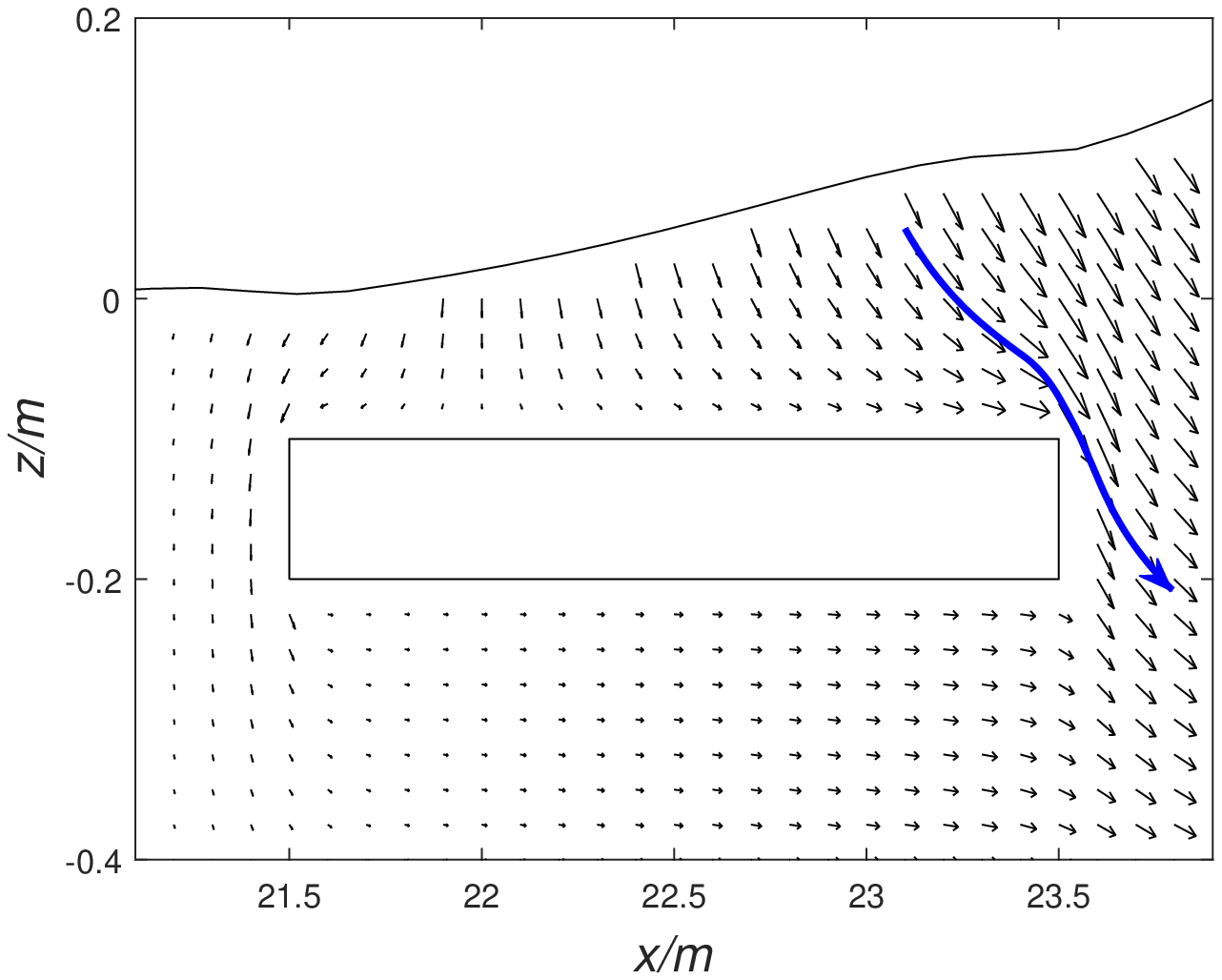}
\includegraphics[scale=0.32]{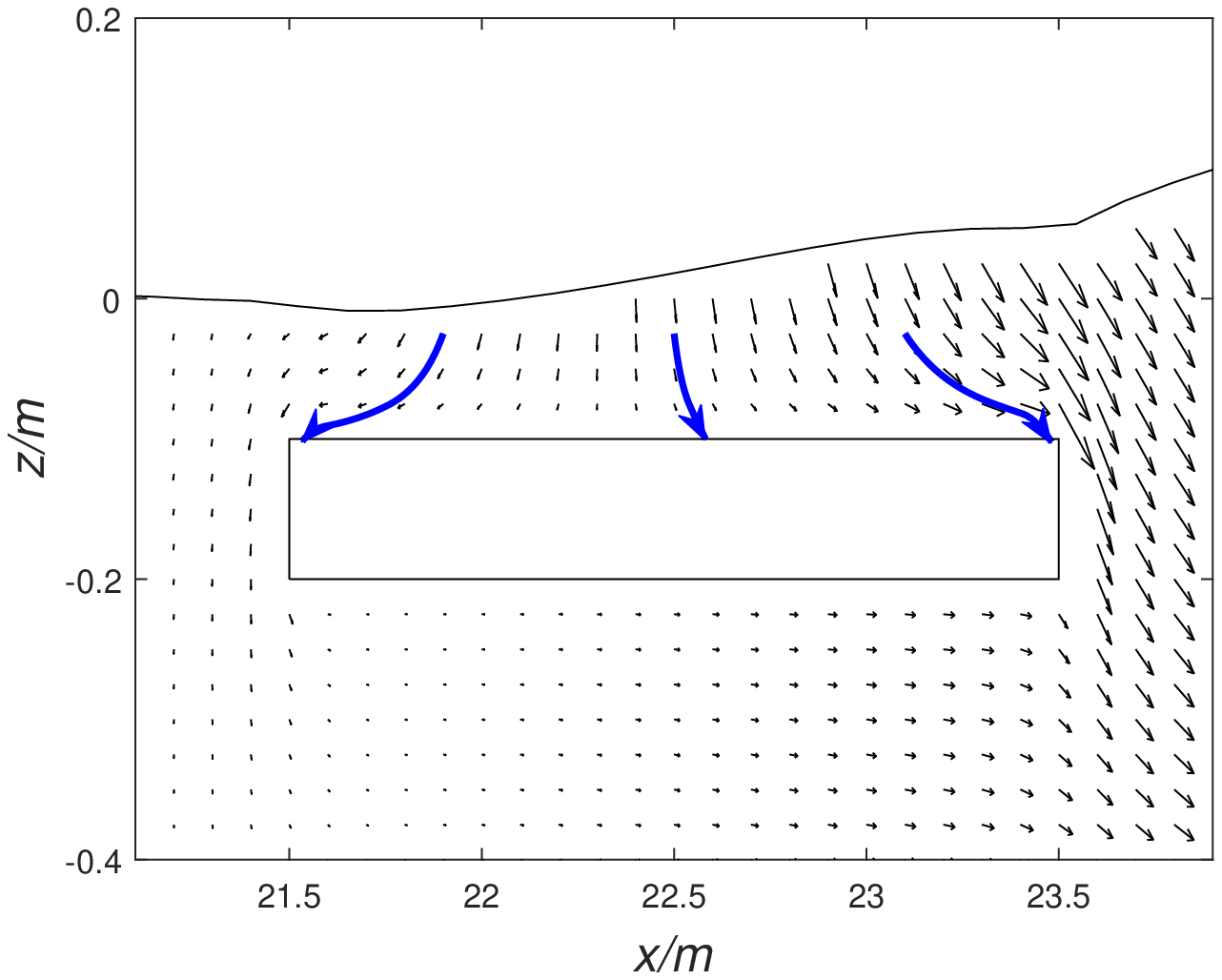}
\includegraphics[scale=0.32]{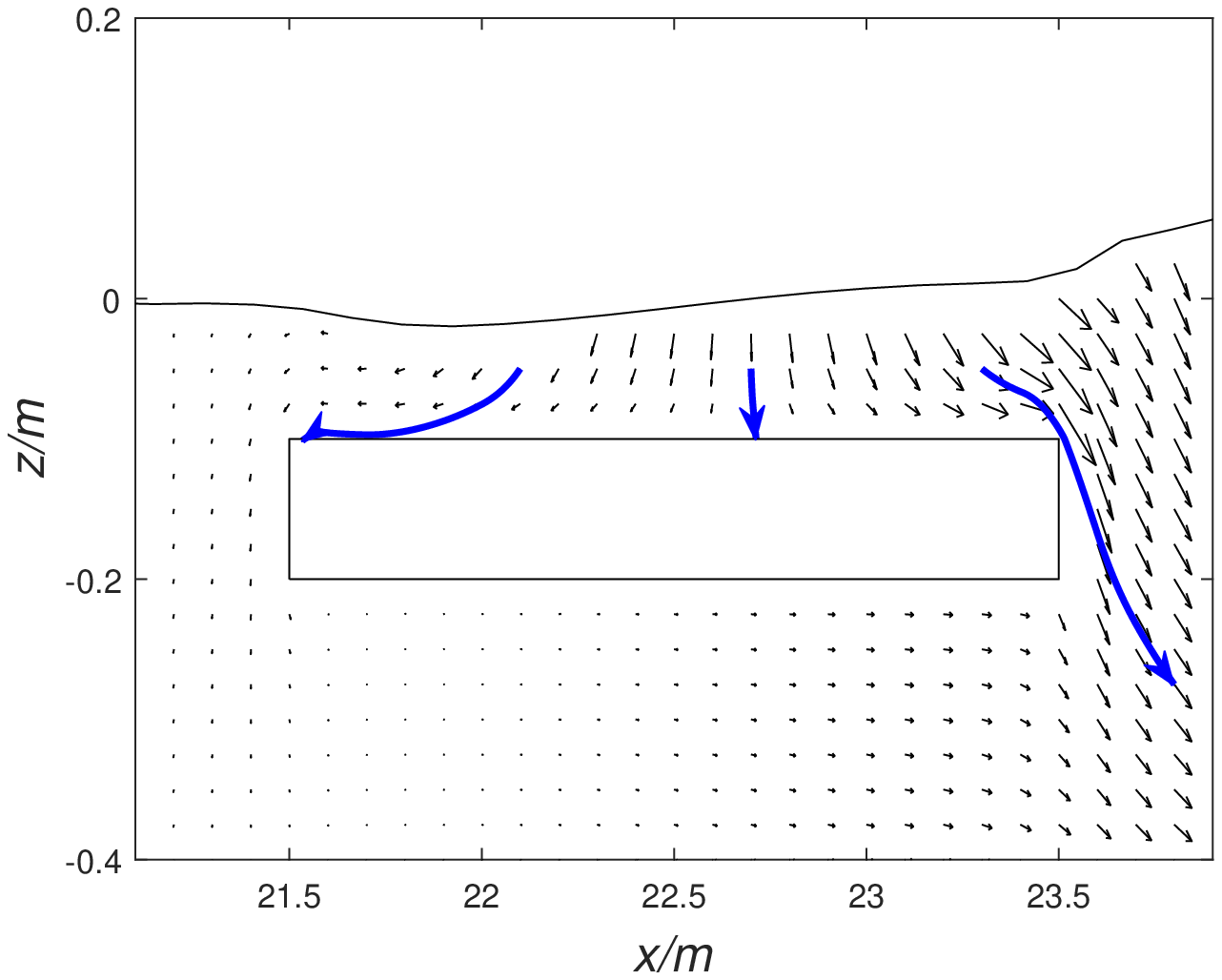}
\includegraphics[scale=0.32]{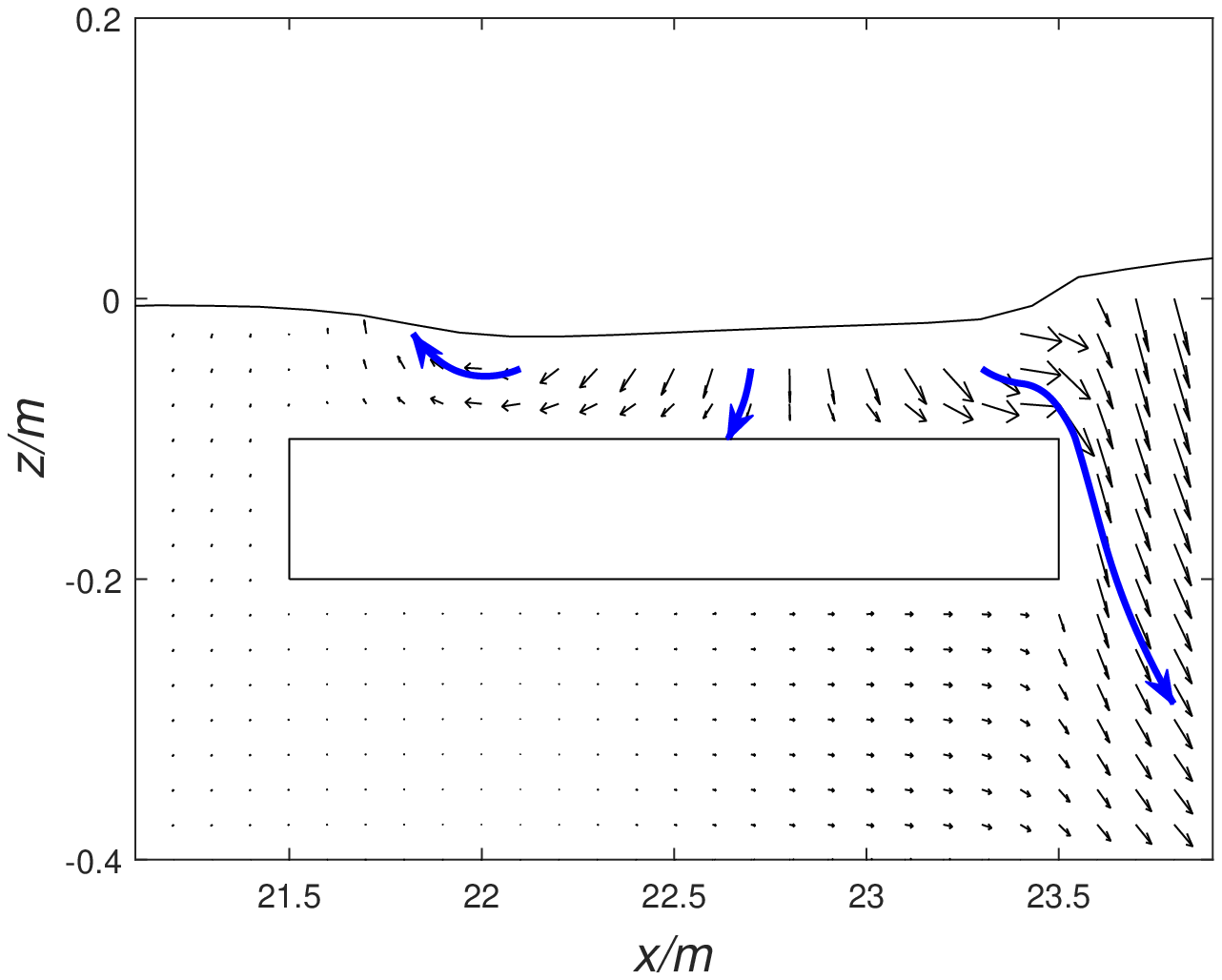}
\includegraphics[scale=0.32]{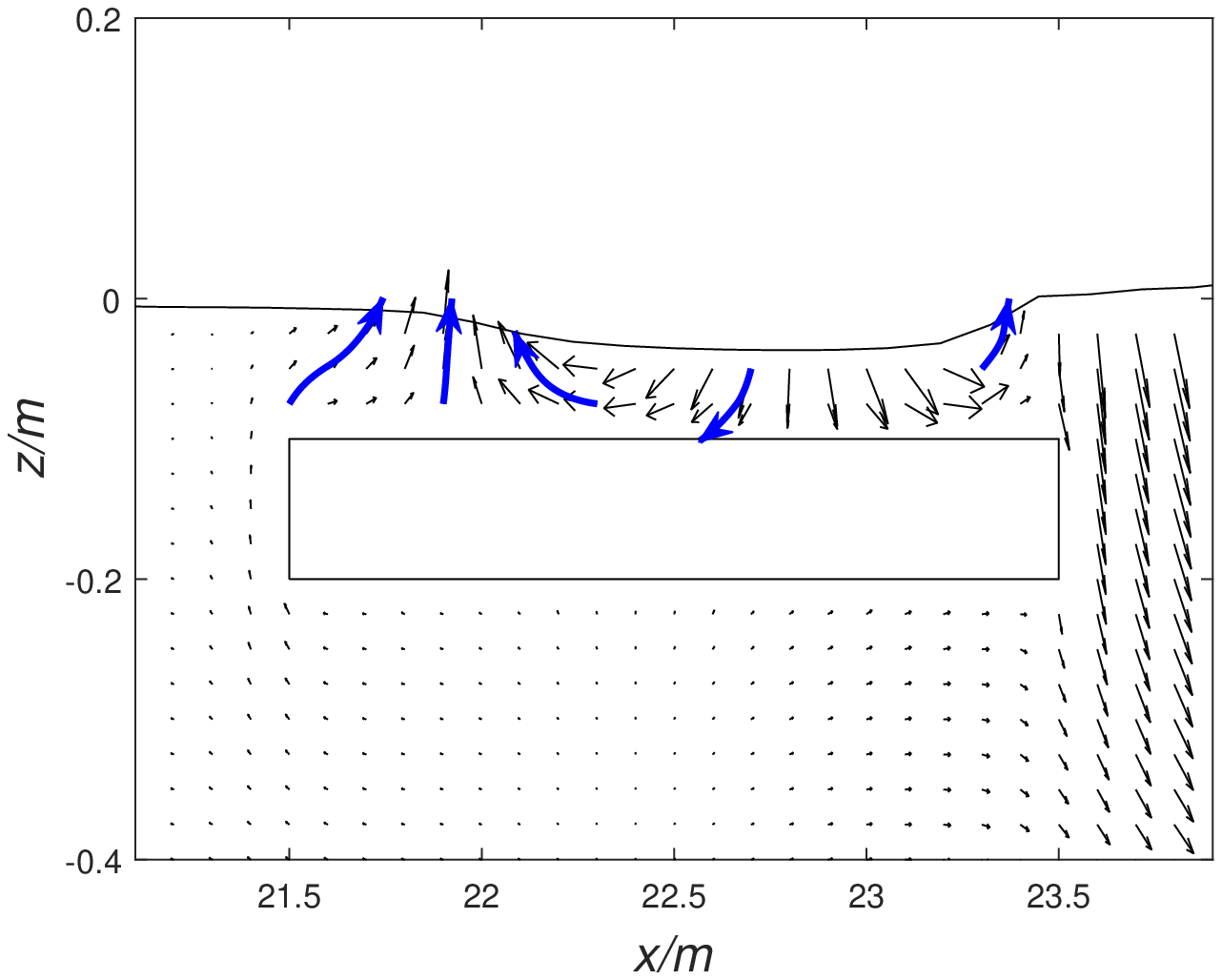}
\includegraphics[scale=0.32]{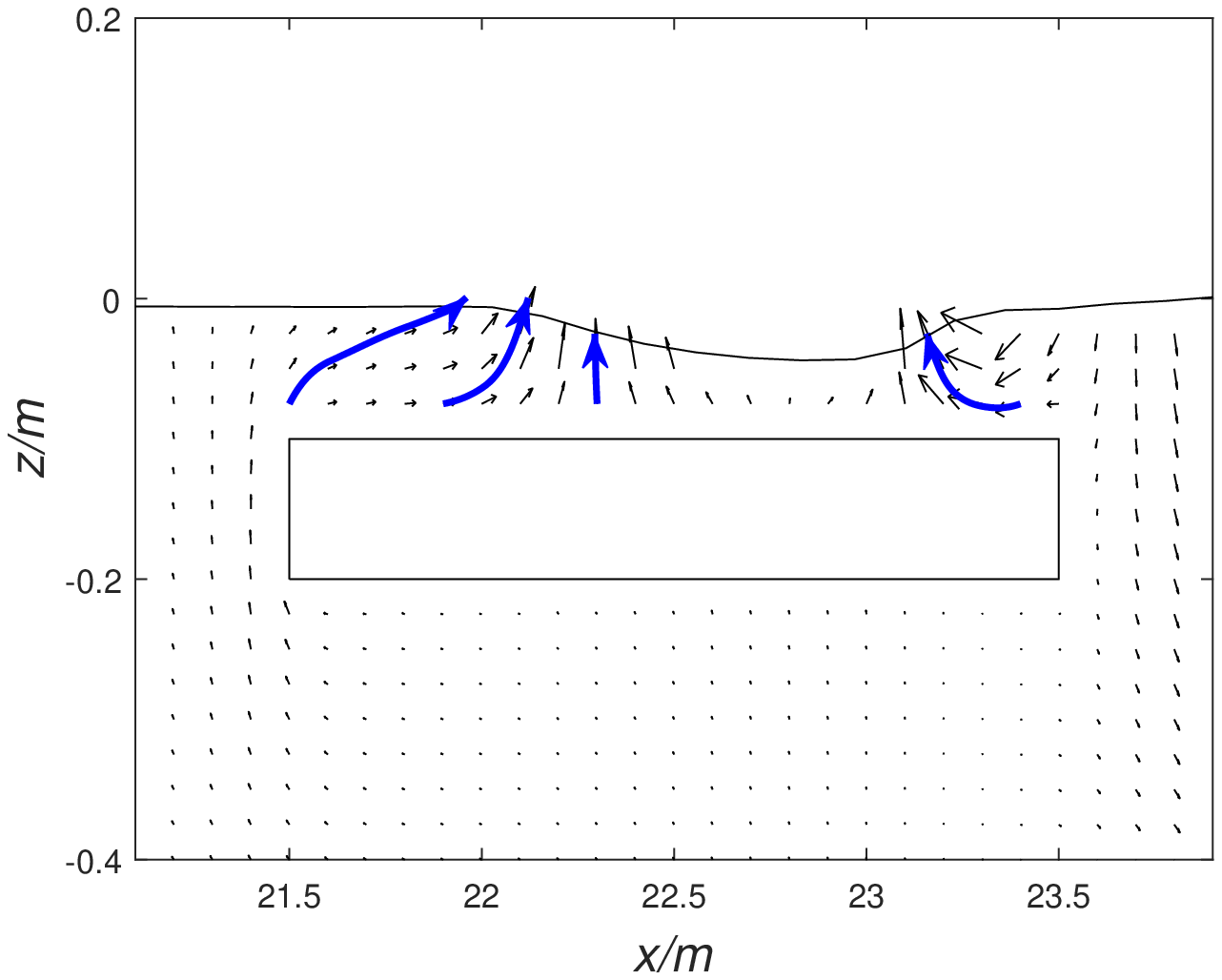}
\includegraphics[scale=0.32]{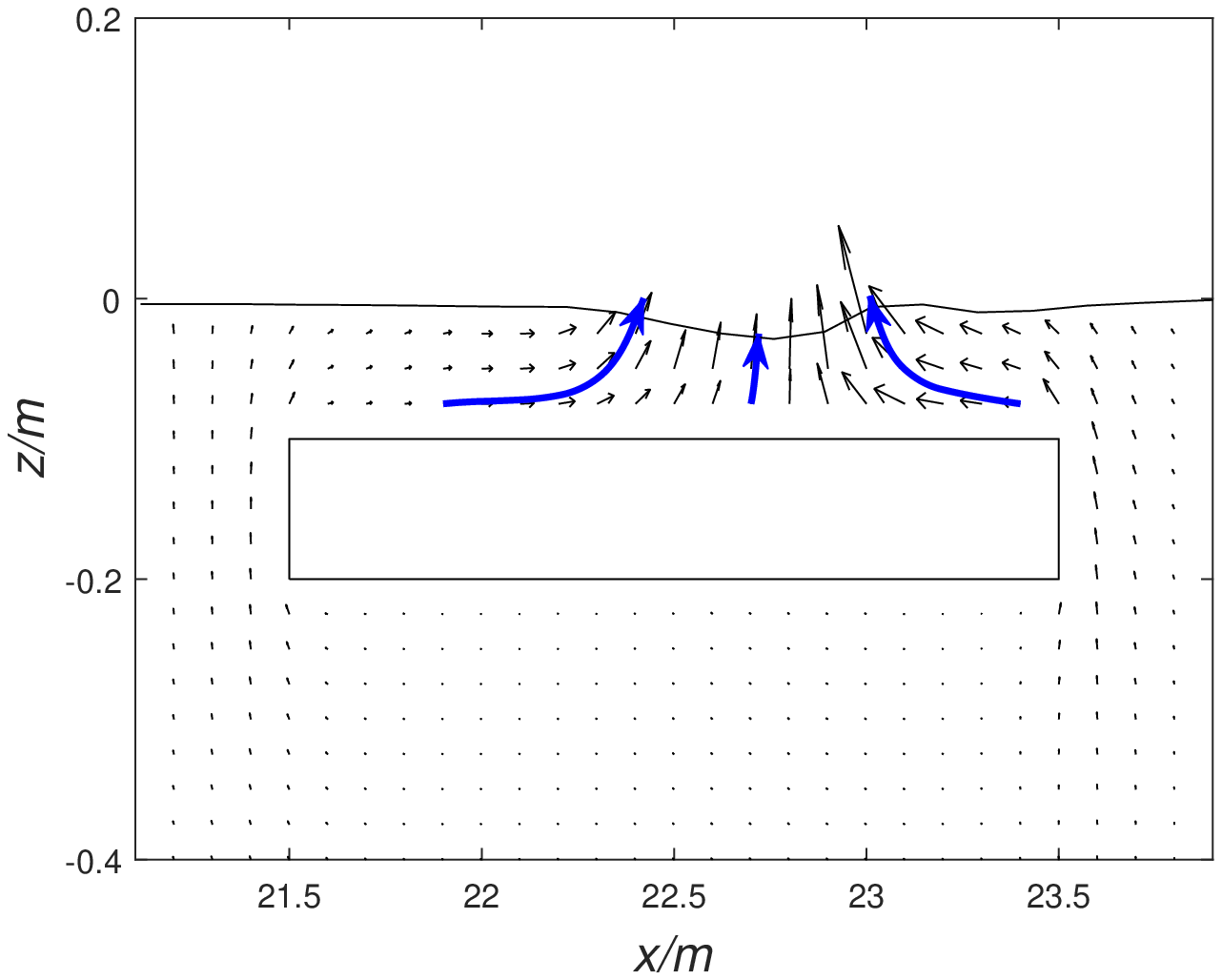}
\includegraphics[scale=0.32]{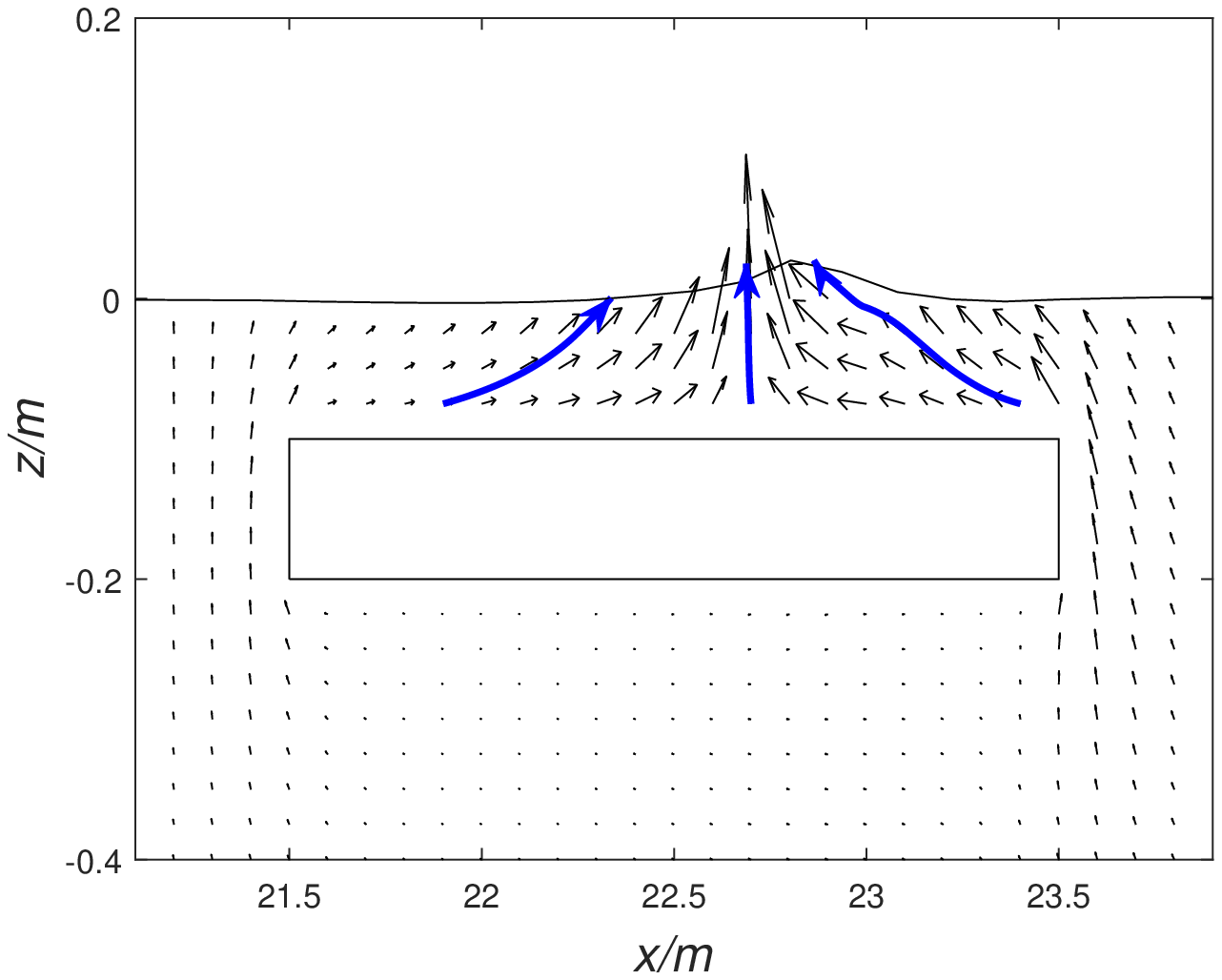}
\includegraphics[scale=0.32]{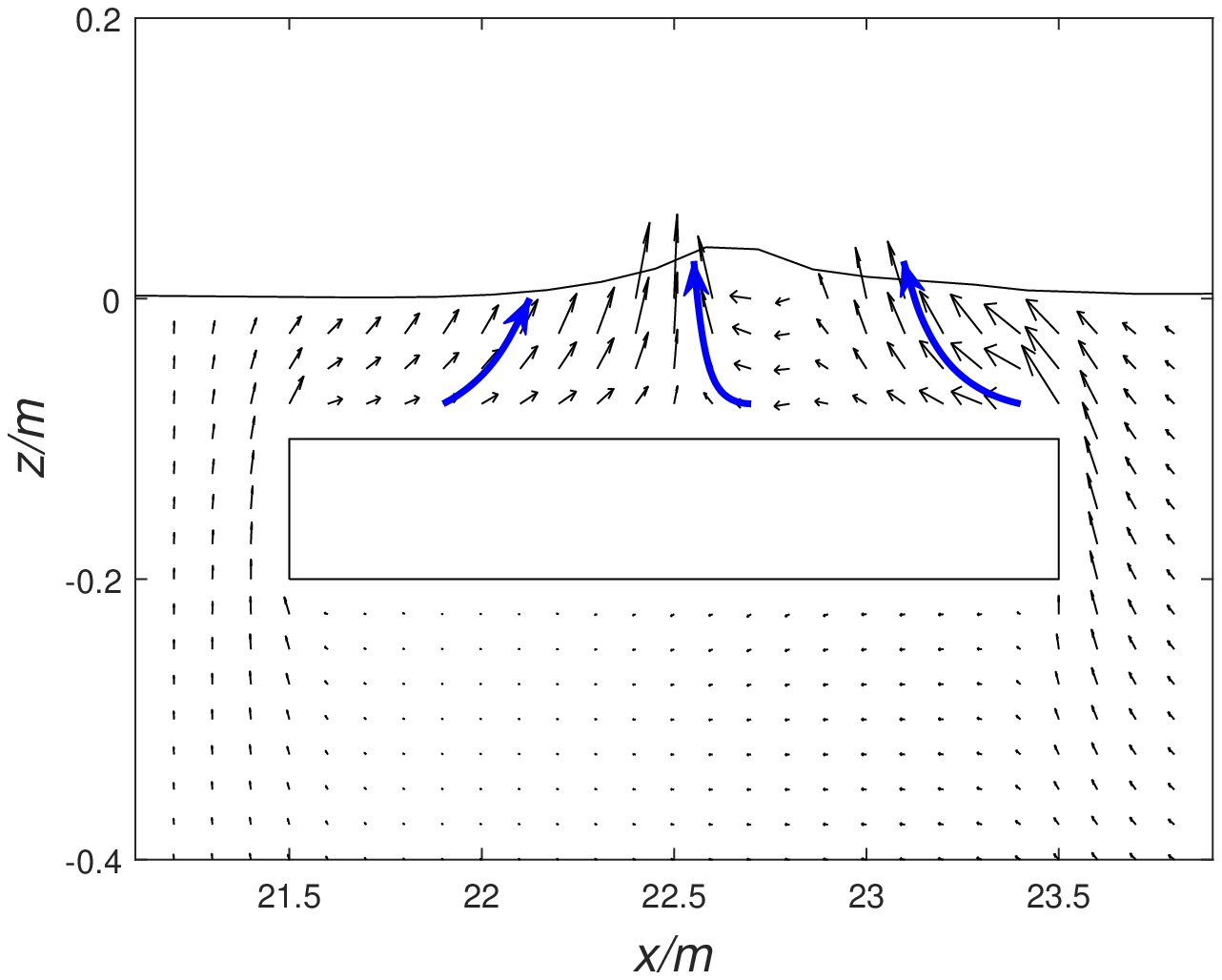}
\includegraphics[scale=0.32]{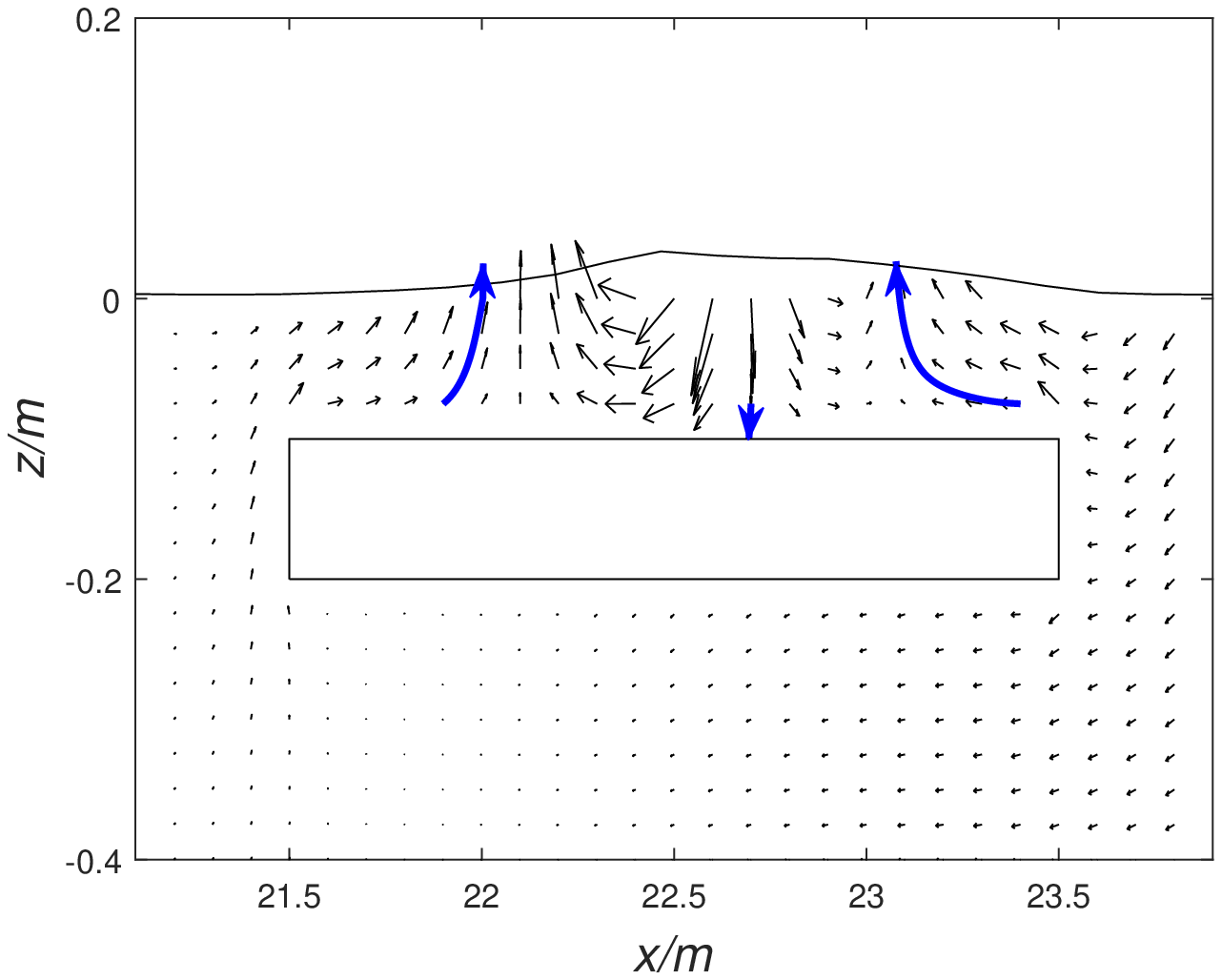}
\includegraphics[scale=0.32]{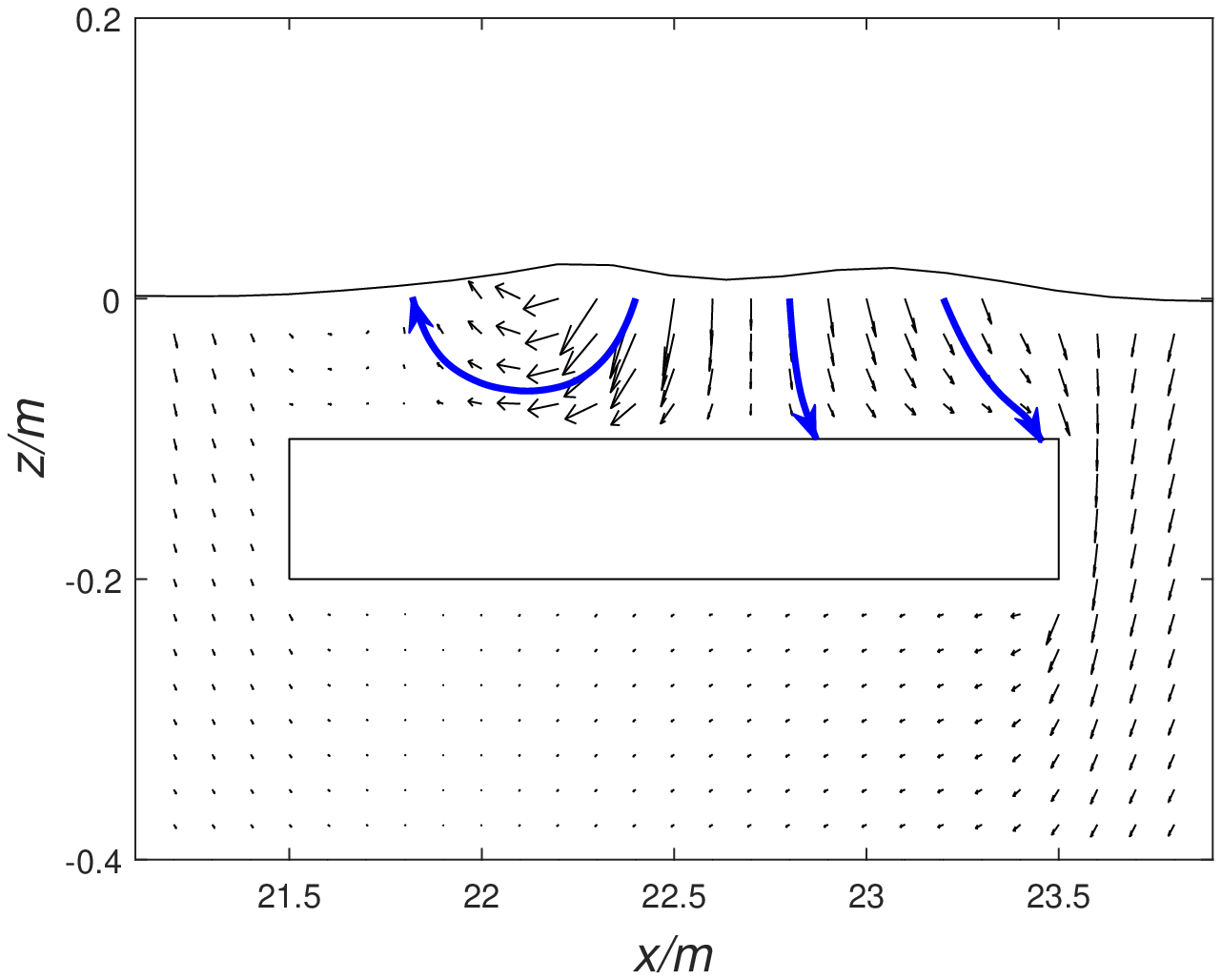}
\includegraphics[scale=0.32]{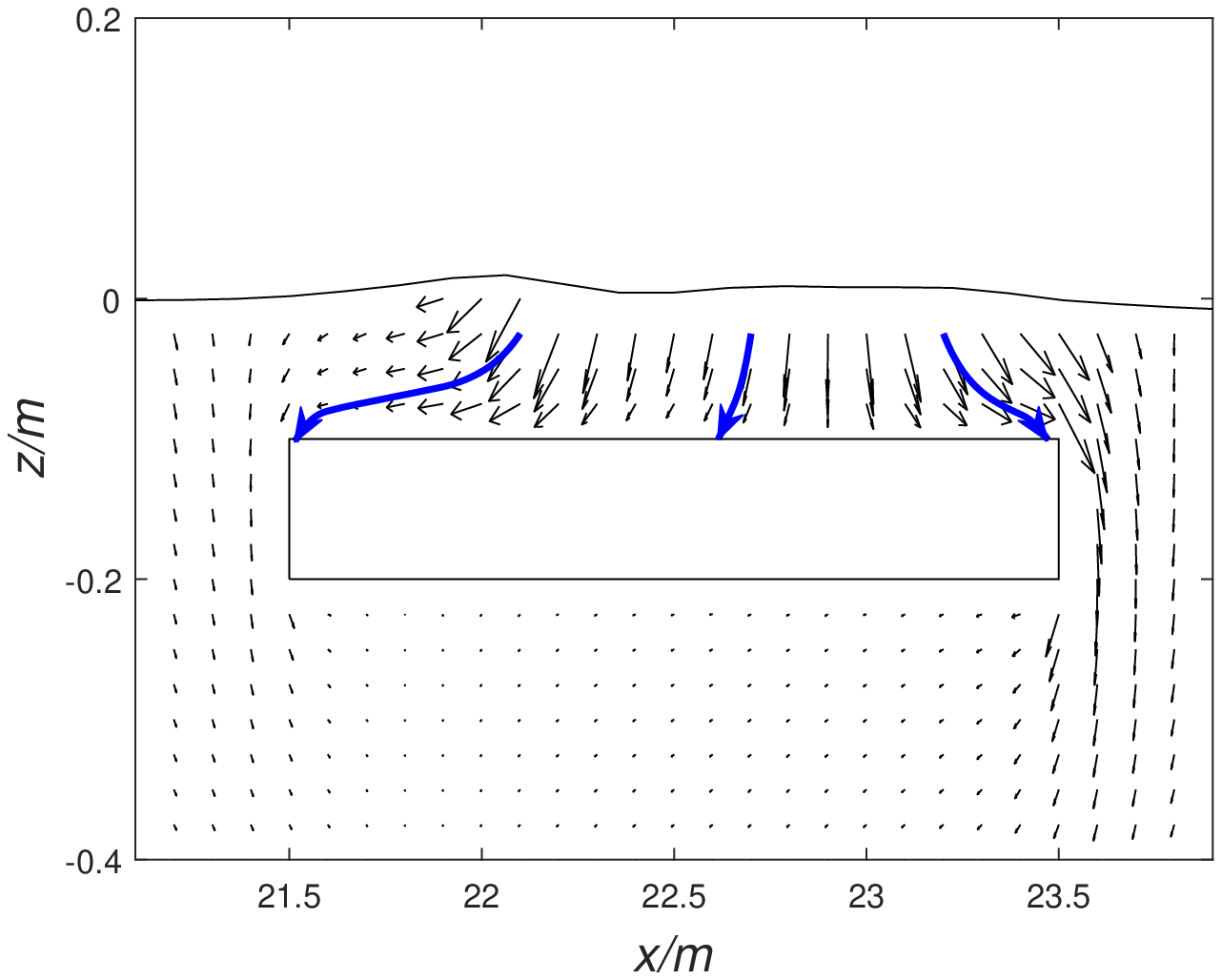}
\includegraphics[scale=0.32]{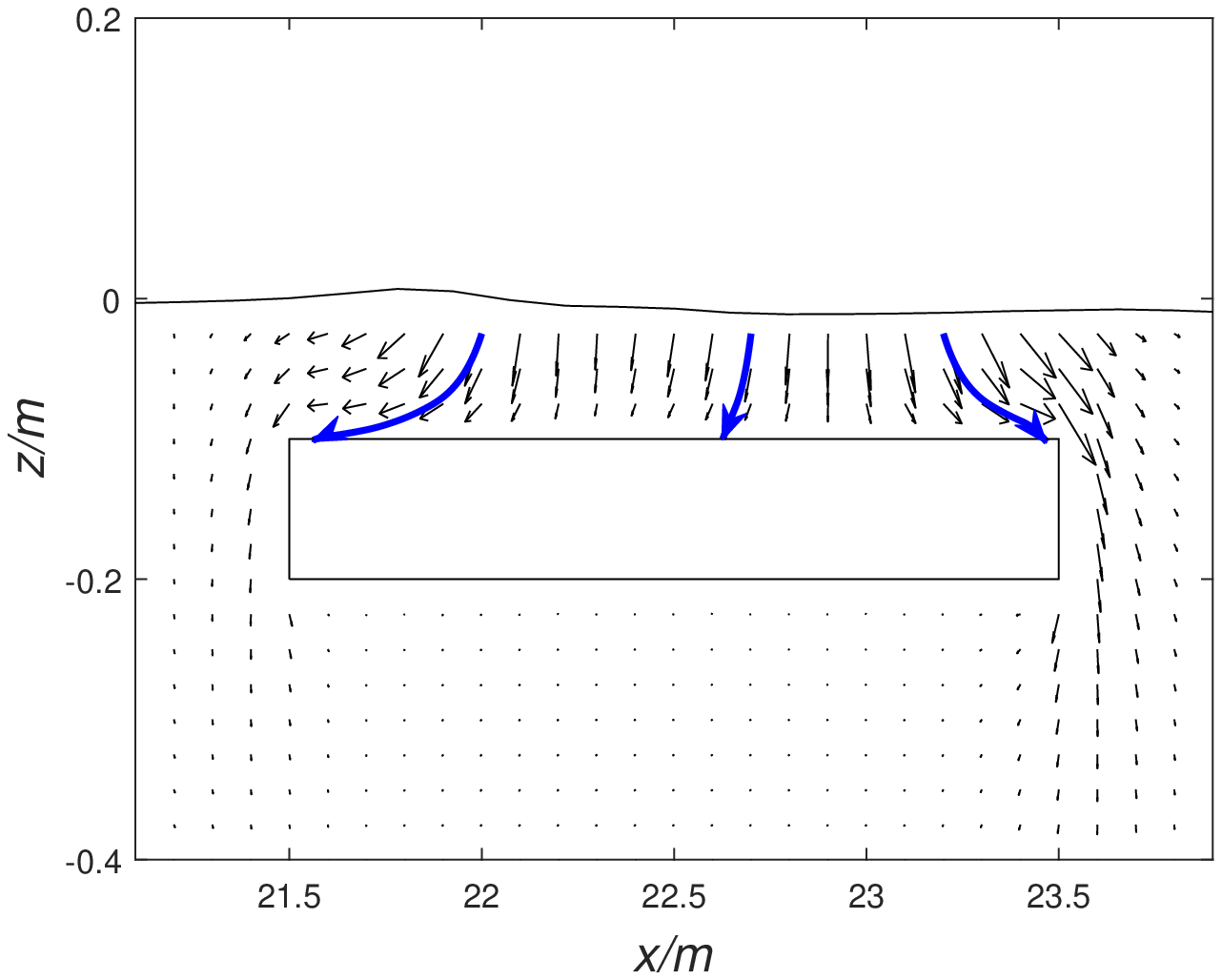}
\caption{Projection of the velocity field in the $(x,z)-$plane corresponding to the middle plane of the wave tank (from 10 s to 13.4 s -- the interval is 0.2 s) when $B=40$ cm, $h=60$ cm and $H/h=0.3$. The blue lines show some streamlines that are representative of how the flow changes as the solitary wave progresses. }
\label{fig_velmid}
\end{figure*}

The numerical code used in the present study can also be used to compute velocities inside the fluid domain. Fig \ref{fig_velmid} shows the velocity field in the middle line of the tank for the case $h=60$ cm, $H/h=0.3$, $B=40$ cm. After the solitary wave passes over the plate, two separate flows come from opposite directions and focus above the plate. The flow from the left is due to the reflection from the upper surface of the plate. Meanwhile, the flow from the right is caused by the propagation of the bulge mentioned above. Once the focused wave reaches its peak, it spreads again.

\begin{figure*}
\centering
\includegraphics[scale=0.3]{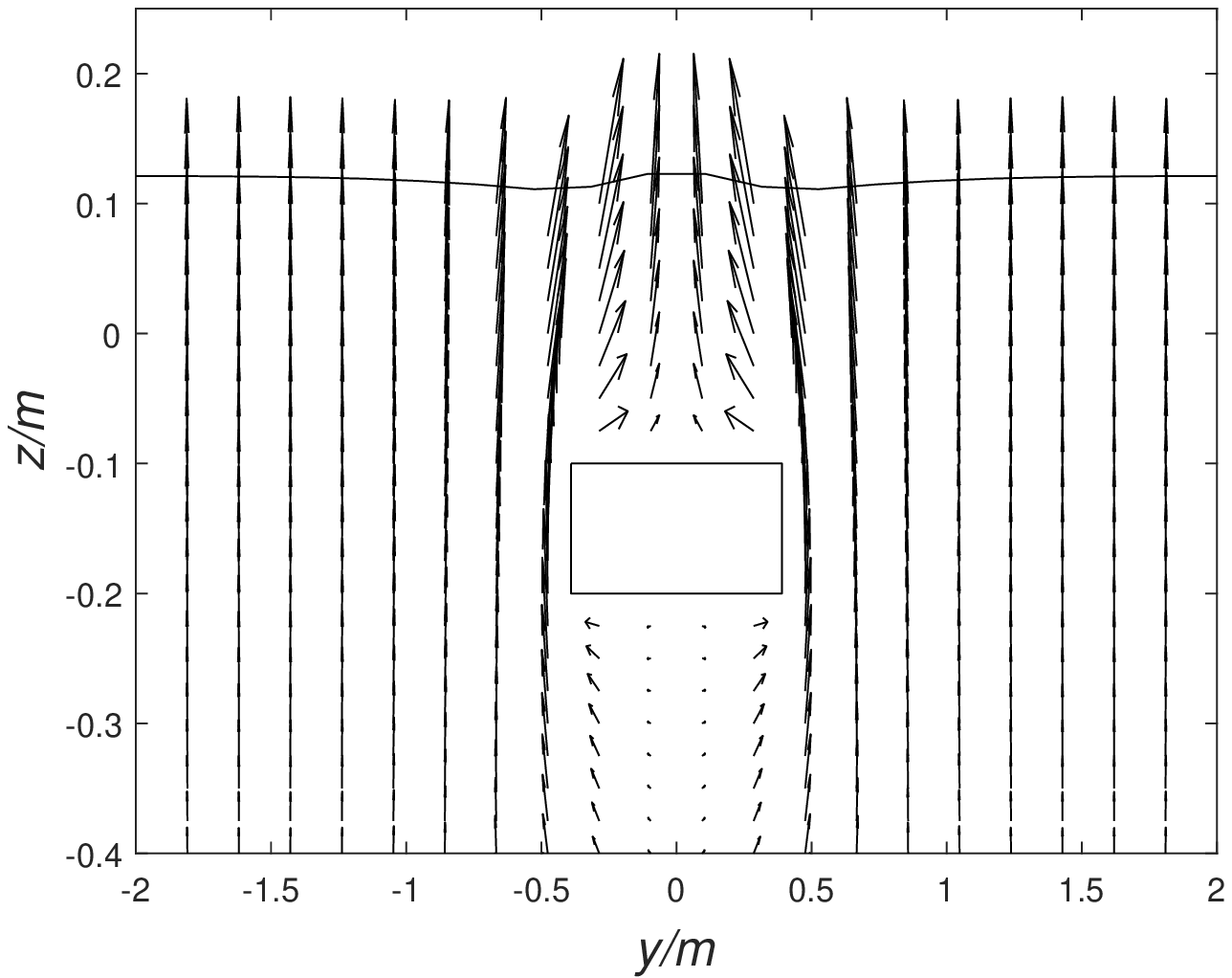}
\includegraphics[scale=0.3]{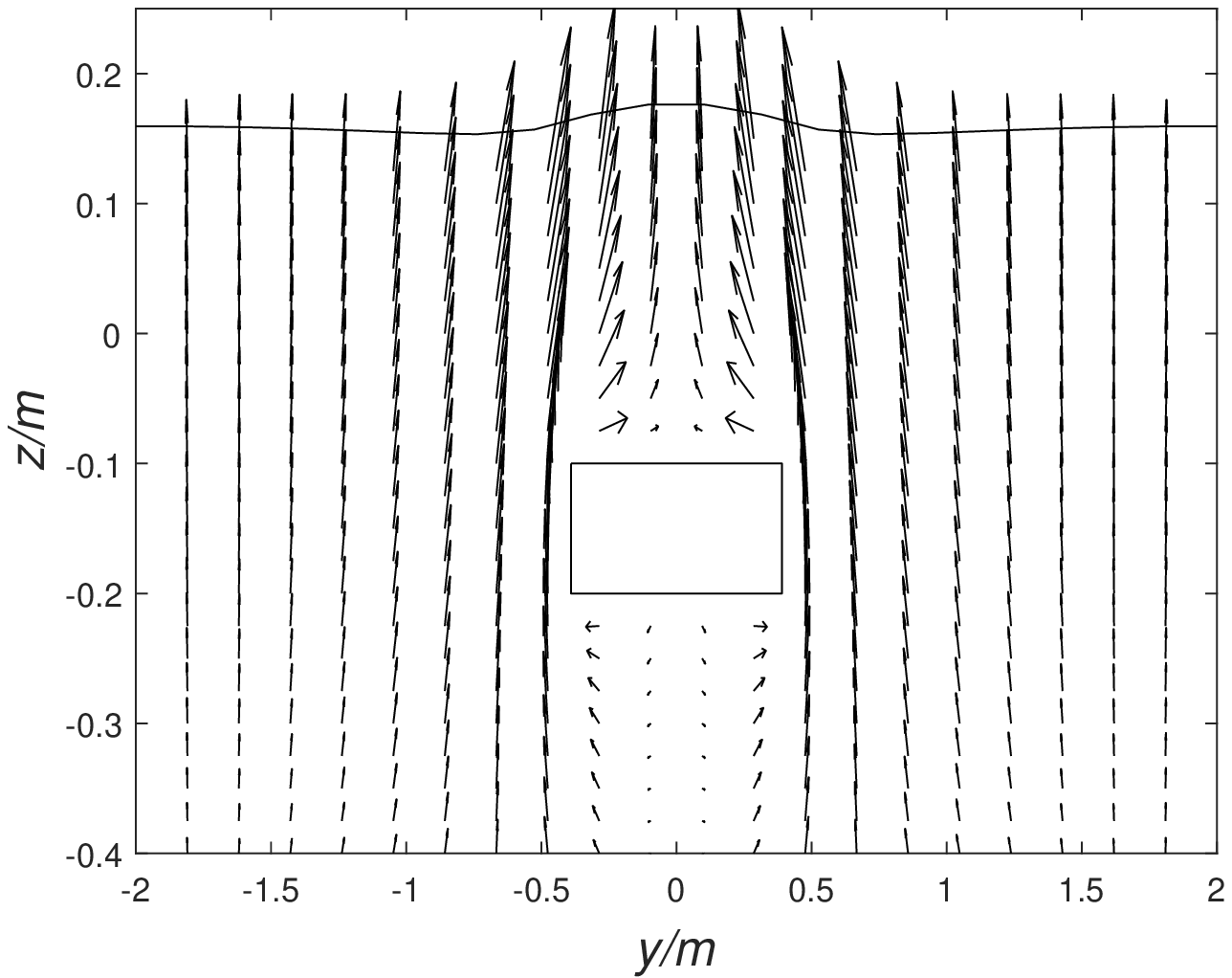}
\includegraphics[scale=0.3]{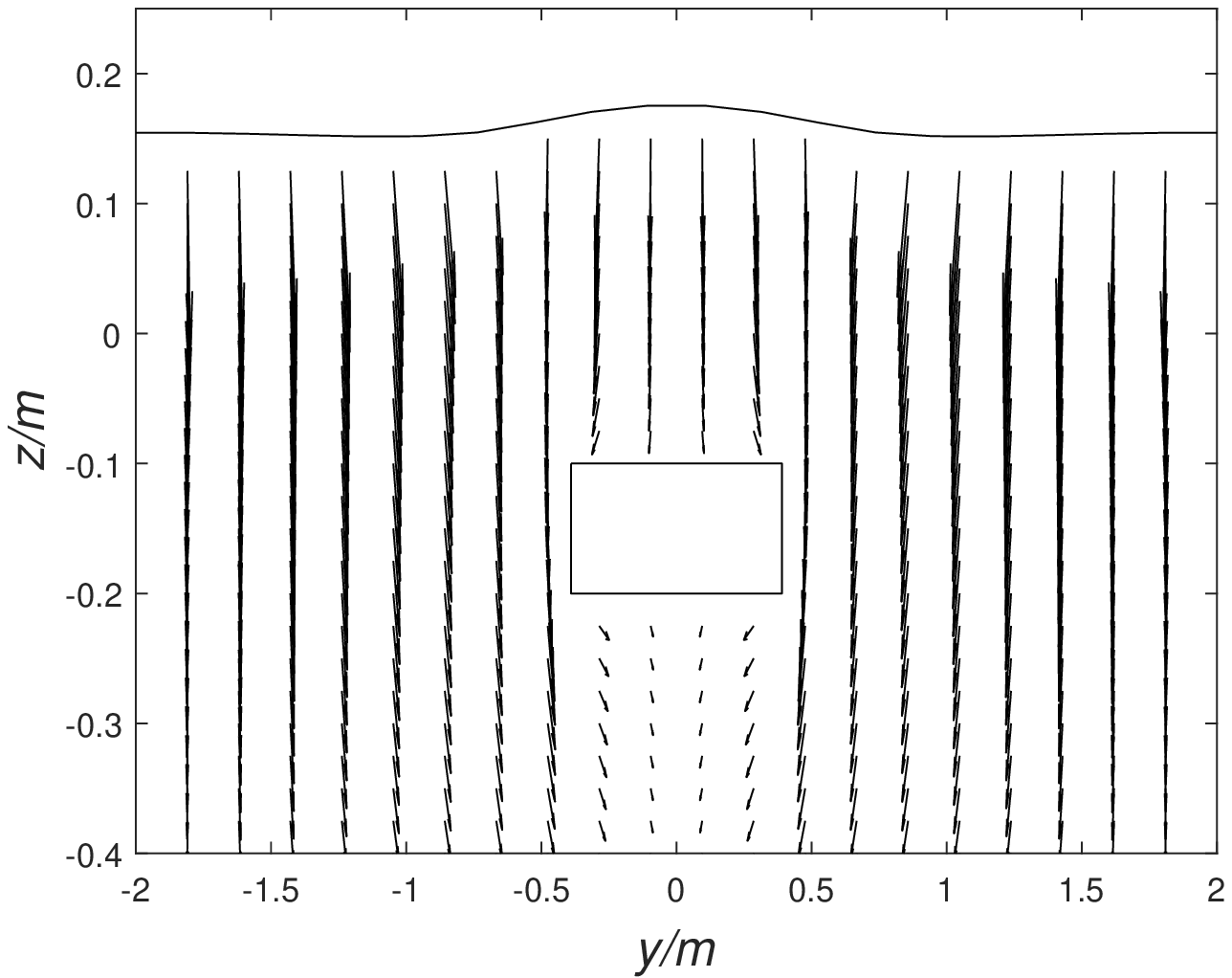}
\includegraphics[scale=0.3]{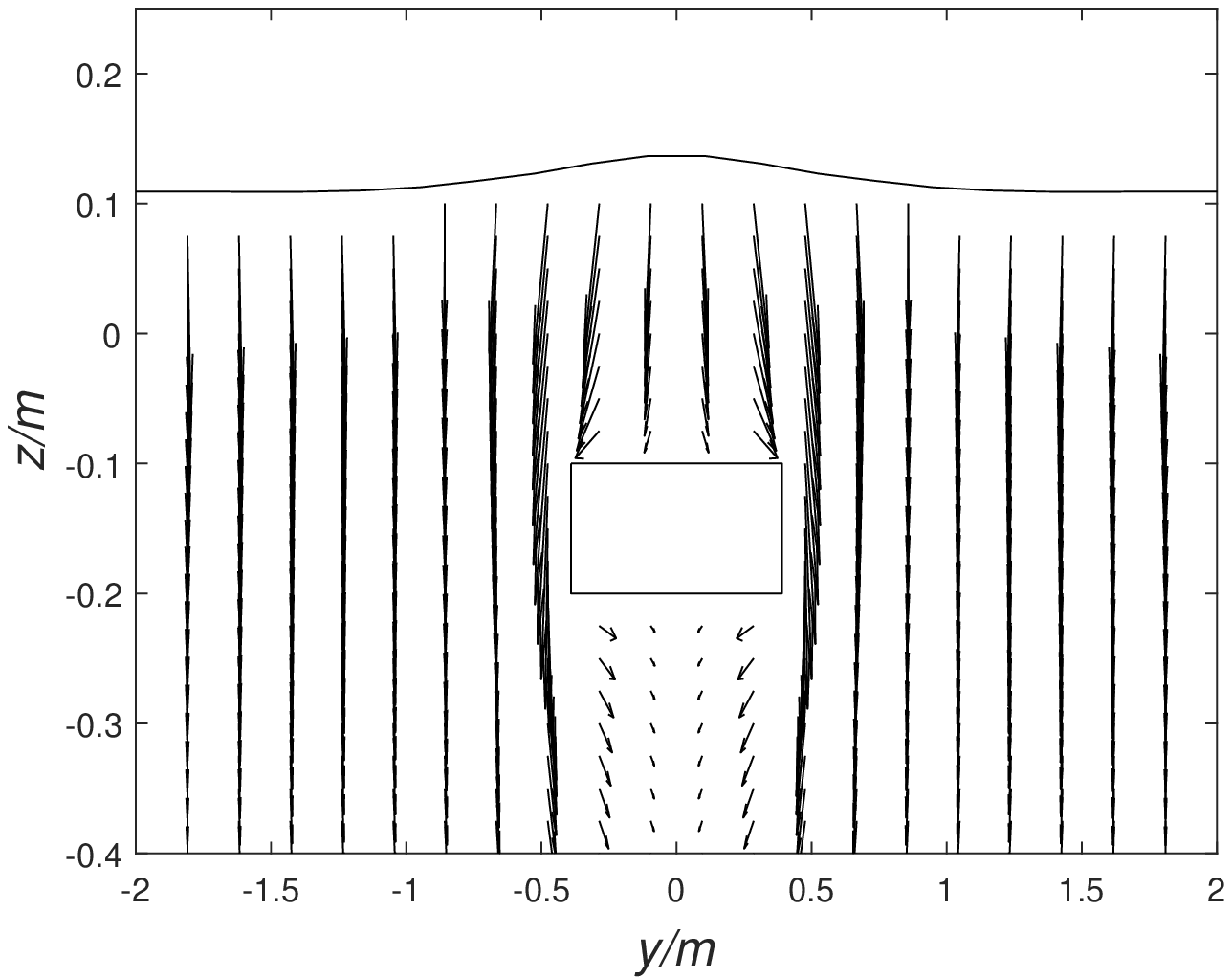}
\includegraphics[scale=0.3]{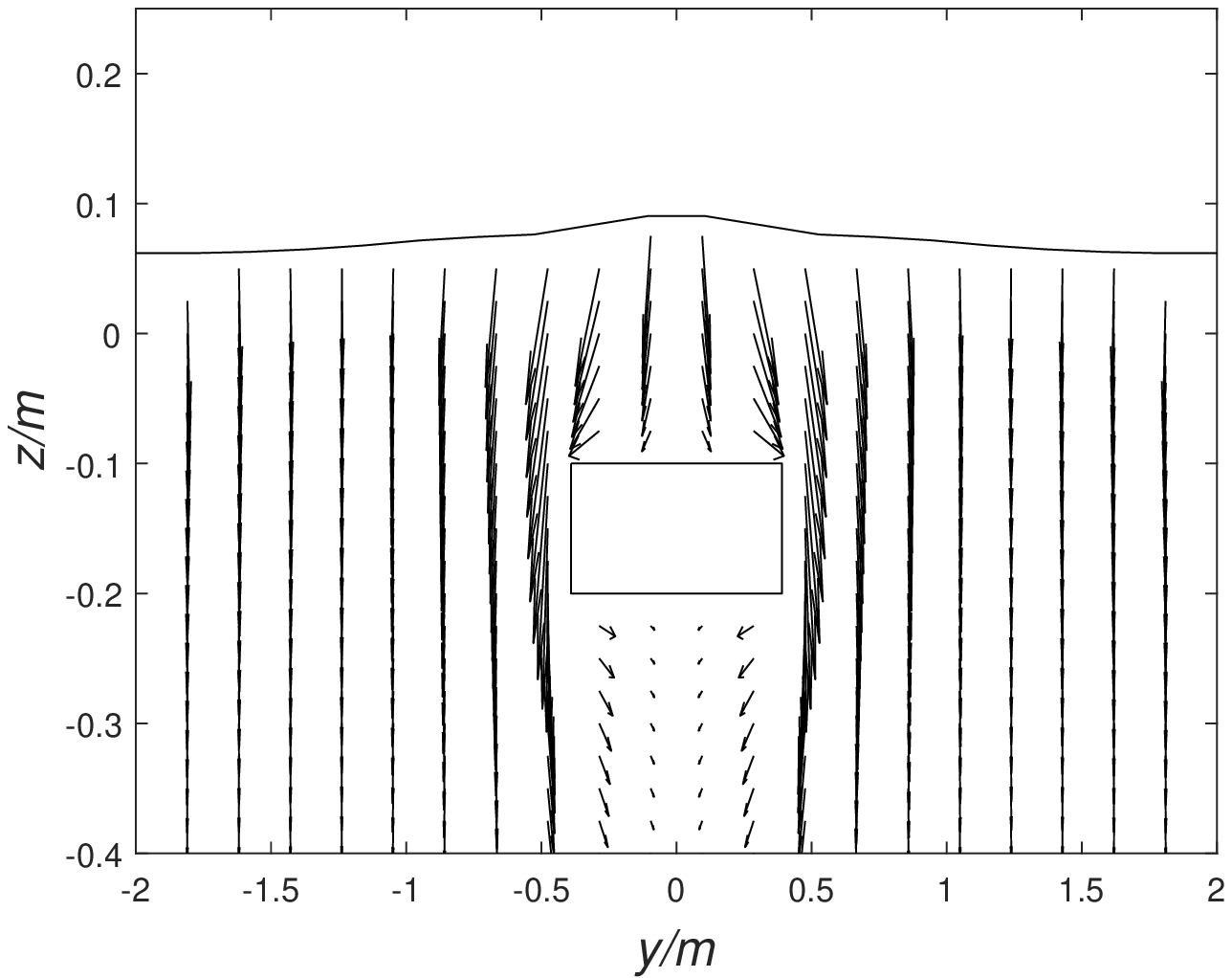}
\includegraphics[scale=0.3]{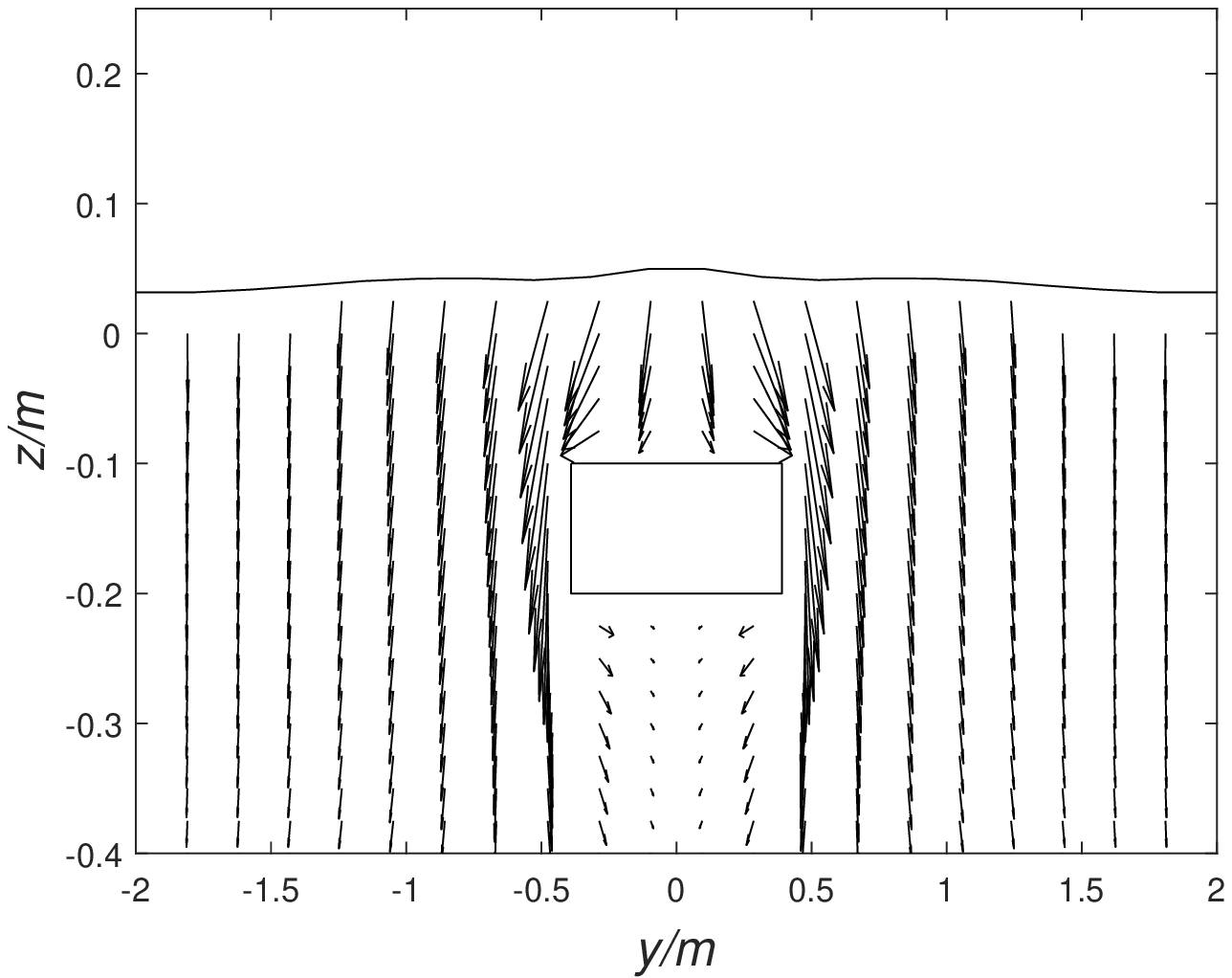}
\includegraphics[scale=0.3]{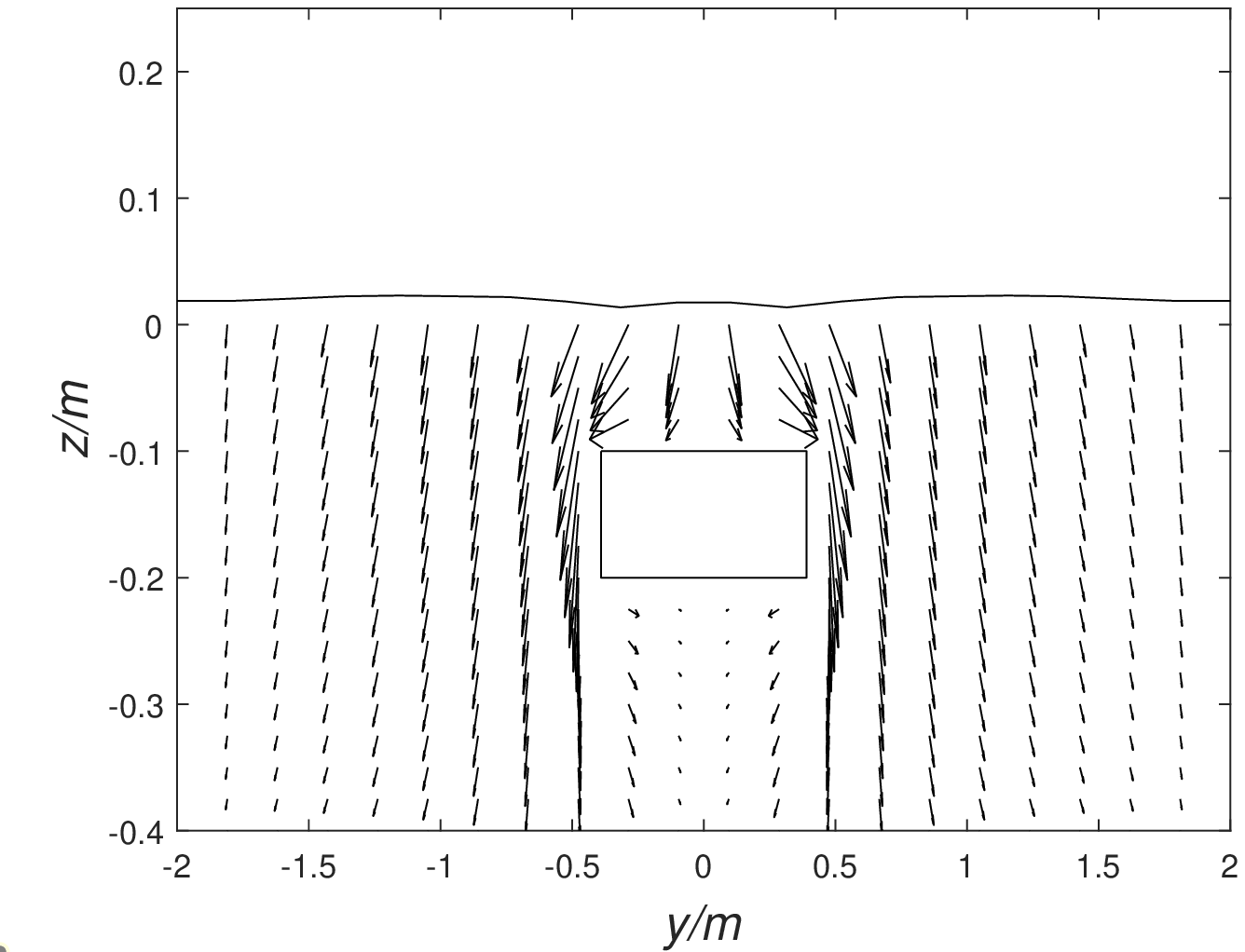}
\includegraphics[scale=0.3]{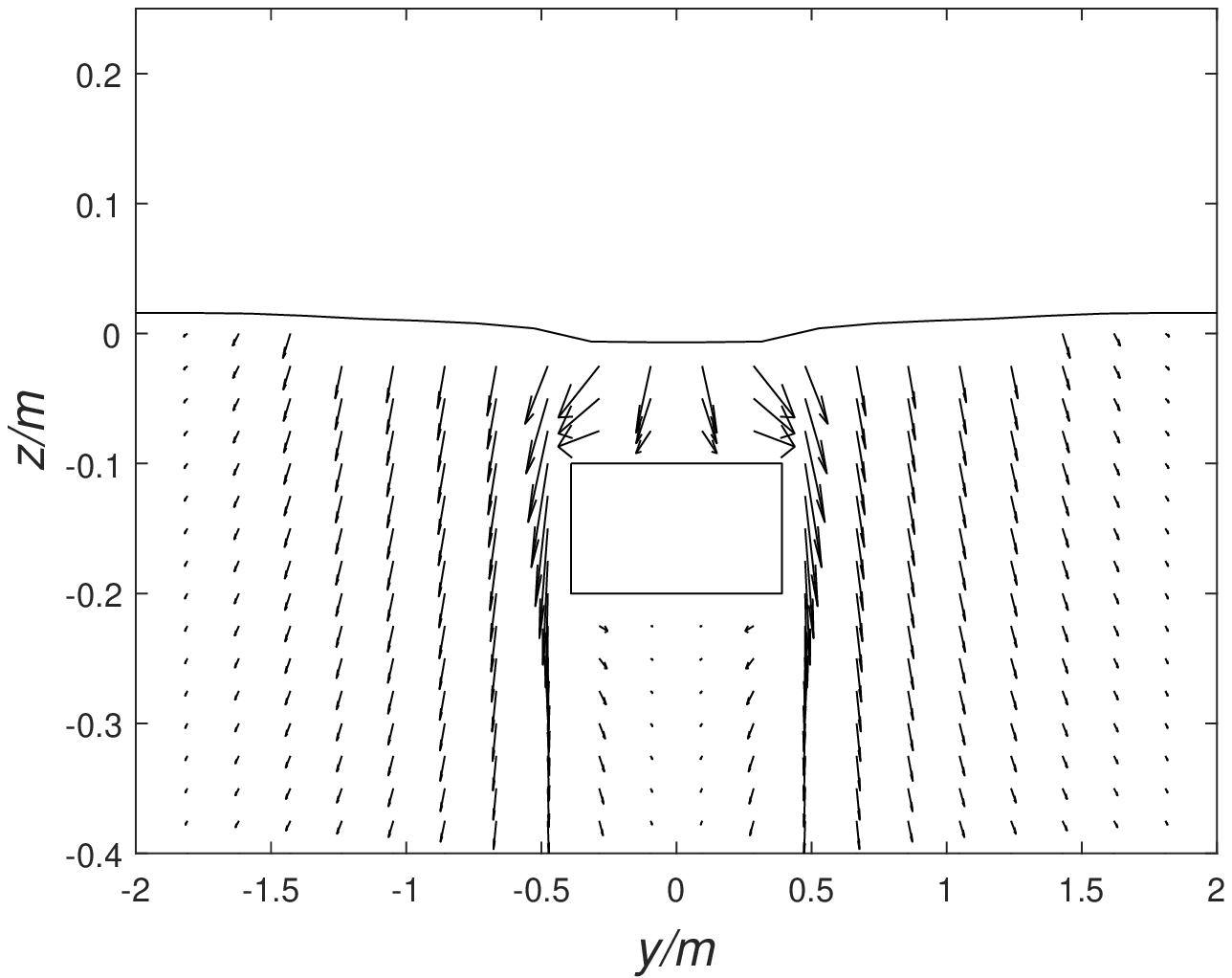}
\includegraphics[scale=0.3]{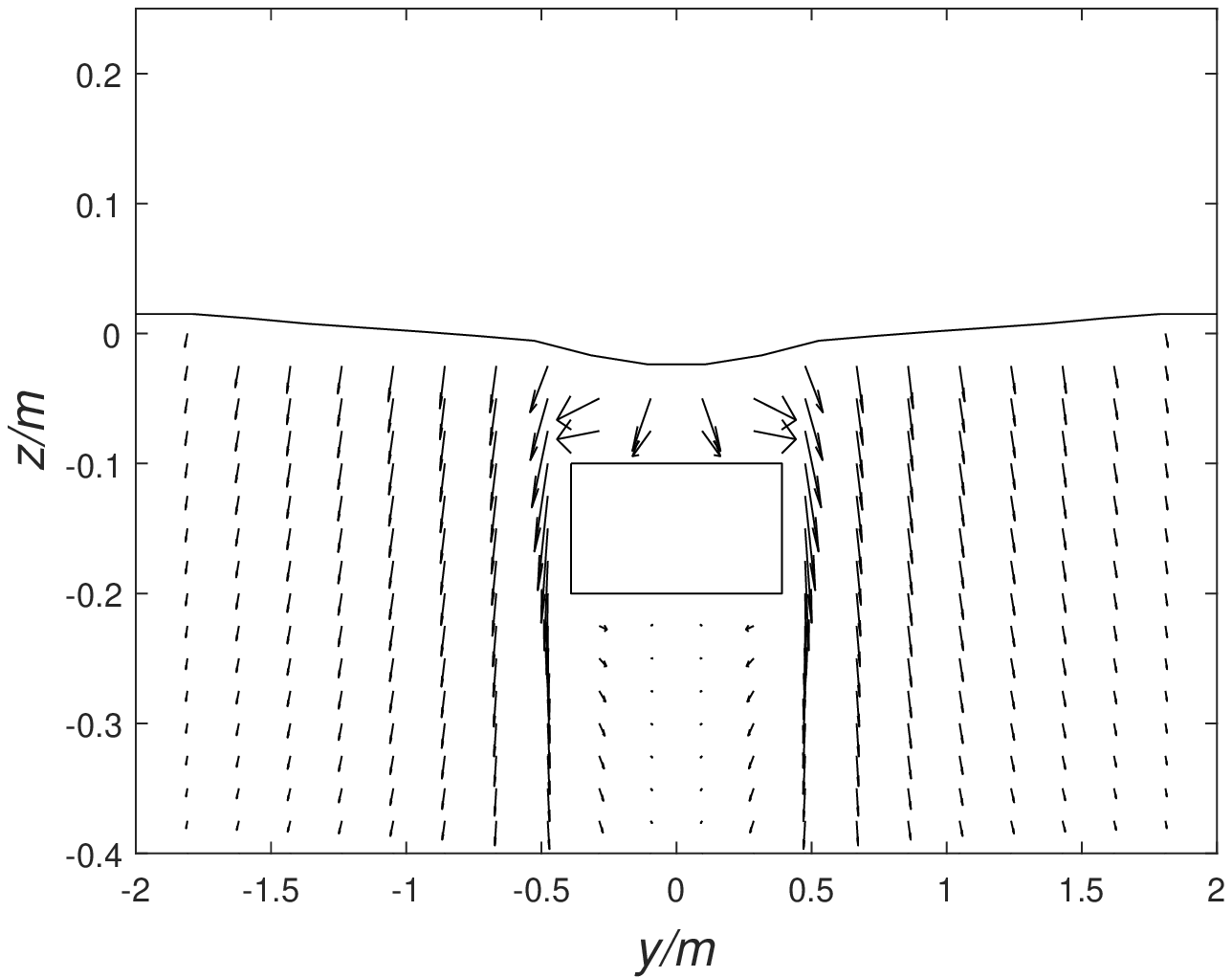}
\includegraphics[scale=0.3]{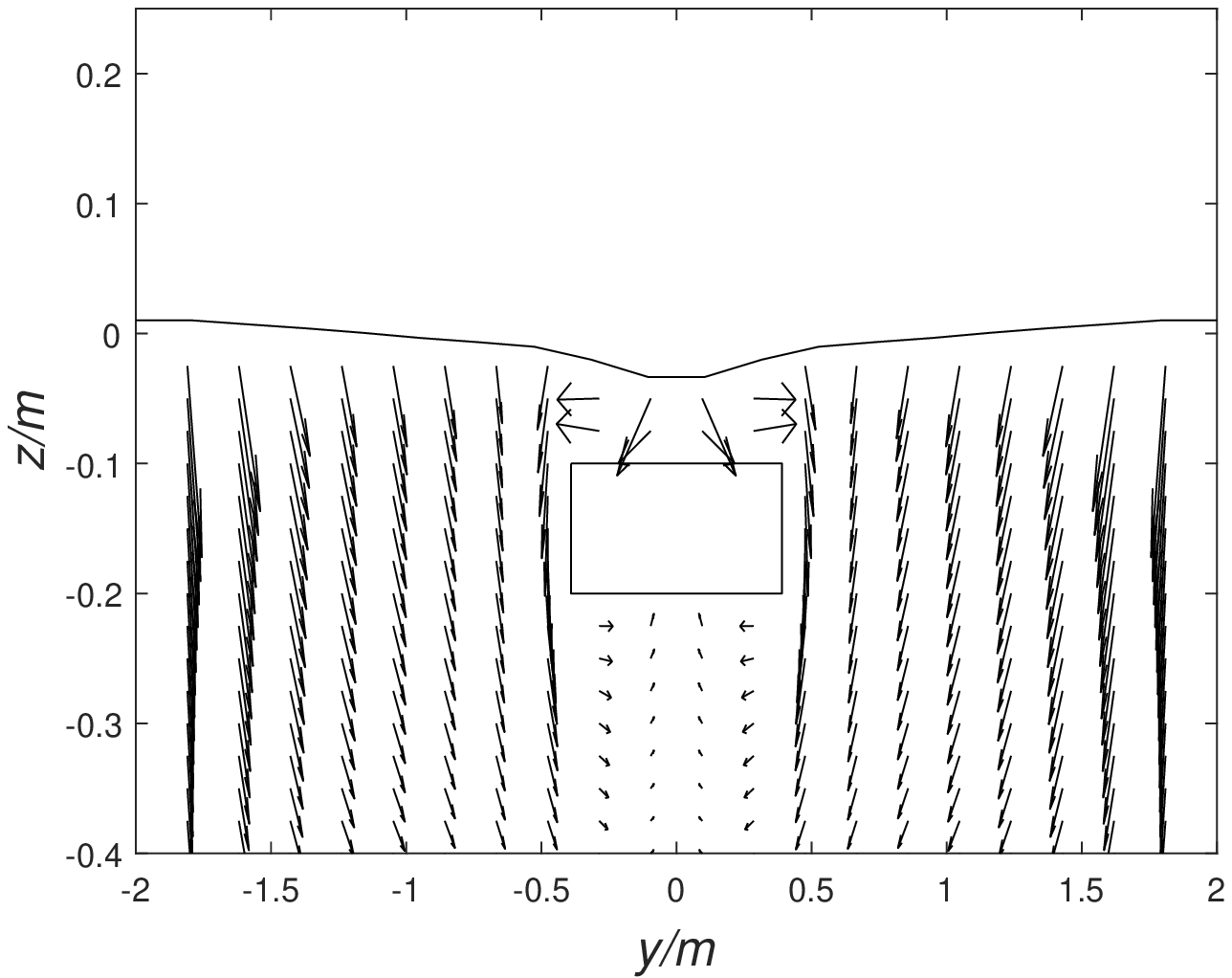}
\includegraphics[scale=0.3]{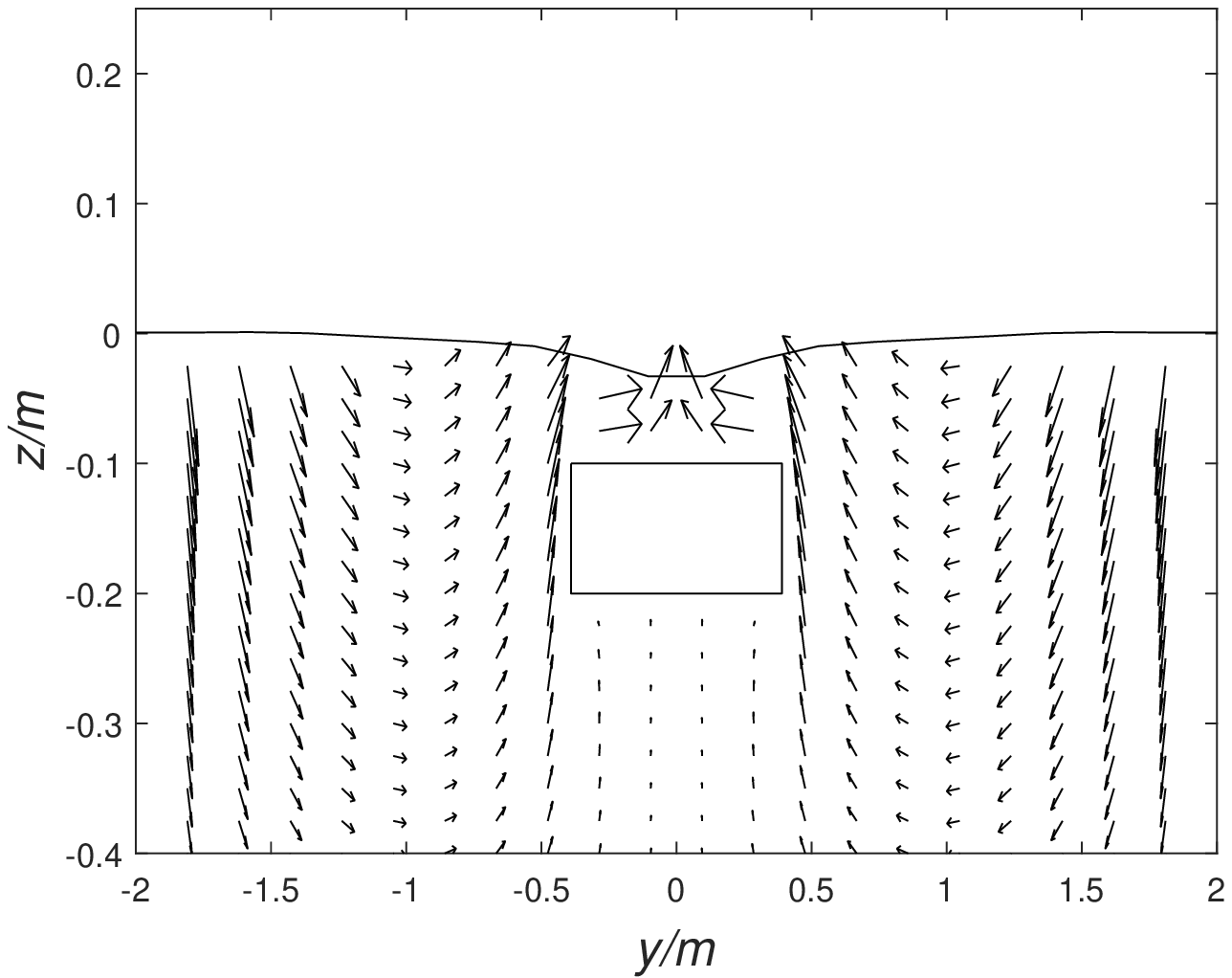}
\includegraphics[scale=0.3]{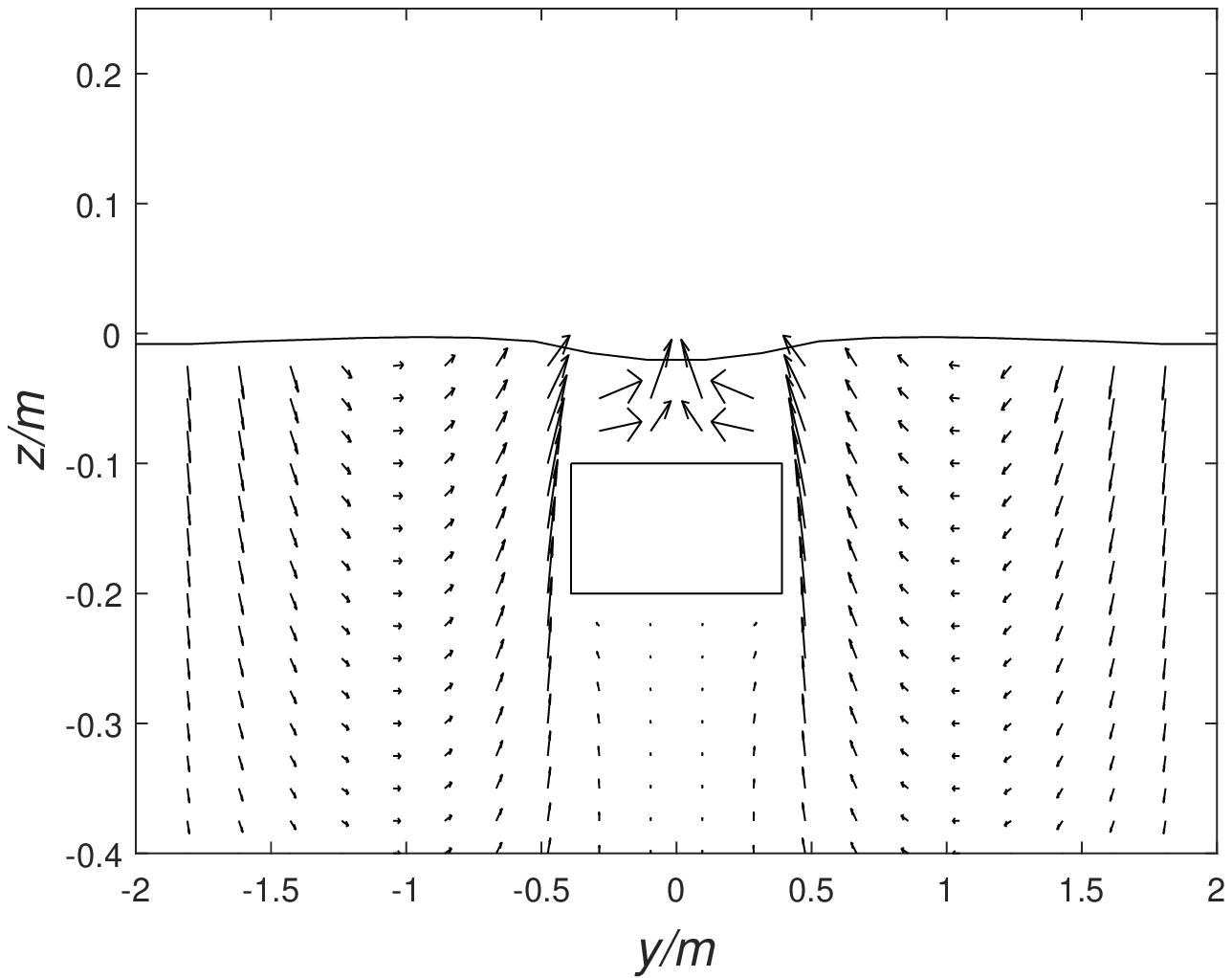}
\includegraphics[scale=0.3]{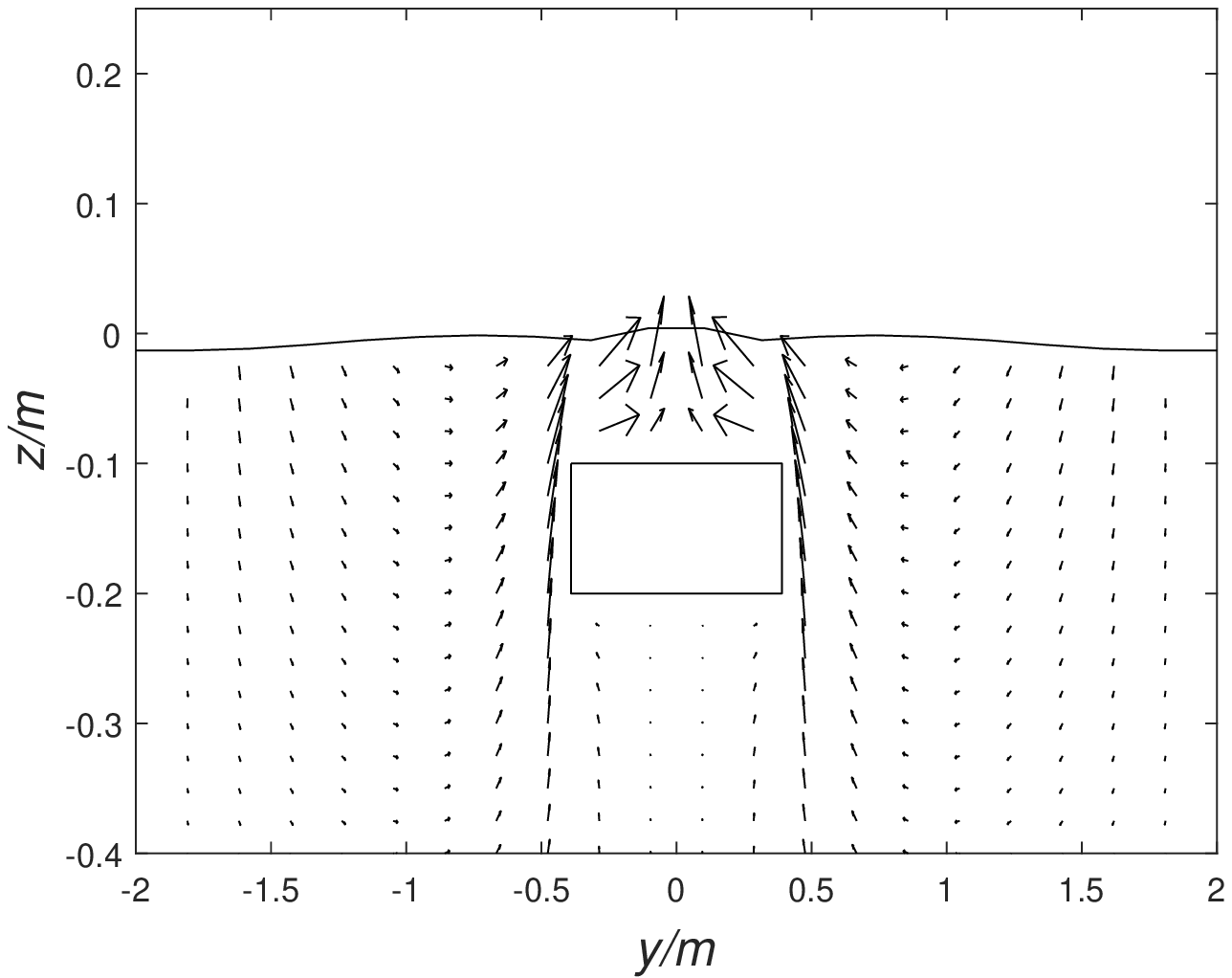}
\includegraphics[scale=0.3]{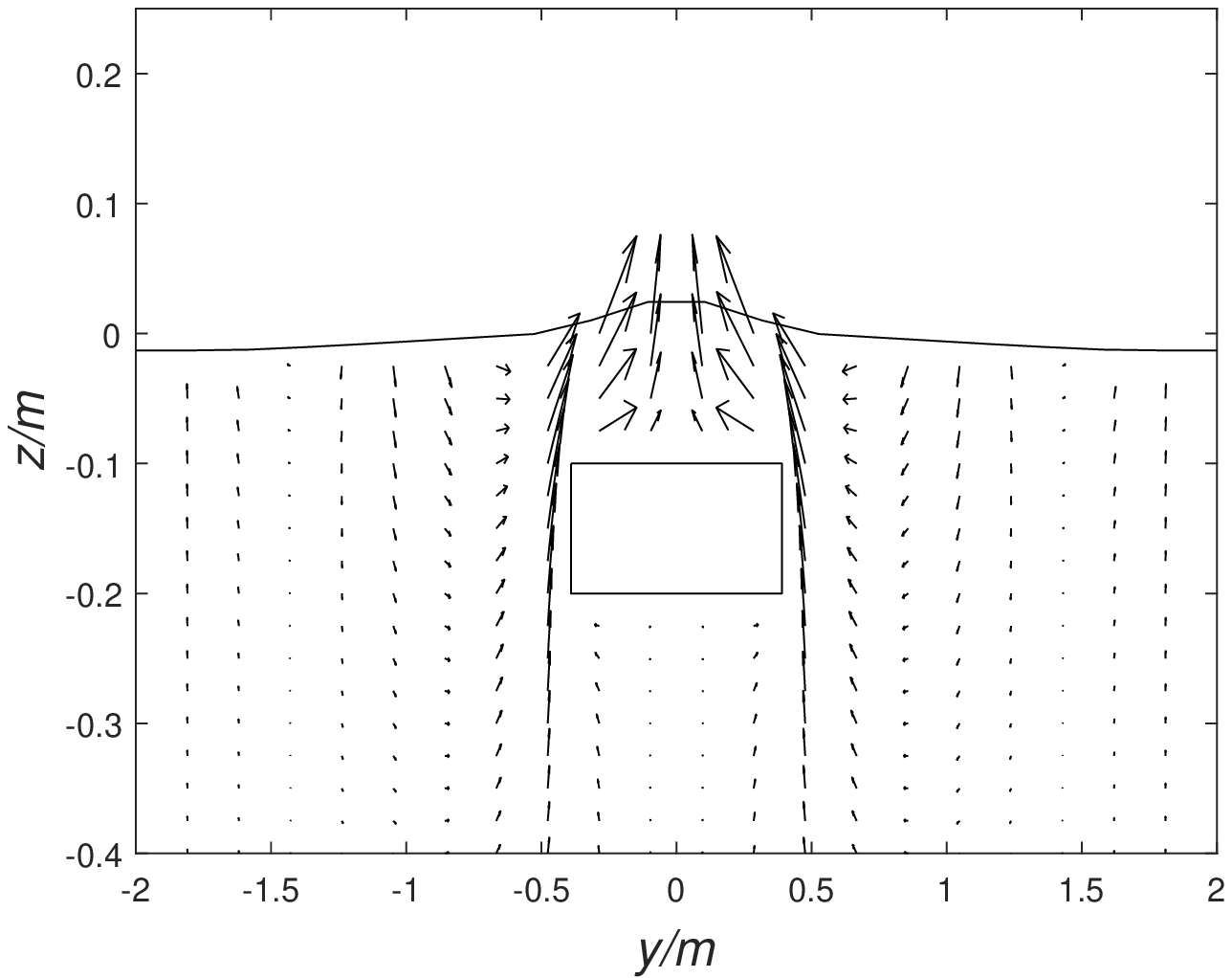}
\includegraphics[scale=0.3]{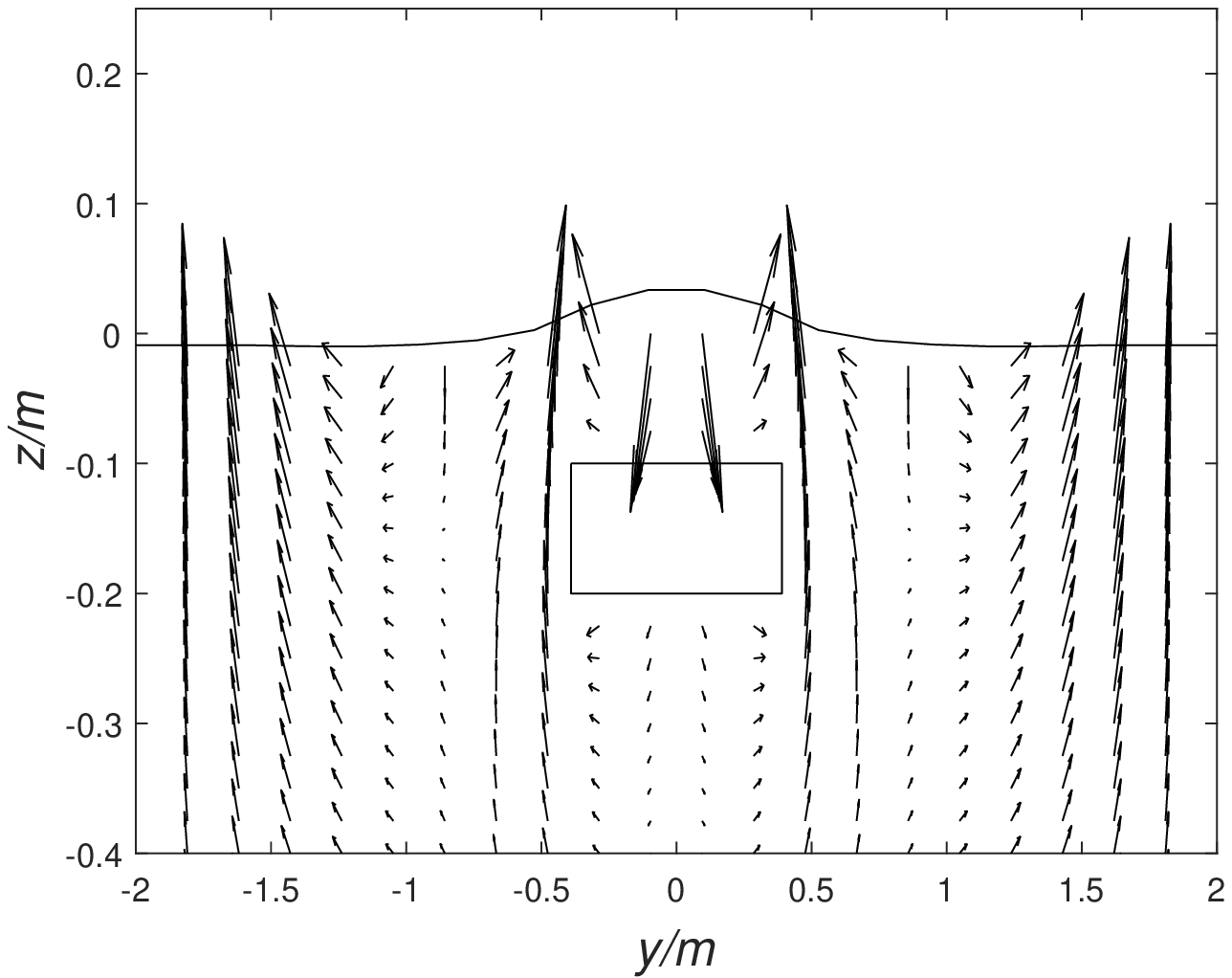}
\includegraphics[scale=0.3]{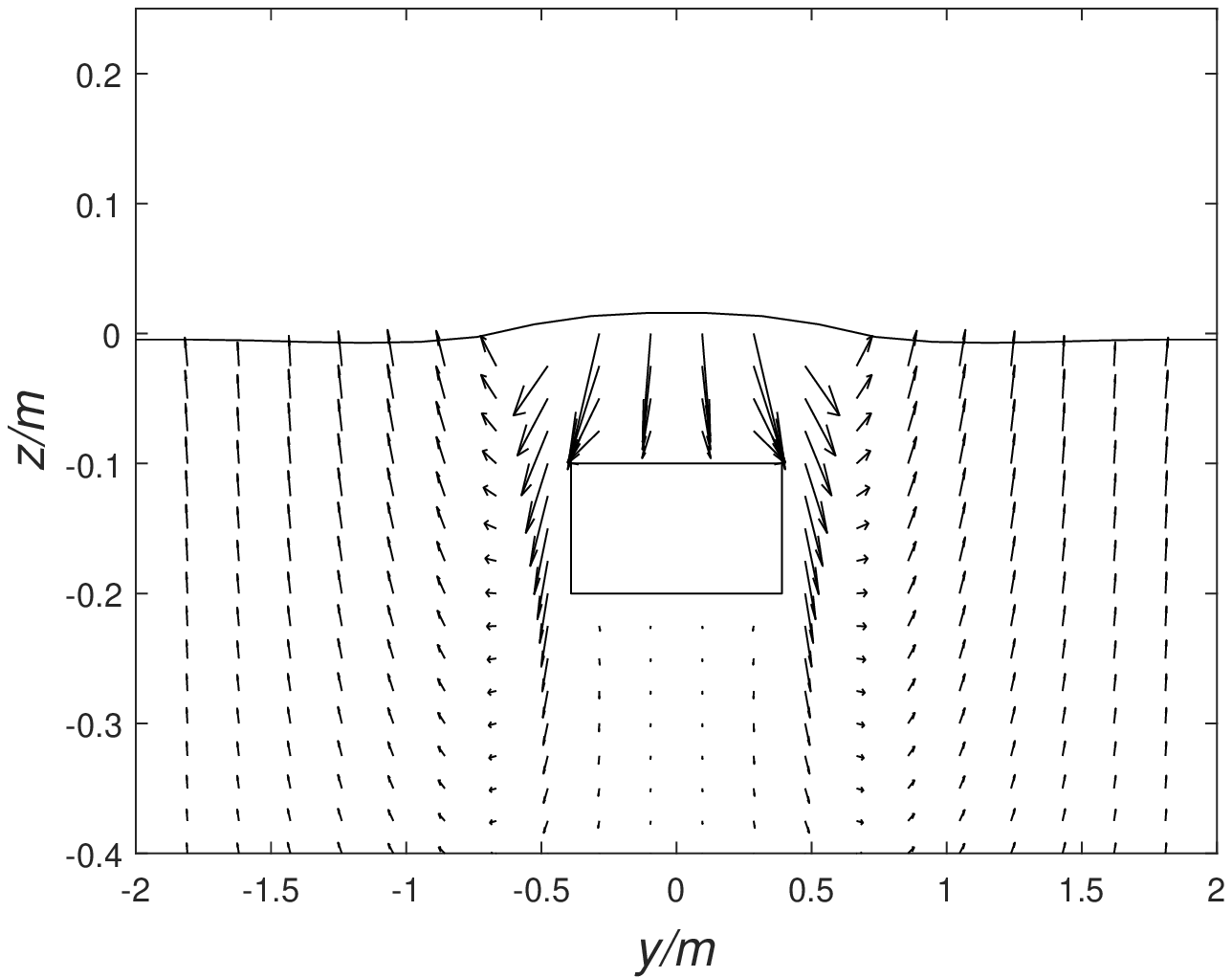}
\includegraphics[scale=0.3]{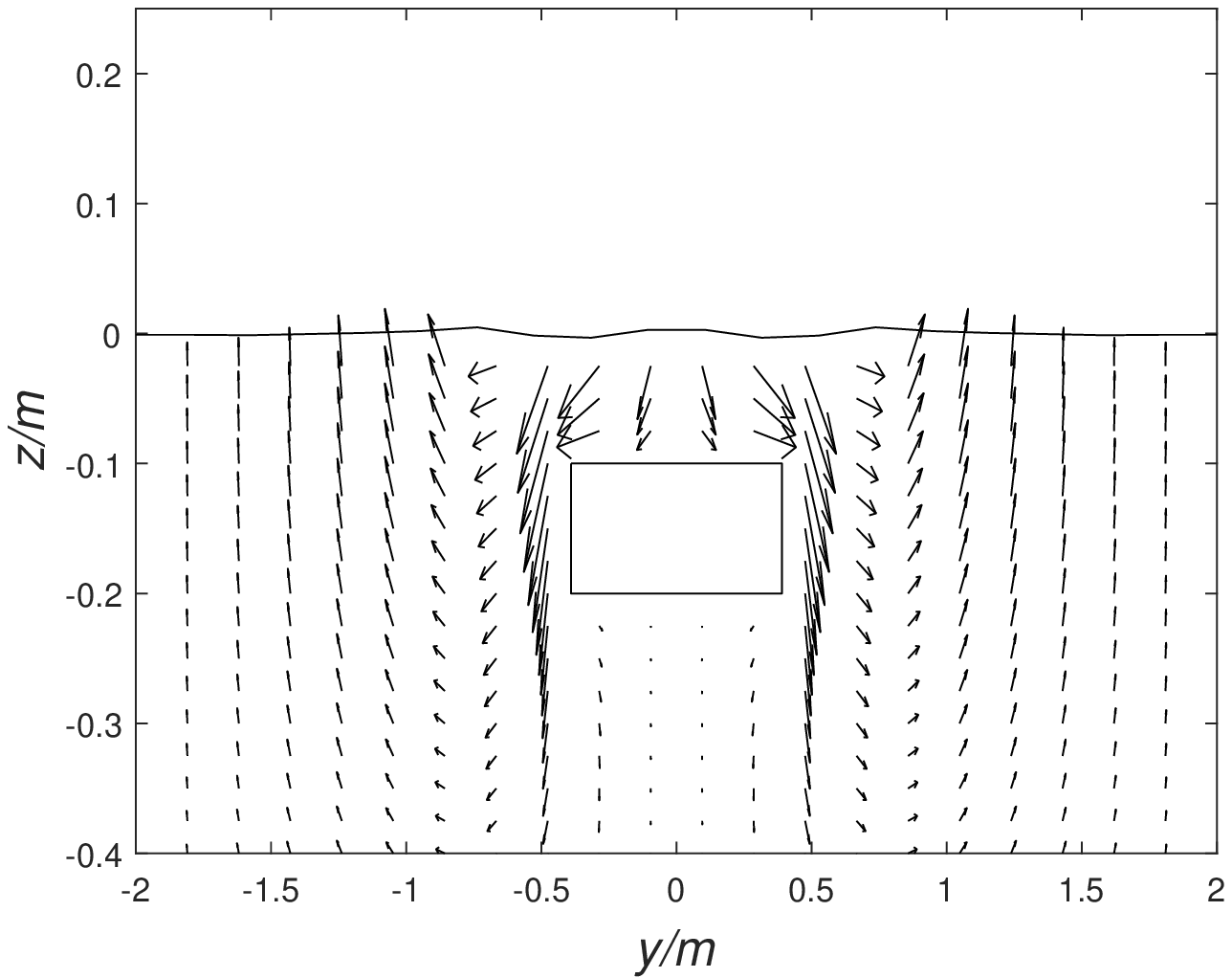}
\includegraphics[scale=0.3]{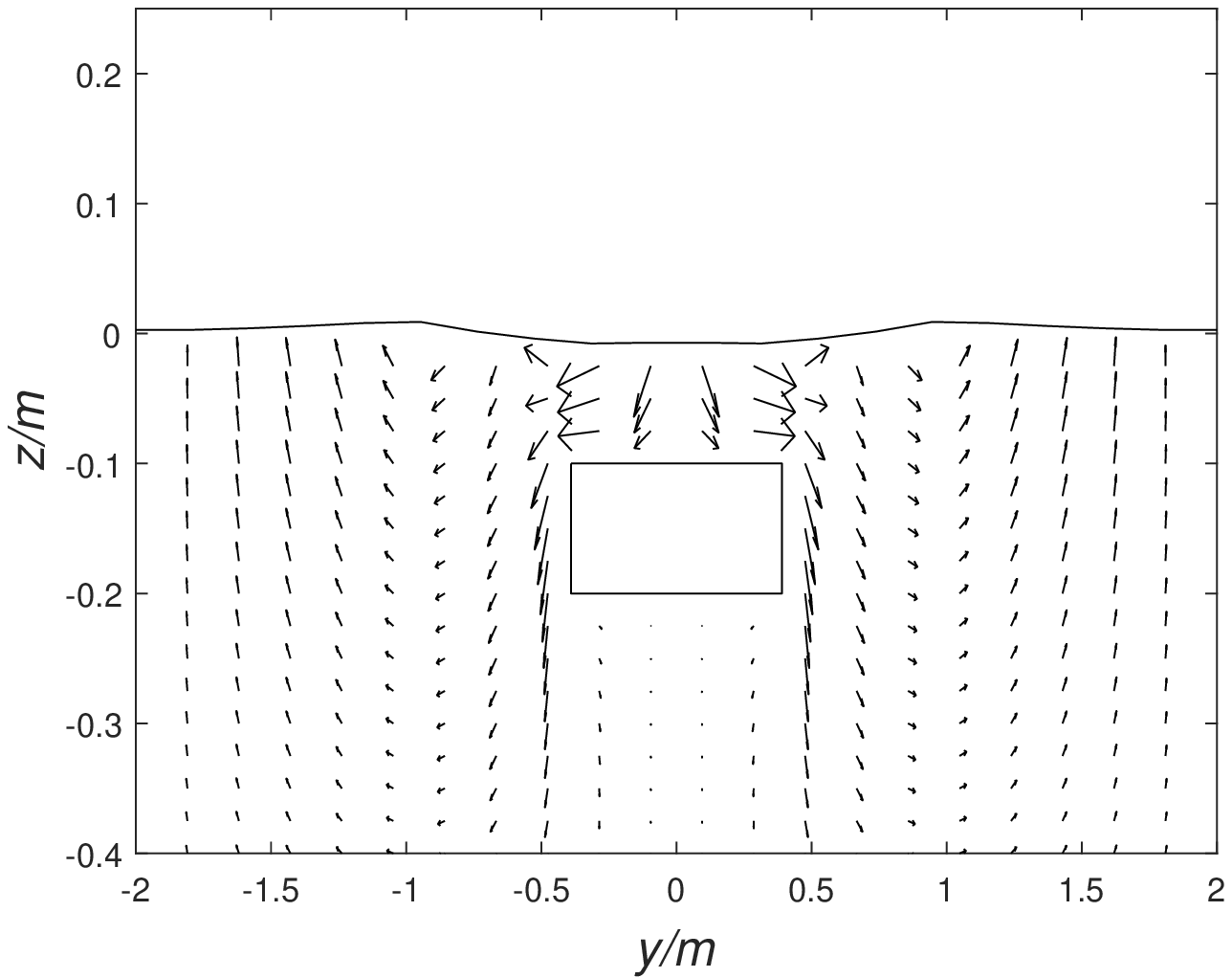}
\caption{Velocity field in the transverse middle section (from 10 s to 13.4 s -- the interval is 0.2 s) when $B=40$ cm, $h=60$ cm and $H/h=0.3$.}
\label{fig_transmid}
\end{figure*}

Fig \ref{fig_transmid} shows the velocity field in the transverse middle section. The transverse flow along the plate and the reflection from the lateral wall can be observed. Combined with what we find in Fig \ref{fig_velmid}, the focused wave is caused by multiple factors, including the reflection from the upper surface of the plate, shoaling and the transverse sloshing mode.

\section{Influence of the thickness of the plate}

Having shown the validity of the numerical method, several purely numerical experiments are conducted to study the effects of the thickness of the plate. As shown in Fig \ref{fig_sideview}, once we change the thickness we may also change $G$ and $B$, which are the submerged depth and the distance between the plate and the bottom, respectively. In the following discussion, we decided to fix the water depth once for all to $h=60$ cm and the dimensionless wave height to $H/h=0.3$. We first take $G=20$ cm and consider three different cases for the thickness: $\delta=10$ cm, $\delta=20$ cm and $\delta=30$ cm.

\begin{figure}
\centering
\includegraphics[width=0.45\columnwidth]{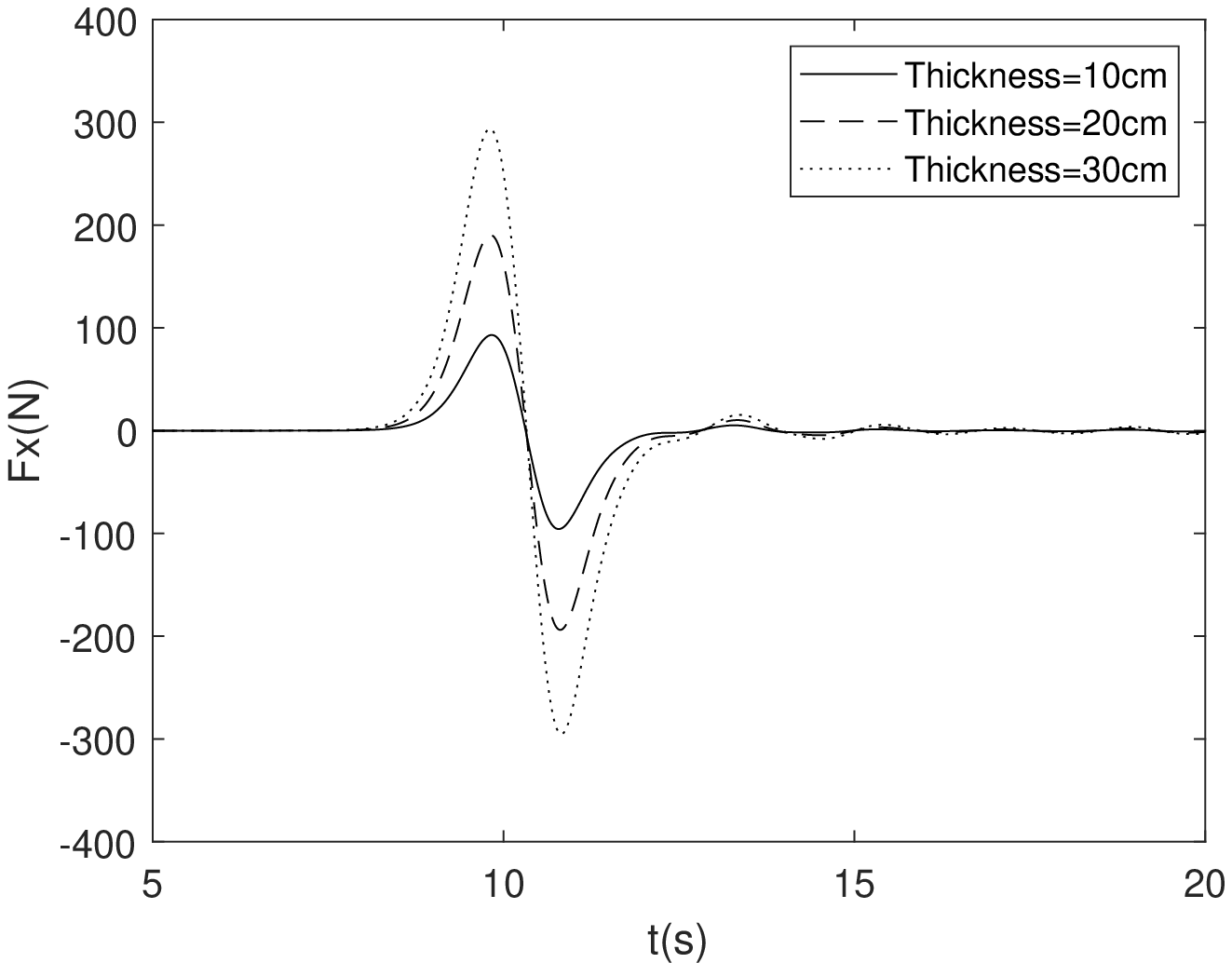} 
\includegraphics[width=0.45\columnwidth]{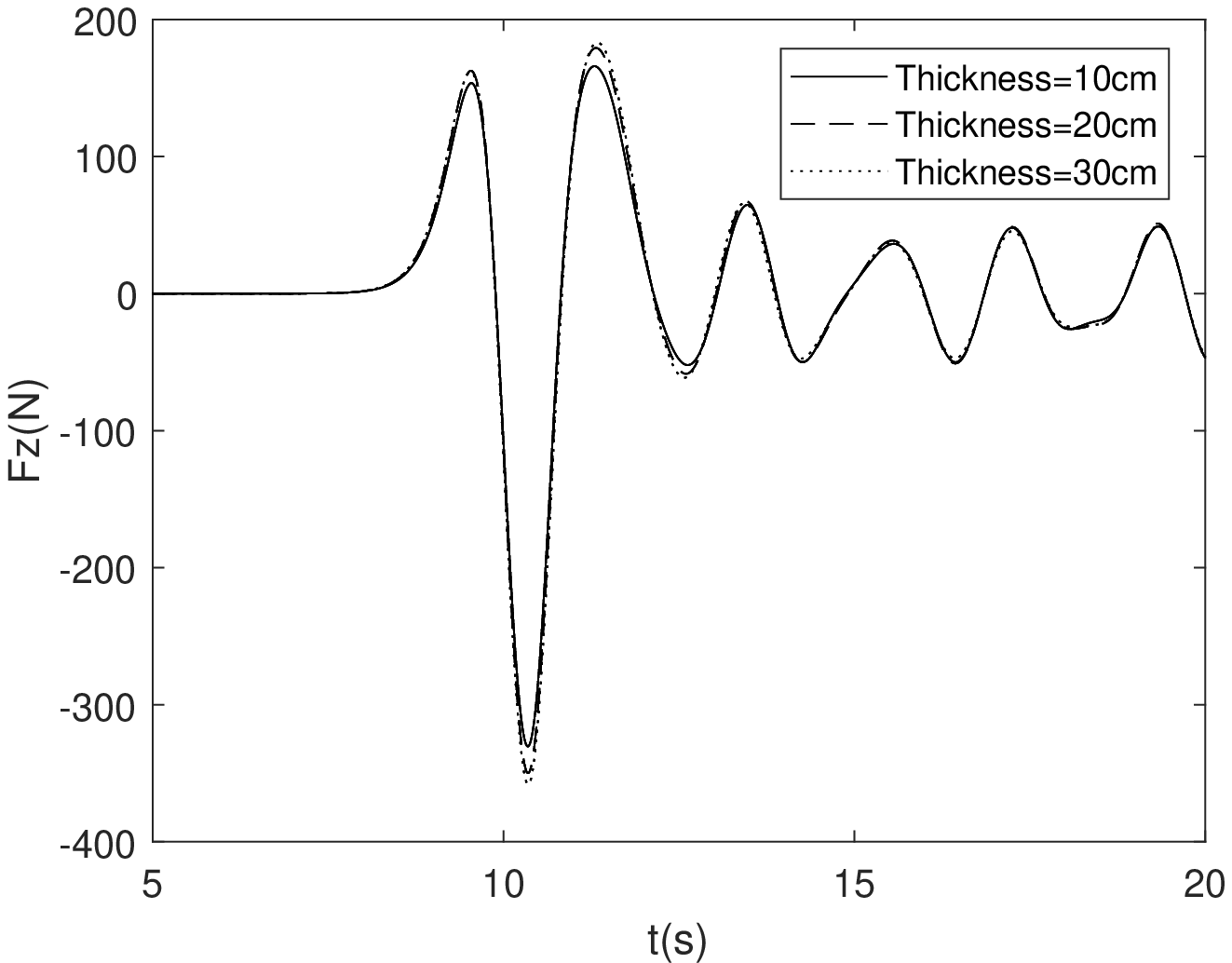}
\caption{Horizontal and vertical force with three different thicknesses when $G=20$ cm, $h=60$ cm and $H/h=0.3$.}
\label{fig_G20f}
\end{figure}

\begin{figure}
\centering
\includegraphics[width=0.45\columnwidth]{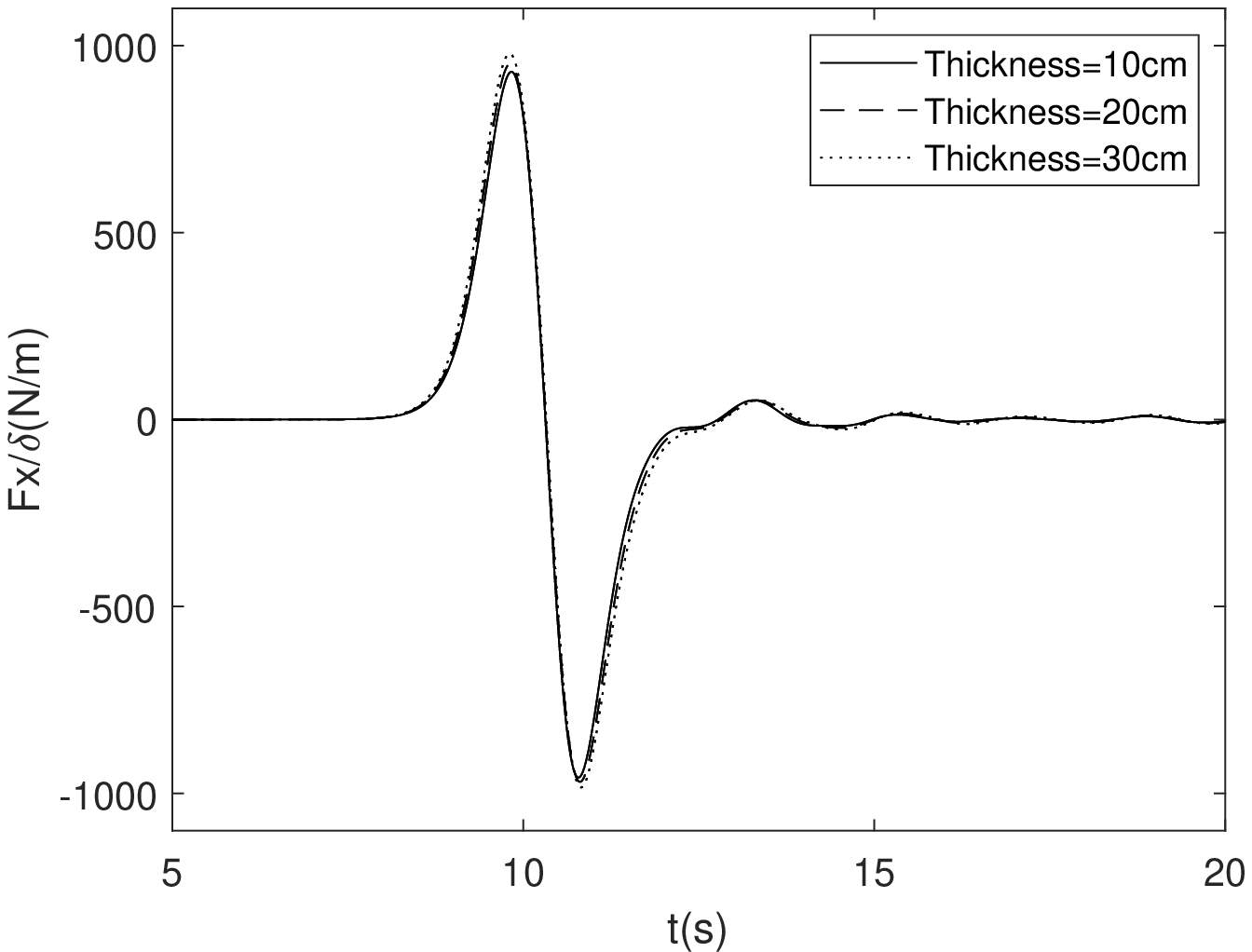}
\caption{Horizontal force divided by the thickness with three different thicknesses when $G=20$ cm, $h=60$ cm and $H/h=0.3$.}
\label{fig_G20ndfx}
\end{figure}

The horizontal and vertical forces are shown in Fig \ref{fig_G20f}. The vertical force only increases slightly with the thickness. The horizontal force increases with the thickness, which is not surprising. We can normalize the horizontal force with the thickness as shown in Fig \ref{fig_G20ndfx}. We also checked that the free-surface elevations at three wave gauges along the middle line are the same within graphical accuracy. Therefore we can conclude that under the same submerged depth $G$, the thickness of the plate has negligible effects on both the hydrodynamic loads and the free-surface elevation around the plate. For the latter one, Lo and Liu \cite{Lo2013} have shown that in the two-dimensional case varying the thickness changes the shapes of both the reflected and transmitted waves.

\begin{figure}
\centering
\includegraphics[width=0.45\columnwidth]{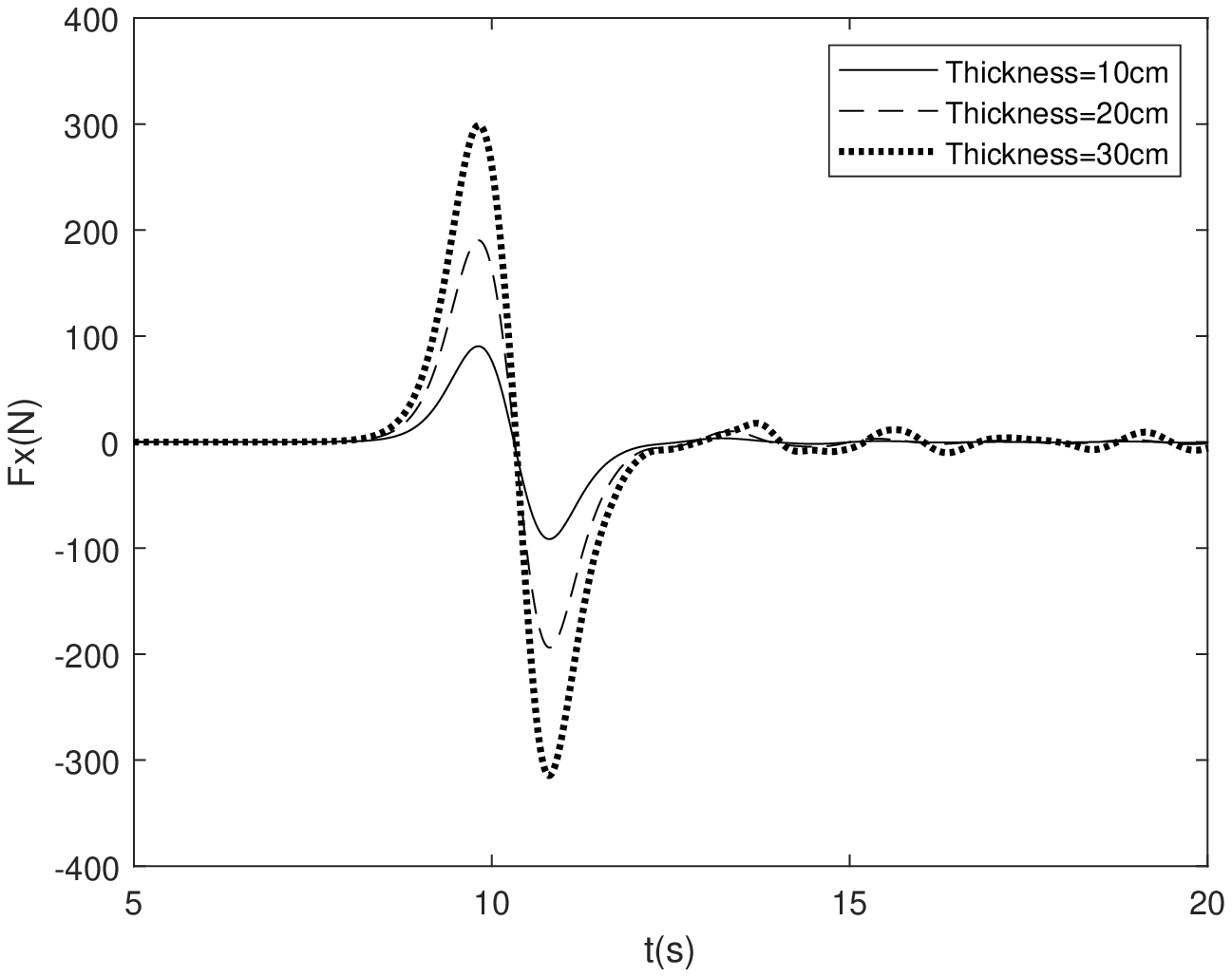} 
\includegraphics[width=0.45\columnwidth]{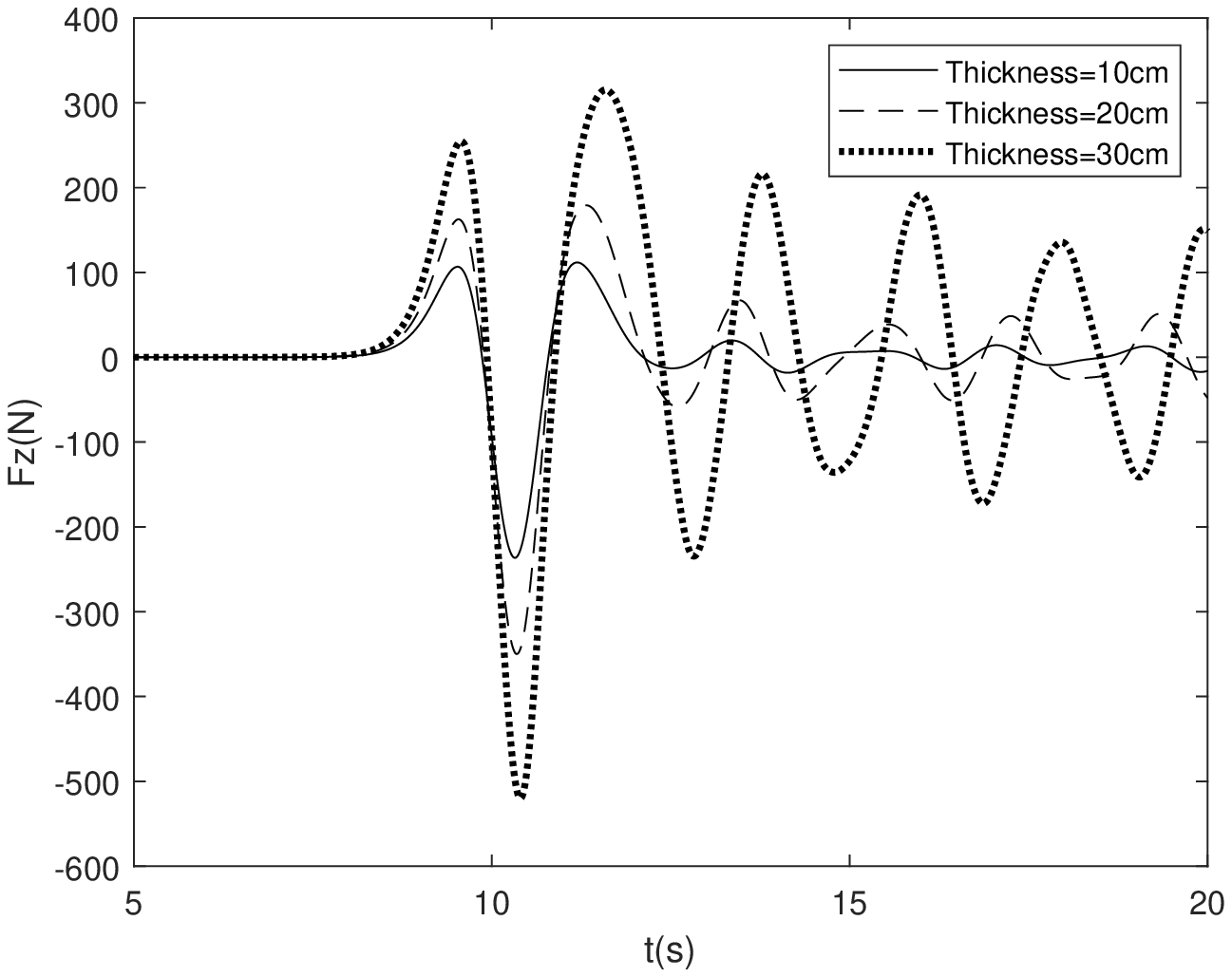} \\
\includegraphics[width=0.45\columnwidth]{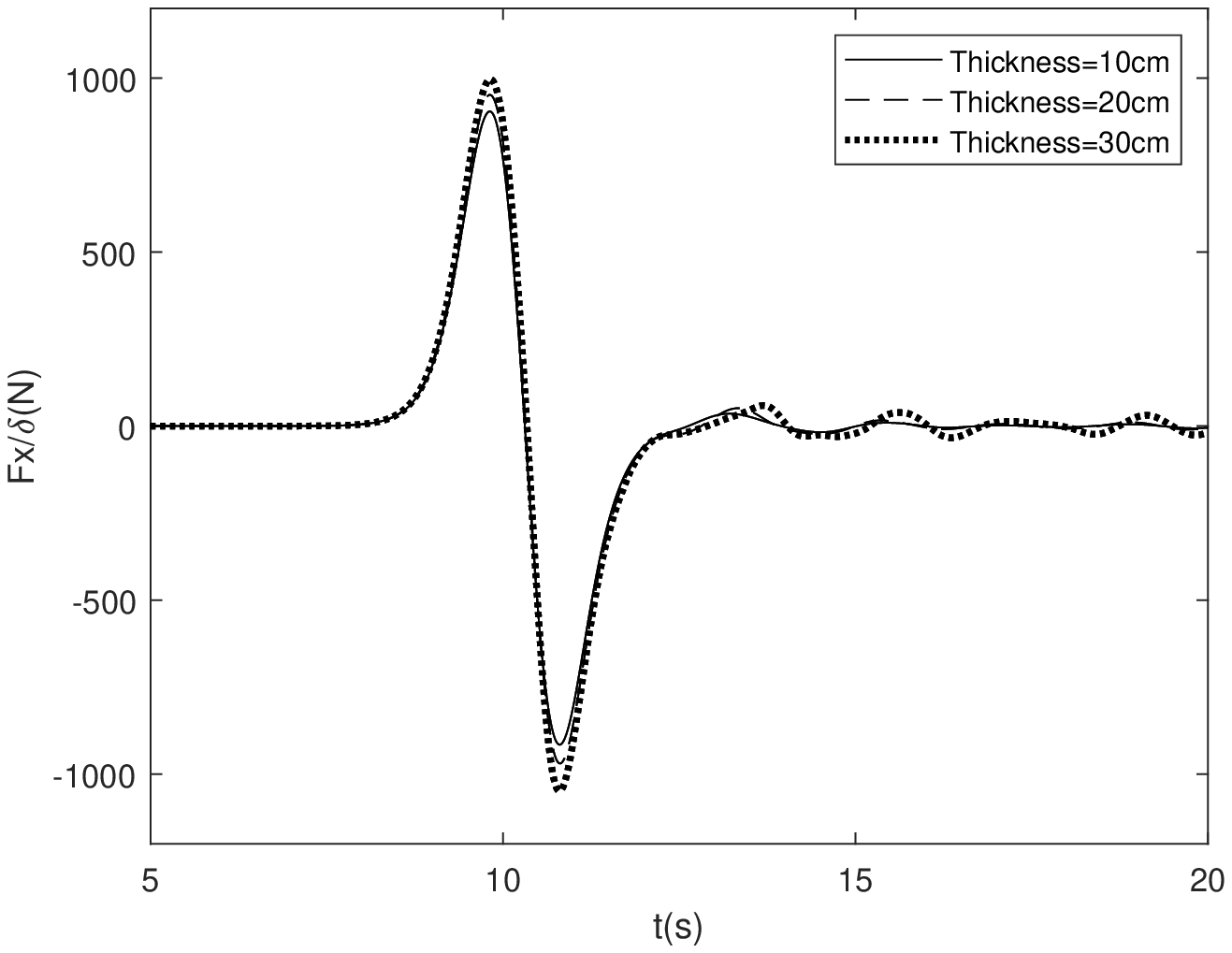}
\caption{Horizontal, vertical and normalized horizontal forces with three different thicknesses when $B=20$ cm, $h=60$ cm and $H/h=0.3$.}
\label{fig_B20f}
\end{figure}

\begin{figure}
\centering
\includegraphics[width=0.45\columnwidth]{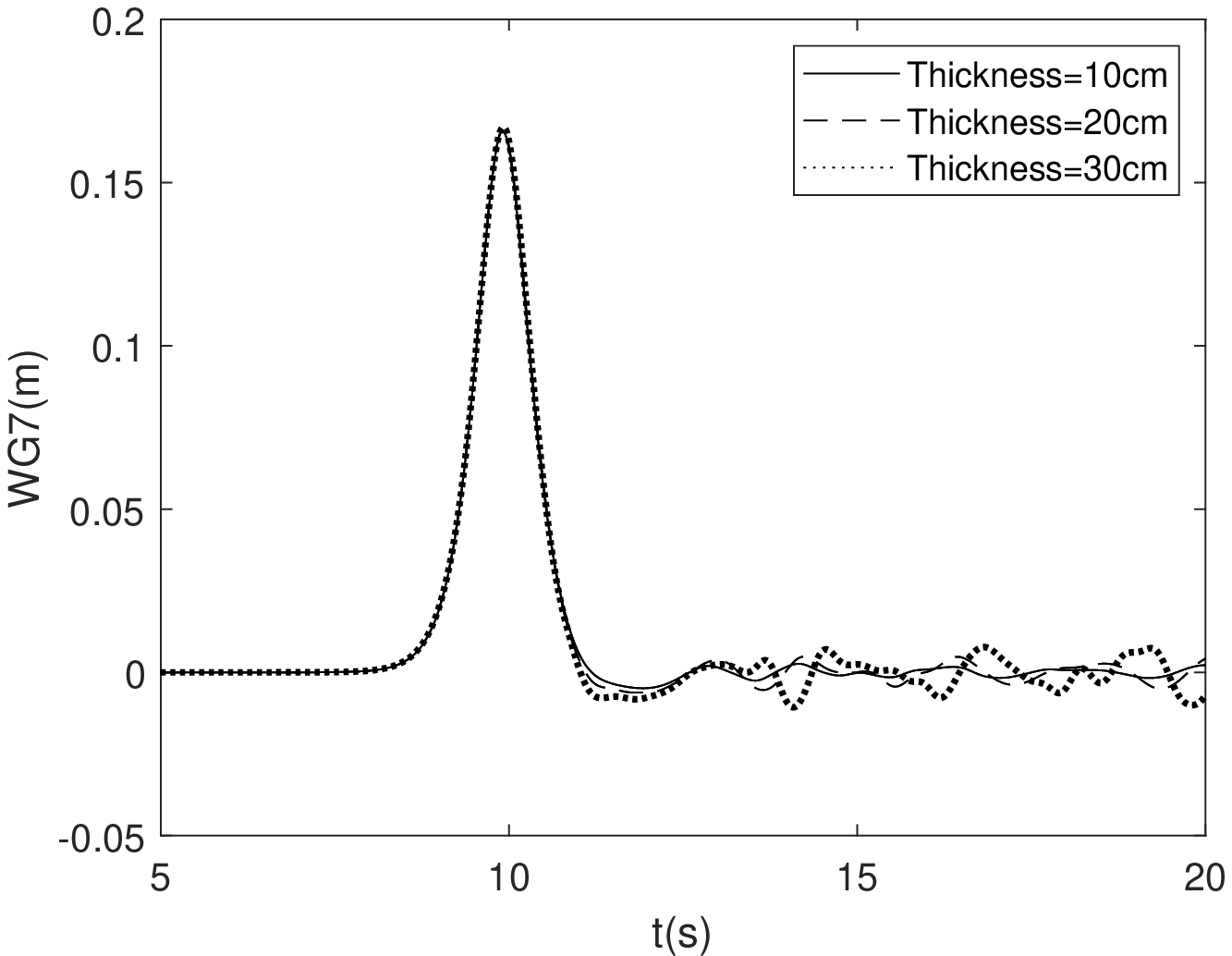} 
\includegraphics[width=0.45\columnwidth]{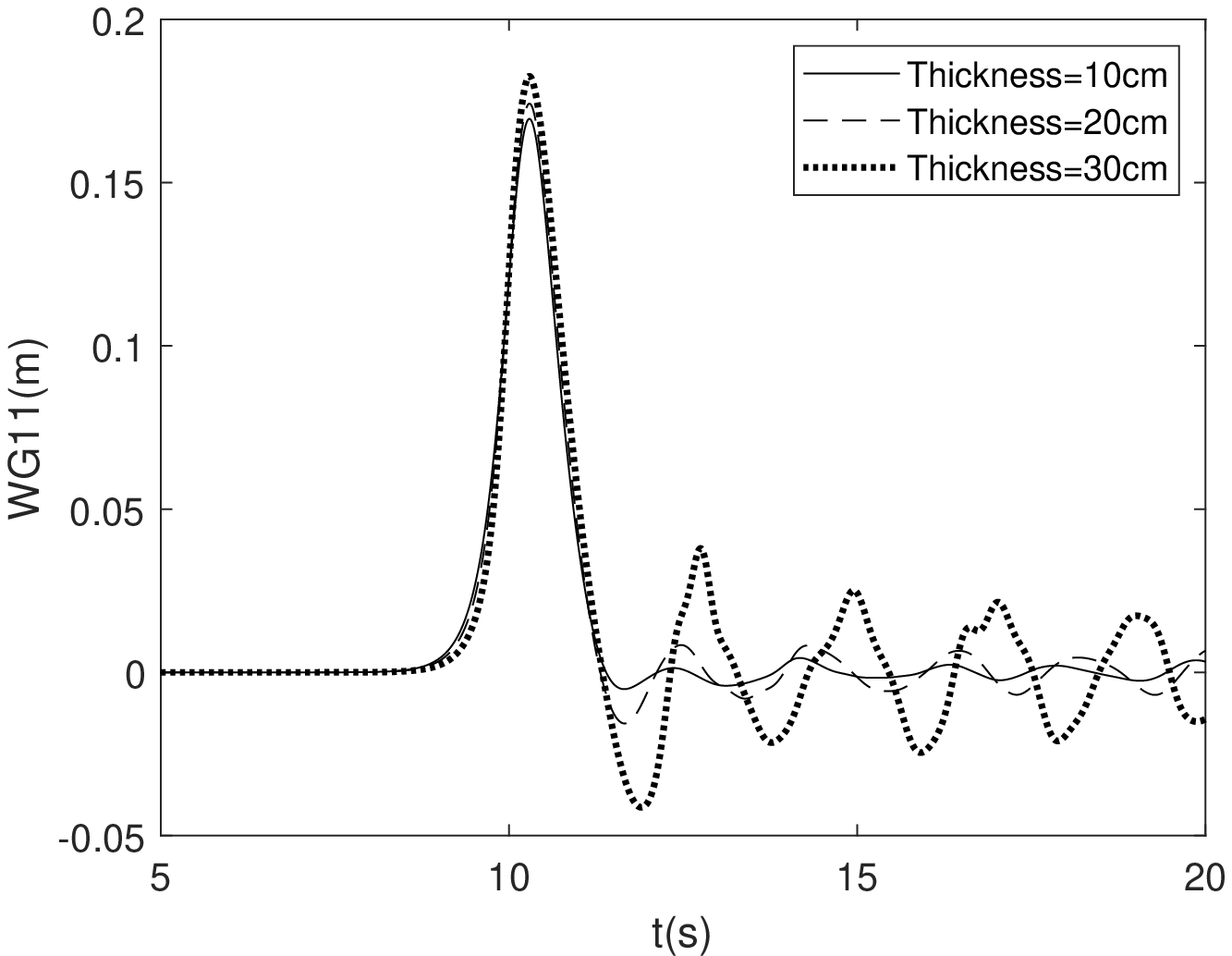} \\
\includegraphics[width=0.45\columnwidth]{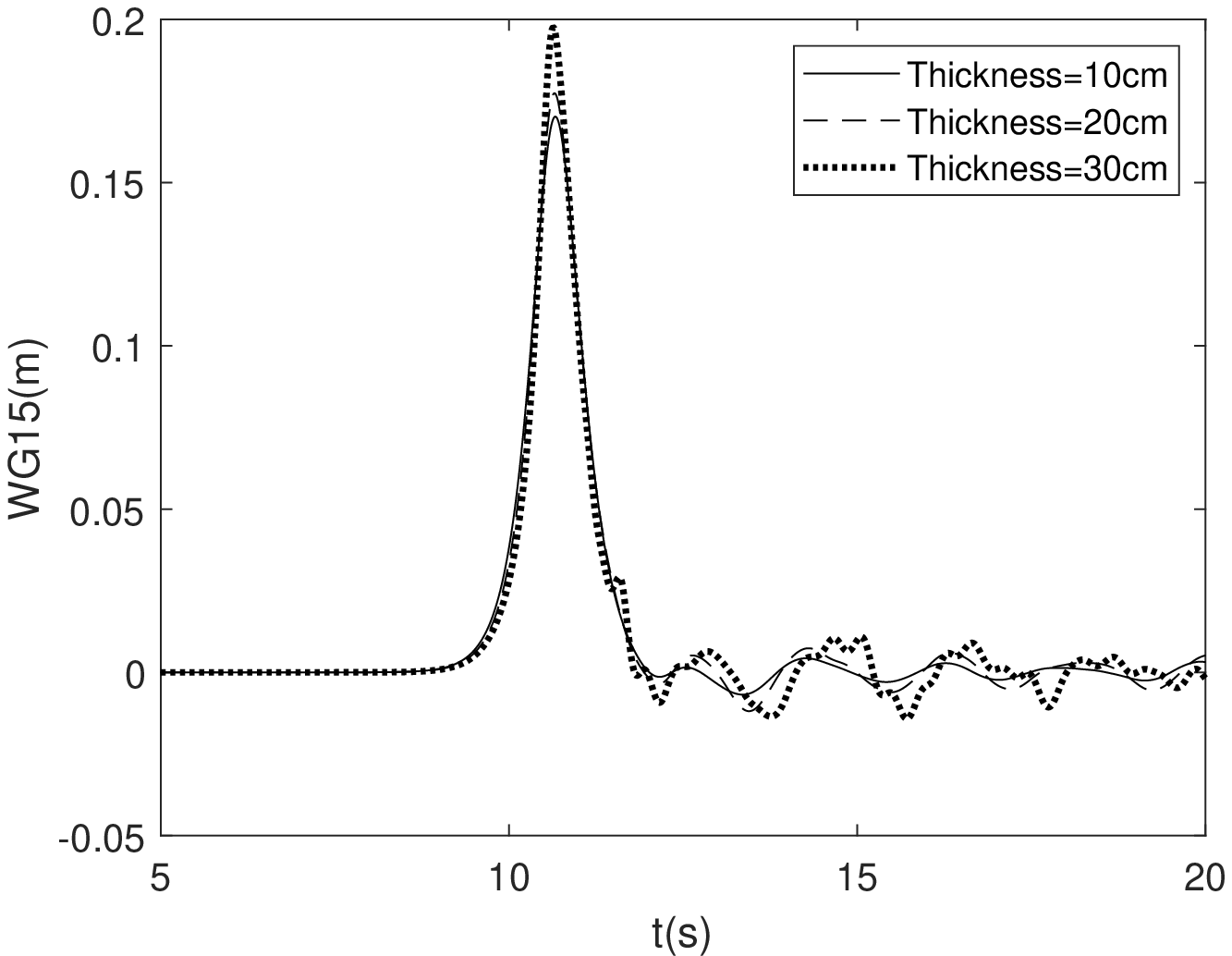}
\caption{Wave elevation at three wave gauges along the middle line with three different thicknesses when $B=20$ cm, $h=60$ cm and $H/h=0.3$.}
\label{fig_B20WG}
\end{figure}

After discussing the effects of the thickness with fixed $G$, we now fix $B=20$ cm and change the thickness in three different cases: $\delta=10$ cm, $\delta=20$ cm and $\delta=30$ cm. In other words the corresponding depths of submergence are $G=30$ cm, $G=20$ cm and $G=10$ cm. The horizontal, vertical and normalized horizontal forces are shown in Fig \ref{fig_B20f}. The wave elevations at three selected wave gauges are shown in Fig \ref{fig_B20WG}. They are substantially different in the three cases. To check whether or not the differences are caused by the difference in thickness or the difference in depth of submergence, we computed the vertical force with $G=10$ cm, $B=40$ cm, $\delta=10$ cm and $G=10$ cm, $B=20$ cm, $\delta=30$ cm. We found that the main feature of the vertical force is maintained with different thicknesses under the same depth of submergence. Therefore, the characteristics of the hydrodynamic loads and wave elevations around the plate are dominated by the depth of submergence.

\section{Concluding remarks}

The interaction between a solitary wave and a fully submerged three-dimensional horizontal plate is investigated numerically and the results are compared with those of three-dimensional laboratory experiments described in \cite{Wang2020}. The wave focusing phenomenon that was observed experimentally is also observed numerically. The free-surface elevation increases gradually along the center line of the plate and exceeds the incoming wave amplitude. The presence of the three-dimensional plate modifies the shape of the solitary wave. A wave amplification is produced by the local shoaling effect of the plate and the shoaling-induced wave refraction. A larger amplitude of the solitary wave leads to a stronger focusing up to the end of the plate.

The horizontal wave force is characterized by a peak followed by a trough. The vertical wave force is characterized by a series of peaks and troughs. The loading process is described based on the peaks of the vertical force and the pitching moment. The process is linked with the oscillations of the free-surface elevation. When the wave approaches or leaves the plate, the channel flow under the plate contributes to the positive peaks of the vertical force. As the wave crest approaches the center of the plate, the negative force caused by the dynamic pressure on the top side of the plate dominates. The strong focusing of large-amplitude waves reduces the second positive peak. The pitching moment is mainly generated by the vertical force, and the observation of surface elevation indicates that the time for the peak of pitching moment depends on the occurrence of the maximum surface elevation.

Then the velocity field is illustrated. Unfortunately, there are no experimental results to compare it with. Finally, the influence of the plate thickness is found to be negligible. 
 
In the introduction, we asked the question: can the problem of a solitary wave impacting on a submerged horizontal plate only be solved by Computational Fluid Dynamics (CFD) or can fully nonlinear potential flow theory still be applied to this problem? In the experiments of Wang \textit{et al}. \cite{Wang2020} that we used for the comparisons, the Reynolds number was of the order of $10^{5}$. Overall, the potential flow model gives good results. However, there are some discrepancies, especially for the pitching moment. In the future, it is suggested to compare the experimental results also with a Navier-Stokes solver, especially when breaking waves have been observed in the experiments. The present problem could become an excellent benchmark to study the performance of codes that can handle wave breaking.

\section*{Supplementary material}
Results for all the cases are available in the supplementary material.

\begin{acknowledgments}
This work was funded by the China Scholarship Council (first author), by the National Natural Science Foundation of China (Grant No. 11632012 and Grant No. 41861144024) (second author) and by the European Research Council (ERC) under the European Union’s Horizon 2020 research and innovation programme (Grant agreement No. 833125 - HIGHWAVE) (third author).
\end{acknowledgments}

\section*{Data availability}
The data that support the findings of this study are available from the corresponding author upon reasonable
request.


\end{document}